\documentclass[a4paper,11pt]{article}
\pdfoutput=1 

\usepackage{jcappub} 

\newcommand\Mpch{\, \mathrm{Mpc}/h}
\newcommand\Mpchc{\, (\mathrm{Mpc}/h)^{3}}
\newcommand\hMpc{\, h/\mathrm{Mpc}}
\newcommand\eV{\, \mathrm{eV}}
\newcommand\FW{\tilde{W}}
\newcommand\window{\mathcal{W}}
\newcommand\vc[1]{\mathbf{#1}}
\newcommand\vk{\vc{k}}
\newcommand\vr{\vc{r}}
\newcommand\rn[1]{r_{#1}}
\newcommand\vrn[1]{\vc{r}_{#1}}
\newcommand\vs{\vc{s}}
\newcommand\vd{\vc{d}}
\newcommand\vD{\vc{\Delta}}
\newcommand\vx{\vc{x}}
\newcommand\vy{\vc{y}}

\newcommand\loss{\hat{\vc{\eta}}_{\vs}}
\newcommand\losD{\hat{\vc{\eta}}_{\vD}}
\newcommand\losrn[1]{\hat{\vc{r}}_{#1}}
\newcommand\Os{\Omega_{s}}
\newcommand\OD{\Omega_{\Delta}}
\newcommand\Ok{\Omega_{k}}
\newcommand\ep{\mathrm{ep}}
\newcommand\ic{\mathrm{ic}}
\newcommand\cic{\mathrm{cic}}
\newcommand\glo{\mathrm{glo}}
\newcommand\rad{\mathrm{rad}}
\newcommand\ang{\mathrm{ang}}
\newcommand\aver[1]{\langle#1\rangle}
\newcommand\abs[1]{\vert#1\vert}
\newcommand\Leg[1]{\mathcal{L}_{#1}}
\newcommand\Diracn[1]{\delta^{#1D}}
\newcommand\apar{\alpha_{\parallel}}
\newcommand\aper{\alpha_{\perp}}

\newcommand\Eq[1]{eq.~\eqref{eq:#1}}
\newcommand\Fig[1]{figure~\ref{fig:#1}}
\newcommand\Tab[1]{table~\ref{tab:#1}}
\newcommand\Sec[1]{section~\ref{sec:#1}}
\newcommand\App[1]{appendix~\ref{app:#1}}

\title{\boldmath Integral constraints in spectroscopic surveys}

\author{Arnaud de Mattia}
\author{and Vanina Ruhlmann-Kleider}
\affiliation{IRFU, CEA, Universit\'e Paris-Saclay, F-91191 Gif-sur-Yvette, France}

\emailAdd{arnaud.de-mattia@cea.fr}
\emailAdd{vanina.ruhlmann-kleider@cea.fr}

\abstract{Clustering analyses of spectroscopic surveys are based upon density fluctuations, which are estimated by comparing the observed tracer density field to a selection function accounting for the survey density and geometry. However, this survey selection function is commonly partly inferred from the observed data itself, leading to so-called integral constraints, for which we propose a complete derivation. We discuss the normalisation of the introduced window functions, the shot noise contribution to the integral constraint corrections and wide-angle effects. Using this formalism, we review the well-known global integral constraint, arising when the expected mean galaxy density is taken to be the measured one. Another, stronger, constraint is imposed when the radial selection function itself is estimated from the data redshift distribution, as is often the case in the literature. We find that the impact of such a radial integral constraint can be as significant as the window function effect at large scales, depending on the survey geometry. Equations for this radial integral constraint are derived within our general formalism.
We assess the validity of our approach by performing a Redshift Space Distortions (RSD) analysis on mock catalogues and emphasise that our results may be even more useful for analyses focusing on larger scales. Finally, as a further application, we show that unknown angular systematics can be mitigated by nulling the density fluctuations on a chosen angular scale. The induced loss of clustering is modelled by an angular integral constraint which can be combined with the radial one.}

\begin{document}
\maketitle
\flushbottom

\section{Introduction}

Spectroscopic surveys measure redshift-space positions of galaxies (or any other tracer of the matter density field). These surveys are not exhaustive and are characterised by a selection function $W(\vr)$, giving the expected density of observed galaxies at any redshift-space position $\vr$ in the absence of clustering. In case one has a full knowledge of this selection function, density fluctuations can be fairly estimated by the difference of the observed density of galaxies to $W(\vr)$. In particular, density fluctuations averaged on the whole survey footprint can be non-zero, due to large-scale clustering modes.

However, the true survey selection function is difficult to determine in practice. Its norm, i.e. the expected mean density of galaxies, may be unknown. It is commonly taken to be the actually observed mean data density. Thus, the integral of the inferred density fluctuations over the whole survey footprint is set to zero, leading to a so-called integral constraint (IC)~\cite{Peacock1991:10.1093/mnras/253.2.307,Wilson2015:1511.07799v2}, referred to as global IC in the following. Similarly, the radial survey selection function may be estimated from the data itself, as is currently done in the clustering analyses of the Baryon Oscillation Spectroscopic Survey (BOSS) (e.g.~\cite{Gil-Marin:1509.06386v2,Beutler2016:1607.03150v1,Alam:1607.03155v1}) and its extended program eBOSS (e.g.~\cite{Ata2017:1705.06373v2,Zarrouk2018:1801.03062v1}). We propose to treat this effect as another integral constraint, dubbed radial IC.

Section~\ref{sec:ic_general} introduces the general formalism we develop to calculate these integral constraints. The computation of window functions required by our derivation is discussed in \Sec{computing_window}.

We detail a Redshift Space Distortions (RSD) analysis of realistic BOSS-like mocks in \Sec{analysis_radial}, showing that inferring the radial selection function from data can significantly alter clustering measurements, and that the induced bias is completely removed when using our modelling of the radial integral constraint.

As another application of our formalism, we suggest in~\Sec{analysis_angular} to mitigate potential unknown angular systematics by imposing an angular integral constraint, which can be combined with the radial one.

We conclude in \Sec{conclusions}.

\section{Modelling integral constraints}
\label{sec:ic_general}

We detail the impact of global and radial integral constraints on the observed density fluctuations in sections~\ref{sec:delta_global} and~\ref{sec:delta_radial}, showing that both can be accounted for by a general formalism, which we use to derive the effect of integral constraints on 2-point statistics in \Sec{ic_correlation}. We discuss and compare our findings to previous results in \Sec{ic_discussion}.

\subsection{Impact of the global integral constraint on the observed density fluctuations}
\label{sec:delta_global}

Predicting the mean galaxy density of a survey is complex in practice. For example, following~\cite{Feldman1993:astro-ph/9304022v1} the Yamamoto power spectrum estimator~\cite{Yamamoto2005:astro-ph/0505115v2} makes use of the FKP field:
\begin{equation}
F(\vr) = n_{g}(\vr) - \alpha n_{s}(\vr)
\label{eq:fkp_field}
\end{equation}
where $n_{g}(\vr)$ and $n_{s}(\vr)$ denote the density of observed and random galaxies, respectively. Random galaxies come from a Poisson-sampled synthetic catalogue accounting for the survey selection function\footnote{We assume throughout the paper that the density of this synthetic catalogue is sufficiently high to avoid systematic bias from undersampling the survey selection function.}. Observed and random densities may include weights, e.g. corrections for systematics effects or a redshift weighting scheme (such as FKP weights, see \App{FKP_weights}). The observed galaxy density is $n_{g}(\vr) = W(\vr) \left\lbrace 1+\delta(\vr) \right\rbrace$, with $W(\vr)$ the survey selection function and $\delta(\vr)$ the density contrast. We recall that $W(\vr)$ describes the expected density of observed, possibly weighted, galaxies in the absence of clustering. We assume the shape of the survey selection function is known, and sampled by the (weighted) synthetic catalogue: $n_{s}(\vr) \varpropto W(\vr)$. The scaling $\alpha$ is defined by:
\begin{equation}
\alpha  = \frac{\int d^{3} x n_{g}(\vx)}{\int d^{3} x n_{s}(\vx)} = \frac{\sum_{i=1}^{N_{g}} w_{g,i}}{\sum_{i=1}^{N_{s}} w_{s,i}}.
\label{eq:alpha_global}
\end{equation}
with $N_{g}$, $N_{s}$ and $w_{g}$, $w_{s}$ the number and weights of observed and random galaxies, respectively. Then, the observed window-convolved\footnote{In configuration space, we actually mean \emph{window-multiplied}.}, integral-constraint-corrected (subscript $\cic$) density fluctuations are~\cite{Peacock1991:10.1093/mnras/253.2.307,Beutler2014:1312.4611v2}:
\begin{equation}
\begin{split}
\delta^{\cic}(\vr) & = n_{g}(\vr) - \alpha n_{s}(\vr) \\
& = W(\vr) \left\lbrace 1+\delta(\vr) \right\rbrace - W(\vr) \frac{\int d^{3} x W(\vx) \left\lbrace 1+\delta(\vx) \right\rbrace}{\int d^{3} x W(\vx)} \\
& = W(\vr) \left\lbrace \delta(\vr) - \int d^{3} x W_{\glo}(\vx)\delta(\vx) \right\rbrace,
\end{split}
\label{eq:delta_global_cic}
\end{equation}
with $W_{\glo}(\vr) = \frac{W(\vr)}{\int d^{3} x W(\vx)}$.
 
We find two terms, $\delta^{\mathrm{c}}(\vr) = W(\vr)\delta(\vr)$ corresponding to the density contrast weighted by the selection function, and the integral constraint term $\int d^{3} r W_{\glo}(\vr)\delta(\vr)$. The normalisation of $W_{\glo}(\vr)$ ensures that $\int d^{3} r\delta_{\cic}(\vr) = 0$ over the entire footprint: modes larger than the survey size are suppressed.

\subsection{Extension to the radial integral constraint}
\label{sec:delta_radial}

The true radial selection function of a spectroscopic survey is often a complex function of the luminosity function, sky lines, spectrograph efficiency and redshift determination algorithm~\cite{Blake2010:1003.5721v1}. As an example, in BOSS, the radial distribution of the synthetic catalogue is directly inferred from the data. Various techniques exist: random redshifts can be picked from the whole data redshift distribution (the so-called \emph{shuffled} scheme~\cite{Samushia2012:1102.1014,Ross2012:1203.6499v3,Reid2015:1509.06529v2}), or, assuming the true radial selection function should be somewhat smooth, it can be fitted by a spline from which random redshifts are drawn~\cite{Samushia2012:1102.1014,Ross2012:1203.6499v3}. A third possibility (\emph{binned} scheme) is to weight an arbitrary initial random redshift distribution by $\alpha(r)$ (with $r=\abs{\vr}$) to match the data radial density in redshift or comoving distance bins of size $\delta r$, below which variations of $W(r)$ are neglected. Here, $\alpha(r)$ is given by:
\begin{equation}
\alpha(r) = \frac{\int d^{3} x n_{g}(\vx) \epsilon_{\rad}(r,x)}{\int d^{3} x n_{s}(\vx) \epsilon_{\rad}(r,x)},
\label{eq:alpha_radial}
\end{equation}
where $\epsilon_{\rad}(r,x)$ is $1$ if $r$ and $x$ belong to the same radial bin, $0$ otherwise. By construction, the radially-normalised random density and true selection function match:
\begin{equation}
\frac{n_{s}(\vr)}{\int d^{3} x n_{s}(\vx) \epsilon_{\rad}(r,x)} = \frac{W(\vr)}{\int d^{3} x W(\vx) \epsilon_{\rad}(r,x)}.
\end{equation}
Then, the following density fluctuations are observed:
\begin{equation}
\begin{split}
\delta^{\cic}(\vr) & = n_{g}(\vr) - \alpha(r) n_{s}(\vr) \\
& = W(\vr) \left\lbrace 1+\delta(\vr) \right\rbrace - W(\vr) \frac{\int d^{3} x W(\vx) \left\lbrace 1+\delta(\vx) \right\rbrace \epsilon_{\rad}(r,x)} {\int d^{3} x W(\vx) \epsilon_{\rad}(r,x)} \\
& = W(\vr) \left\lbrace \delta(\vr) - \int d^{3} x W_{\rad}(\vx)\delta(\vx) \epsilon_{\rad}(r,x) \right\rbrace,
\end{split}
\label{eq:delta_radial_cic}
\end{equation}
with $W_{\rad}(\vr) = \frac{W(\vr)}{\int d^{3} x W(\vx) \epsilon_{\rad}(r,x)}$.
Note that the global integral constraint is also automatically imposed.

Equations~\eqref{eq:delta_global_cic} for the global IC and~\eqref{eq:delta_radial_cic} for the radial IC can be recast in the following form:
\begin{equation}
\delta^{\cic}(\vr) = W(\vr) \left\lbrace \delta(\vr) - \int d^{3} x W_{\ic}(\vx) \delta(\vx) \epsilon_{\ic}(\vr,\vx) \right\rbrace,
\label{eq:delta_cic}
\end{equation}
with $\epsilon_{\ic}(\vr,\vx)$ some generic window function ($\epsilon_{\glo}(\vr,\vx) = 1$ for the global integral constraint, $\epsilon_{\rad}(r,x)$ for the radial one), and: 
\begin{equation}
W_{\ic}(\vr) = \frac{W(\vr)}{\int d^{3} x W(\vx)\epsilon_{\ic}(\vr,\vx)}.
\label{eq:window_ic}
\end{equation}
The general expression for the observed density fluctuations~\eqref{eq:delta_cic} will be used in the following section to calculate the effect of integral constraints on 2-point statistics.

\subsection{Correlation function in the local plane-parallel approximation}
\label{sec:ic_correlation}

Pursuing our calculation in configuration space, we want to express the even multipoles $\xi^{\cic}_{\ell}(s)$ of the correlation function $\xi^{\cic}(\vs)$ of the observed density fluctuations~\eqref{eq:delta_cic}. $\xi^{\cic}(\vs)$ would correspond to the quantity estimated with the Landy-Szalay estimator~\cite{Landy1993} if the division by the window function represented by the random pair counts $RR(\vs)$ was omitted in the estimator.

Here we assume the local plane-parallel approximation~\cite{Beutler2014:1312.4611v2}, namely that the pair separation $s$ between two galaxies at positions $\vx$ and $\vx+\vs$ is small compared to the distances $\abs{\vx}$, $\abs{\vx+\vs}$ of the galaxies to the observer. Then, the cosine angle betwen $\vs$ and $\vx$ or $\vx+\vs$ can be well approximated by $\loss\cdot\hat{\vs}$, with $\vc{\eta}_{\vs} = \vx + \vs/2$ the mid-point line-of-sight and $\hat{\vs}$ the unit $\vs$ vector.

As $W(\vr)$ is uncorrelated with $\delta(\vr)$, the correlation function of the observed density fluctuations~\eqref{eq:delta_cic} can be expressed as:
\begin{subequations}
\begin{align}
\xi^{\cic}(\vs) & = \int d^{3} x W(\vx)W(\vx+\vs) \xi(\vs) \label{eq:term_delta_delta}\\
& - \int d^{3} \Delta \xi(\vD) \int d^{3} x W(\vx)W(\vx+\vs)W_{\ic}(\vx+\vD) \epsilon_{\ic}(\vx+\vs,\vx+\vD) \label{eq:term_delta_ic}\\
& - \int d^{3} \Delta \xi(\vD) \int d^{3} x W(\vx)W(\vx-\vs)W_{\ic}(\vx+\vD) \epsilon_{\ic}(\vx-\vs,\vx+\vD) \label{eq:term_ic_delta}\\
& + \int d^{3} \Delta \xi(\vD) \int d^{3} x W(\vx)W(\vx+\vs) \nonumber \\
& \int d^{3} y W_{\ic}(\vy)W_{\ic}(\vy+\vD)  \epsilon_{\ic}(\vx,\vy) \epsilon_{\ic}(\vx+\vs,\vy+\vD).
\label{eq:term_ic_ic}
\end{align}
\end{subequations}
Term~\eqref{eq:term_delta_delta} is the true correlation function $\xi(\vs)$ multiplied by the window function $\window^{\delta,\delta}(\vs) = \int d^{3} x W(\vx)W(\vx+\vs)$. We account for the multipoles of~\eqref{eq:term_delta_delta} using the formalism of~\cite{Wilson2015:1511.07799v2}:
\begin{equation}
\xi_{\ell}^{\mathrm{c}}(s) = \sum_{p,q} A_{\ell p}^{q} \frac{2\ell+1}{2q+1} \xi_{p}(s) \window_{q}^{\delta,\delta}(s),
\label{eq:corr_c}
\end{equation}
where we use $\xi_{p}(s)$ the multipoles of $\xi(\vs)$. The window function multipoles are obtained by integration over the Legendre polynomials $\Leg{\ell}$:
\begin{equation}
\window_{\ell}^{\delta,\delta}(s) = \frac{2\ell+1}{4\pi} \int d\Os \int d^{3} x W(\vx)W(\vx+\vs) \Leg{\ell}(\loss\cdot\hat{\vs})
\end{equation}
and coefficient $A_{\ell p}^{q}$ is defined by:
\begin{equation}
\Leg{\ell}(\mu) \Leg{p}(\mu) = \sum_{q=0}^{\ell + p} A_{\ell p}^{q} \Leg{q}(\mu).
\end{equation}
Cross-terms~\eqref{eq:term_delta_ic} and~\eqref{eq:term_ic_delta} account for the correlation between the density field and the integral constraint term of \Eq{delta_cic}. We are only interested in even multipoles with respect to the mid-point line-of-sight, so term $IC_{\ell}^{\ic,\delta}$~\eqref{eq:term_ic_delta} is equal to $IC_{\ell}^{\delta,\ic}$~\eqref{eq:term_delta_ic}. We are thus left with evaluating the multipoles of term~\eqref{eq:term_delta_ic}:
\begin{equation}
IC_{\ell}^{\delta,\ic}(s) = \frac{2\ell+1}{4\pi} \int d\Os \int d^{3} \Delta \xi(\vD) \int d^{3} x W(\vx)W(\vx+\vs)W_{\ic}(\vx+\vD)  \Leg{\ell}(\loss\cdot\hat{\vs}) \epsilon_{\ic}(\vx+\vs,\vx+\vD).
\end{equation}
The redshift-space correlation function is fully described by its multipoles. We thus develop:
\begin{equation}
\xi(\vD) = \sum_{\ell} \xi_{\ell}(\Delta) \Leg{\ell}(\losD\cdot\hat{\vD}),
\end{equation}
leading to:
\begin{equation}
\begin{split}
IC_{\ell}^{\delta,\ic}(s) & = \frac{2\ell+1}{4\pi} \int d\Os \int \Delta^{2} d\Delta \int d\OD \sum_{p} \xi_{p}(\Delta) \\
& \int d^{3} x W(\vx)W(\vx+\vs)W_{\ic}(\vx+\vD) \Leg{\ell}(\loss\cdot\hat{\vs}) \Leg{p}(\losD\cdot\hat{\vD}) \epsilon_{\ic}(\vx+\vs,\vx+\vD).
\end{split}
\end{equation}
Writing the 3-point correlation function of the selection function:
\begin{equation}
\begin{split}
\window_{\ell p}^{\delta,\ic}(s,\Delta) & = \frac{(2\ell+1)(2p+1)}{(4\pi)^{2}} \int d\Os \int d\OD \\ 
& \int d^{3} x W(\vx)W(\vx+\vs)W_{\ic}(\vx+\vD) \Leg{\ell}(\loss\cdot\hat{\vs}) \Leg{p}(\losD\cdot\hat{\vD}) \epsilon_{\ic}(\vx+\vs,\vx+\vD),
\label{eq:window_delta_ic}
\end{split}
\end{equation}
we obtain the simple formula:
\begin{equation}
IC_{\ell}^{\delta,\ic}(s) = \int \Delta^{2} d\Delta 
\sum_{p} \frac{4 \pi}{2p+1} \xi_{p}(\Delta) \window_{\ell p}^{\delta,\ic}(s,\Delta).
\label{eq:ic_delta_ic}
\end{equation}
Finally, term~\eqref{eq:term_ic_ic} accounts for the auto-correlation of the integral term in \Eq{delta_cic}. Similarly, its multipoles are equal to:
\begin{equation}
IC_{\ell}^{\ic,\ic}(s) = \int \Delta^{2} d\Delta 
\sum_{p} \frac{4 \pi}{2p+1} \xi_{p}(\Delta) \window_{\ell p}^{\ic,\ic}(s,\Delta),
\label{eq:ic_ic_ic}
\end{equation}
with:
\begin{equation}
\begin{split}
\window_{\ell p}^{\ic,\ic}(s,\Delta) &= \frac{(2\ell+1)(2p+1)}{(4\pi)^2} \int d\Os \int d\OD \int d^{3} x W(\vx)W(\vx+\vs) \\
& \int d^{3} y W_{\ic}(\vy) W_{\ic}(\vy+\vD) \Leg{\ell}(\loss\cdot\hat{\vs}) \Leg{p}(\losD\cdot\hat{\vD}) \epsilon_{\ic}(\vx,\vy) \epsilon_{\ic}(\vx+\vs,\vy+\vD).
\end{split}
\label{eq:window_ic_ic}
\end{equation}
Putting all terms together, the observed correlation function multipoles are:
\begin{equation}
\xi_{\ell}^{\cic}(s) = \xi_{\ell}^{\mathrm{c}}(s) - IC_{\ell}^{\delta,\ic}(s) - IC_{\ell}^{\ic,\delta}(s) + IC_{\ell}^{\ic,\ic}(s).
\label{eq:corr_cic}
\end{equation}
The observed window-convolved, integral-constraint-corrected power spectrum multipoles can be inferred from \Eq{corr_cic} by a simple Hankel transform~\cite{Wilson2015:1511.07799v2}:
\begin{equation}
P_{\ell}^{\cic}(k) = 4 \pi (-i)^{\ell} \int s^{2} ds j_{\ell}(ks) \xi_{\ell}^{\cic}(s),
\label{eq:hankel}
\end{equation}
with $j_{\ell}$ the spherical Bessel function of order $\ell$. $P^{\cic}_{\ell}(k)$ corresponds to the quantity measured by the Yamamoto estimator~\cite{Yamamoto2005:astro-ph/0505115v2} in the local plane-parallel approximation, to some global normalisation factor discussed in \Sec{normalisation}. We report the reader to \App{wide-angle} for a full derivation in the case of the end-point line-of-sight for the Yamamoto estimator, taking into account wide-angle corrections to the plane-parallel approximation.
In the specific case of the global integral constraint ($\epsilon_{\ic}(\vr,\vx) = 1$), the window function~\eqref{eq:window_ic_ic} is separable and:
\begin{equation}
IC_{\ell}^{\glo,\glo}(s) = \frac{\int \Delta^{2} d\Delta \xi^{\mathrm{c}}_{0}(\Delta)}{\int \Delta^{2} d\Delta \window_{0}^{\delta,\delta}(\Delta)} \window_{\ell}^{\delta,\delta}(s).
\label{eq:ic_glo_glo}
\end{equation}
We will see how this compares to previous results in the next section.

\subsection{Discussion}
\label{sec:ic_discussion}

The global integral constraint was first discussed in~\cite{Peacock1991:10.1093/mnras/253.2.307,Beutler2014:1312.4611v2,Wilson2015:1511.07799v2}, which expressed the observed window-convolved, integral-constraint-corrected power spectrum as:
\begin{equation}
P^{\cic}(\vk) = P^{\mathrm{c}}(\vc{k}) - P^{\mathrm{c}}(\vc{0}) \abs{\FW(\vk)}^{2}
\label{eq:wilson_ic}
\end{equation}
with $P^{\mathrm{c}}(\vc{k})$ the window-convolved power spectrum. $\FW(\vk)$ is the (Fourier transform of) the survey selection function, rescaled by imposing $\vert \FW(\vk) \vert^{2} = 1$ when $k \ll 1 \hMpc$, so that $P^{\cic}(k) \rightarrow 0$ as $k \rightarrow 0$. Only one correction term is assumed, $P^{\mathrm{c}}(\vc{0}) \vert \FW(\vk) \vert^{2}$, while our complete derivation leads to 3 terms~\eqref{eq:corr_cic}.

The multipoles of the integral constraint correction $P^{\mathrm{c}}(\vc{0}) \vert \FW(\vk) \vert^{2}$ of \Eq{wilson_ic} are actually exactly equal to our last correction term, $IC_{\ell}^{\glo,\glo}(s)$ after a Hankel transform. However, \Eq{wilson_ic} introduces a minus sign in front of $P^{\mathrm{c}}(\vc{0}) \vert \FW(\vk) \vert^{2}$ to obtain a damping of the power spectrum at large scales. Instead, in our \Eq{corr_cic}, $IC_{\ell}^{\glo,\glo}(s)$, which accounts for the auto-correlation of the density contrast integral over the footprint, leads to an increase of power at large scales. Damping of power comes from the two other terms $-IC_{\ell}^{\delta,\glo}(s)-IC_{\ell}^{\glo,\delta}(s)$ which correspond to the cross-correlations between the density contrast and its integral over the footprint, respectively. Finally, note that our full derivation of the integral constraint effect makes it clear that the correction $P^{\mathrm{c}}(\vc{0}) \vert \FW(\vk) \vert^{2}$, with $W$ describing the full survey footprint, is no longer valid if this footprint is composed of different chunks with separate normalisation factors~\eqref{eq:alpha_global}. Indeed, in this case, $\epsilon_{\glo}(\vr,\vx)$ is $1$ if $\vr$ and $\vx$ belong to the same chunk and integrals of term~\eqref{eq:window_ic_ic} cannot be split as in \Eq{ic_glo_glo}. The resulting global IC effect would be larger.
 
\begin{figure}[t]
\centering
\includegraphics[width=0.45\textwidth]{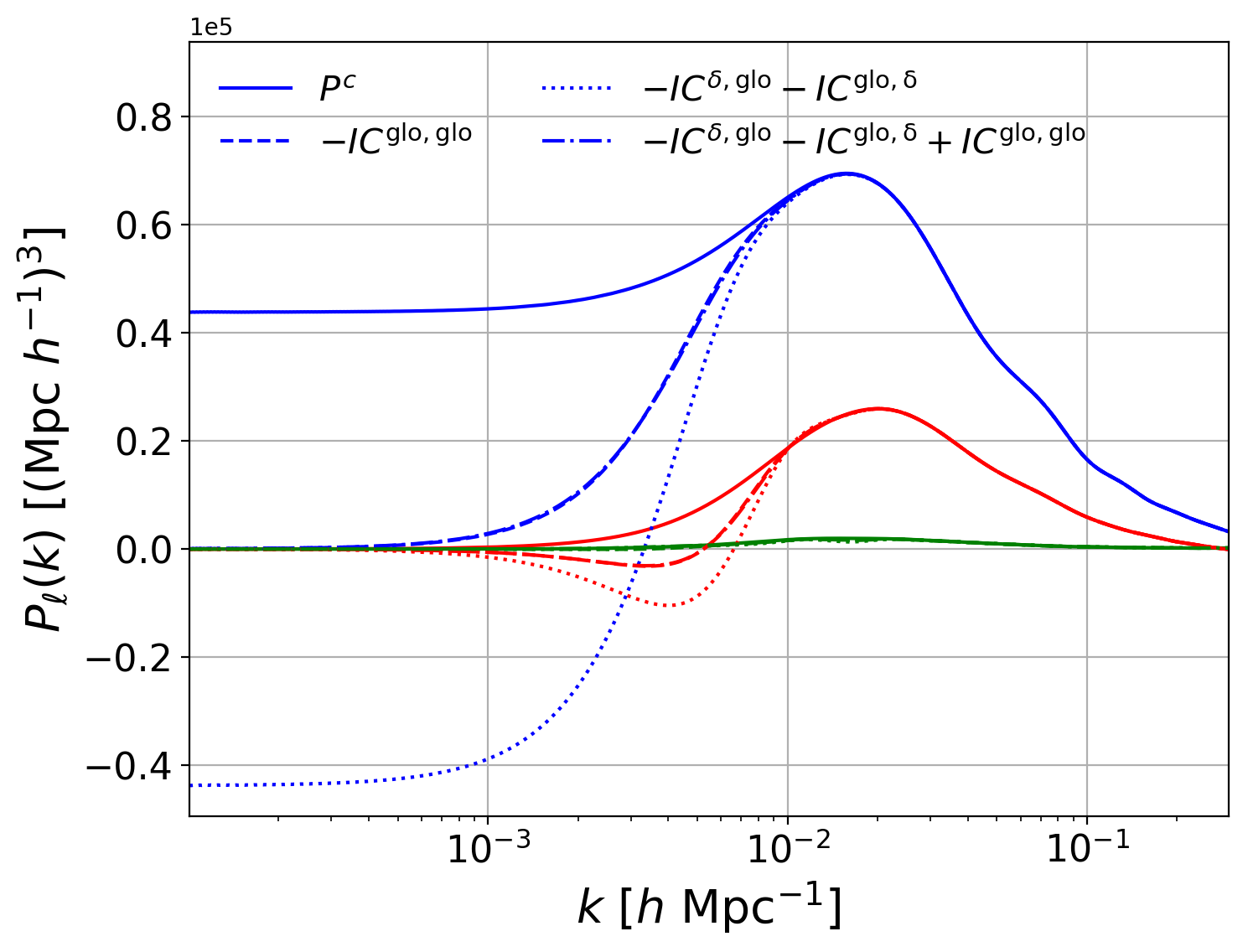}
\includegraphics[width=0.45\textwidth]{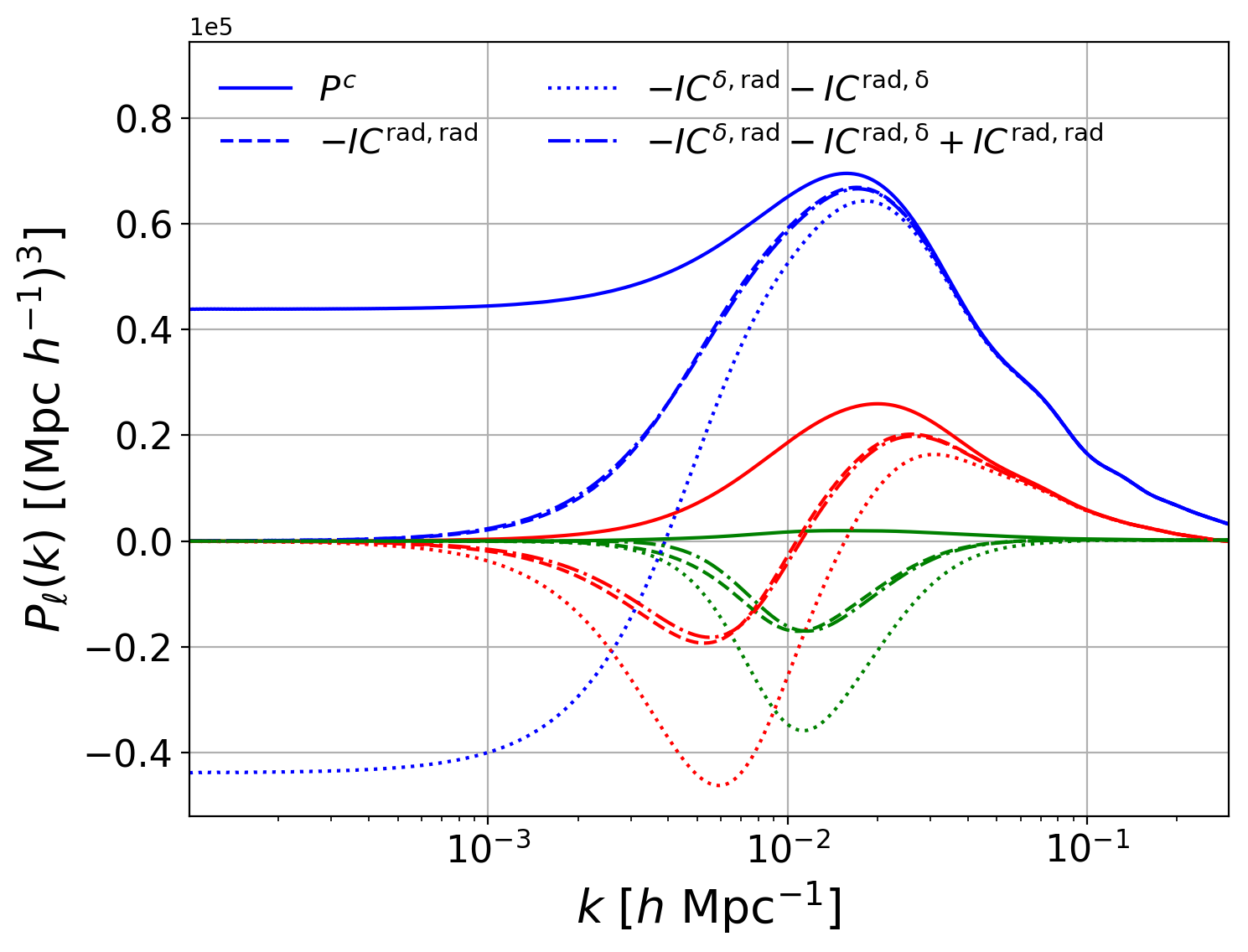}
\caption{Left: power spectrum multipoles (blue: monopole, red: quadrupole, green: hexadecapole) including the different contributions to the global integral constraint (see text). The complete result (dot-dashed curve) cannot be distinguished from the partial correction (dashed). Right: same, with the radial integral constraint. See \Sec{analysis_radial} for more details about the survey configuration assumed in this figure.}
\label{fig:integral_constraint_terms_global_radial}
\end{figure}

We detail the different terms contributing to the global integral constraint on the left panel of \Fig{integral_constraint_terms_global_radial}. As expected, on large scales, the window-convolved, integral-constraint-corrected monopole reaches $0$. The effect of the complete global integral constraint is negligible at the scales involved in a RSD analysis ($k \gtrsim 0.01 \hMpc$), but is significant at large scales. As also shown in this panel, taking only $-IC_{\ell}^{\glo,\glo}$, as in \Eq{wilson_ic} (with chunk-splitting --- see \Sec{analysis_radial}), for the global integral constraint correction~\eqref{eq:corr_cic} appears to be a very legitimate approximation in the illustrated survey case\footnote{We checked that this remains a good approximation for the radial integral constraint, as can be inferred from the right panel of \Fig{integral_constraint_terms_global_radial} (dot-dashed and dashed curves). However, we observed that this approximation fails to account for the angular integral constraint described in \Sec{analysis_angular}.}.

The right panel of \Fig{integral_constraint_terms_global_radial} displays the different contributions to the radial integral constraint. Compared to the global IC, the radial IC has a larger effect on a broader range of wavenumbers, especially in the quadrupole and hexadecapole. We may thus expect a non-negligible impact of the radial integral constraint on clustering measurements in the illustrated case.


\section{Computing window functions}
\label{sec:computing_window}

This section is devoted to the practical computation of window functions $\window_{\ell}^{\delta,\delta}(s)$
and $\window_{\ell p}^{i,j}(s,\Delta)$ ($(i,j) \in \{(\delta,\ic),(\ic,\delta),(\ic,\ic)\}$) required in \Sec{ic_correlation}.
We argue that a reasonable approximation for the true survey selection function can be inferred from data in \Sec{approaching_true_selection}. We then discuss the normalisation of window functions, and detail their response to a Poisson shot noise in sections~\ref{sec:normalisation} and~\ref{sec:shot_noise}. Algorithms required to compute these window functions from a synthetic catalogue sampling the survey selection function are given in \Sec{algorithms}.

\subsection{Approaching the true survey selection function}
\label{sec:approaching_true_selection}

Our derivation of integral constraint corrections makes use of the true, underlying, survey selection function $W(\vr)$. However, as we mentioned previously, one has often only access to a synthetic catalogue whose density $n_{s}(\vr)$ is tuned to match the actual data in certain regions (chunks, or radial bins for the radial IC). Thus, clustering modes inprinted in the data distribution are propagated to this synthetic catalogue. Using $n_{s}(\vr)$ for $W(\vr)$ does not change the value of $W_{\ic}(\vr)$ as it is normalised in regions where the integral constraint is imposed, i.e. where the expected mean of $n_{s}(\vr)$ is unknown. However, if that tuned $n_{s}(\vr)$ is taken for $W(\vr)$, the calculated window effect and integral constraint corrections statistically differ from the truth.

We can provide some estimate for this bias in the dominant term~\eqref{eq:corr_c}. The density $n_{s}(\vr)$ tuned on the actual data can be expressed as (see equations~\eqref{eq:delta_global_cic} and~\eqref{eq:delta_radial_cic}):
\begin{equation}
n_{s}(\vr) = W(\vr) \left\lbrace 1 + \int d^{3} x W_{\ic}(\vx) \delta(\vx) \epsilon_{\ic}(\vr,\vx) \right\rbrace,
\end{equation}
where $W_{\ic}(\vr)$ can be described by $n_{s,\ic}(\vr) = \frac{n_{s}(\vr)}{\int d^{3} x n_{s}(\vx)\epsilon_{\ic}(\vr,\vx)}$. Then, the multipoles of the 2-point window function estimated from $n_{s}(\vr)$ can be written:
\begin{equation}
\mathcal{S}_{\ell}^{\delta,\delta}(s) = \window_{\ell}^{\delta,\delta}(s) + IC_{\ell}^{\ic,\ic}(s).
\label{eq:window_ns_ic}
\end{equation}
Taking $n_{s}(\vr)$ for $W(\vr)$ in $IC_{\ell}^{\ic,\ic}(s)$, \Eq{window_ns_ic} provides an estimate for $\window_{\ell}^{\delta,\delta}(s)$ at first order in the integral constraint corrections.

Note that such a dependence of the estimation of window effect and integral constraint corrections in the observed clustering also leads to a change in the variance of cosmological measurements, whose study is beyond the scope of this paper.

\subsection{Normalisation}
\label{sec:normalisation}

In order to recover the true (unconvolved) power spectrum at small scales, \cite{Wilson2015:1511.07799v2} suggested to normalise the 2-point window function in the limit $s \rightarrow 0$. However, this limit may be ill-defined in practice, since its estimation requires a very fine sampling of the survey selection function.

Similarly, in current Fourier space analyses, power spectrum measurements are normalised by~\cite{Beutler2016:1607.03150v1}:
\begin{equation}
A = \alpha^{2} \int d^{3} r n_{s}^{2}(\vr) = \alpha^{2} \sum_{i=1}^{N_{s}} w_{s,i} n_{s,i},
\label{eq:power_spectrum_normalisation}
\end{equation}
using the notations of \Eq{alpha_global}. This also corresponds to the ill-defined limit $\window_{0}^{\delta,\delta}(0) = \int d^{3} r W^{2}(\vr)$ if $W(\vr)$ is represented by the same synthetic catalogue as in \Eq{power_spectrum_normalisation}.

Besides, the estimation of the density $n_{s}(\vr)$ in \Eq{power_spectrum_normalisation} is in practice non-trivial when accounting for various survey selection effects. For example, $n_{s}(\vr)$ is commonly taken to be the redshift density $n(z)$, computed by binning (weighted) data in redshift slices~\cite{Reid2015:1509.06529v2}, while it can also be a function of the angular position on the sky.

We thus also use term~\eqref{eq:power_spectrum_normalisation} in the normalisation of window functions $\window_{\ell}^{\delta,\delta}(s)$ and $\window_{\ell p}^{i,j}(s,\Delta)$, so that $A$ terms divide both the power spectrum measurements and model, and compensate. Therefore, the estimation of $A$ does not impact the estimation of cosmological parameters. Integral constraint terms are properly normalised using \Eq{window_ic}.

\subsection{Shot noise}
\label{sec:shot_noise}

In principle, formulae given in \Sec{ic_correlation} (and \Eq{window_ns_ic}) fully describe integral constraint corrections. However, they are derived in real space, where accounting for a Poisson shot noise term (a spike at $s = 0$) is numerically challenging.

The shot noise contribution to the correlation function weighted by the window function only shows up in the monopole~\cite{Feldman1993:astro-ph/9304022v1,Yamamoto2005:astro-ph/0505115v2}:
\begin{equation}
\xi_{0}^{\mathrm{c}}(s) \ni \int d^{3} x W(\vx)w(\vx) \Diracn3(s)
\label{eq:shot_noise_delta_delta}
\end{equation}
where $w(\vx)$ is the total (data and random\footnote{The total shot noise estimate (equations~\eqref{eq:shot_noise_delta_delta} and~\eqref{eq:shot_noise_ij}) is valid in the limit of infinite random density, but appears to be a legitimate approximation if the random density is high enough, as shown by our cosmological fits in sections~\ref{sec:fits_radial},~\ref{sec:fits_angular}, and~\ref{sec:fits_radialxangular}.}) weight at position $\vx$. In Fourier space, this translate in a simple offset in the monopole. However, the shot noise contribution to the integral constraint correction leaks on all scales of all multipoles:
\begin{equation}
IC_{\ell}^{i,j}(s) \ni SN_{\ell}^{i,j}(s) \qquad (i,j) \in \{(\delta,\ic),(\ic,\delta),(\ic,\ic)\},
\label{eq:shot_noise_ij}
\end{equation}
with:
\begin{equation}
SN_{\ell}^{\ic,\delta}(s) = \frac{2\ell+1}{4\pi} \int d\Os \int d^{3} x W(\vx)w_{\ic}(\vx) W(\vx+\vs) \Leg{\ell}(\loss\cdot\hat{\vs}) \epsilon_{\ic}(\vx,\vx+\vs)
\end{equation}
and
\begin{equation}
\begin{split}
SN_{\ell}^{\ic,\ic}(s) & = \frac{2\ell+1}{4\pi} \int d\Os \int d^{3} x W(\vx)W(\vx+\vs) \\
& \int d^{3} y W_{\ic}(\vy)w_{\ic}(\vy) \Leg{\ell}(\loss\cdot\hat{\vs}) \epsilon_{\ic}(\vx,\vy) \epsilon_{\ic}(\vx+\vs,\vy),
\end{split}
\end{equation}
where we used $w_{\ic}(\vr) = \frac{w(\vr)}{\int d^{3} x W(\vx) \epsilon_{\ic}(\vr,\vx)}$.

Integrals over $\vs$ of shot noise contributions to $\xi^{\mathrm{c}}_{0}(s)$ and $IC_{\ell}^{i,j}(s)$ are equal, such that total shot noise contributions vanish as $k\rightarrow 0$, as expected. In practice, we choose to normalise terms~\eqref{eq:shot_noise_delta_delta} and~\eqref{eq:shot_noise_ij} by their integral over $\vs$\footnote{In the case $W$ is sampled by a synthetic catalogue, the normalisation for $SN_{\ell}^{\ic,\delta}(s)$ and $SN_{\ell}^{\ic,\ic}(s)$ is the (weighted) number of correlated pairs.}, and multiply them by the shot noise measured by the Yamamoto estimator~\cite{Yamamoto2005:astro-ph/0505115v2,Beutler2014:1312.4611v2}.

\begin{figure}[t]
\centering
\includegraphics[width=0.5\textwidth]{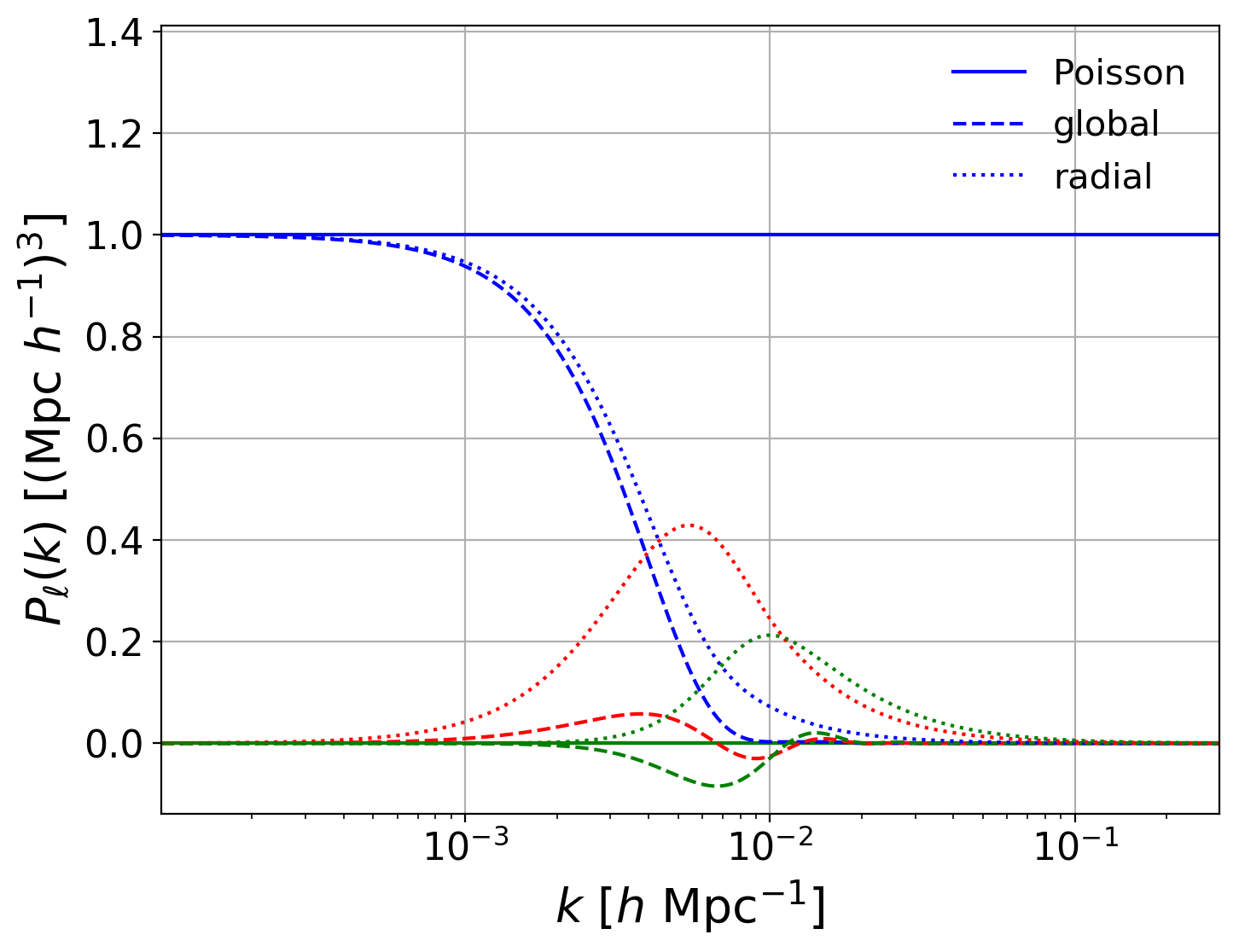}
\caption{Normalised shot noise contributions from the global and radial integral constraints (blue: monopole, red: quadrupole, green: hexadecapole). As expected, the Poisson shot noise in $P^{\mathrm{c}}_{0}(k)$ is cancelled by the integral constraint contribution at large scales. The same survey case as in \Fig{integral_constraint_terms_global_radial} is assumed.}
\label{fig:shot_noise_global_radial}
\end{figure}

Normalised shot noise contributions to the global and radial integral constraints are shown in \Fig{shot_noise_global_radial}. By definition, they reach $1$ as $k \rightarrow 0$ in the monopole. Their impact on the quadrupole and hexadecapole should not be ignored in the case of the radial IC.

\subsection{Calculation from a synthetic catalogue}
\label{sec:algorithms}

The survey selection function entering window functions $\window_{\ell}^{\delta,\delta}(s)$ and $\window_{\ell p}^{i,j}(s,\Delta)$ ($(i,j) \in \{(\delta,\ic),(\ic,\delta),(\ic,\ic)\}$) required in \Sec{ic_correlation} can be randomly sampled by a synthetic catalogue. We use a classic pair-count algorithm to compute the anisotropic 2-point correlation and implement the algorithm from~\cite{Slepian2018:1709.10150v1} to compute the anisotropic 3-point correlation of the synthetic catalogue. As detailed in \App{wide-angle}, lines-of-sight are defined according to the FKP power spectrum estimator, which we use for clustering measurements.

Pairs (or triplets) of random objects are binned in comoving distance bins of typical size $1 \Mpch$ (for $\window_{\ell}^{\delta,\delta}(s)$) to $4 \Mpch$ (for $\window_{\ell p}^{i,j}(s,\Delta)$); the result is divided by the comoving volume of each distance bin to recover density, renormalised to data counts (by multiplying by $\alpha^{2}$), and divided by the normalisation factor $A$ as stated in \Sec{normalisation}.

Window function calculations $\window_{\ell p}^{i,j}(s,\Delta)$ involving $\epsilon_{\ic}(\vr,\vx)$ are performed by correlating random objects at position $\vr$ with random objects at position $\vx$ satisfying $\epsilon_{\ic}(\vr,\vx) \neq 0$, i.e. residing in the same spatial bin.

\begin{figure}[t]
\centering
\includegraphics[width=0.45\textwidth]{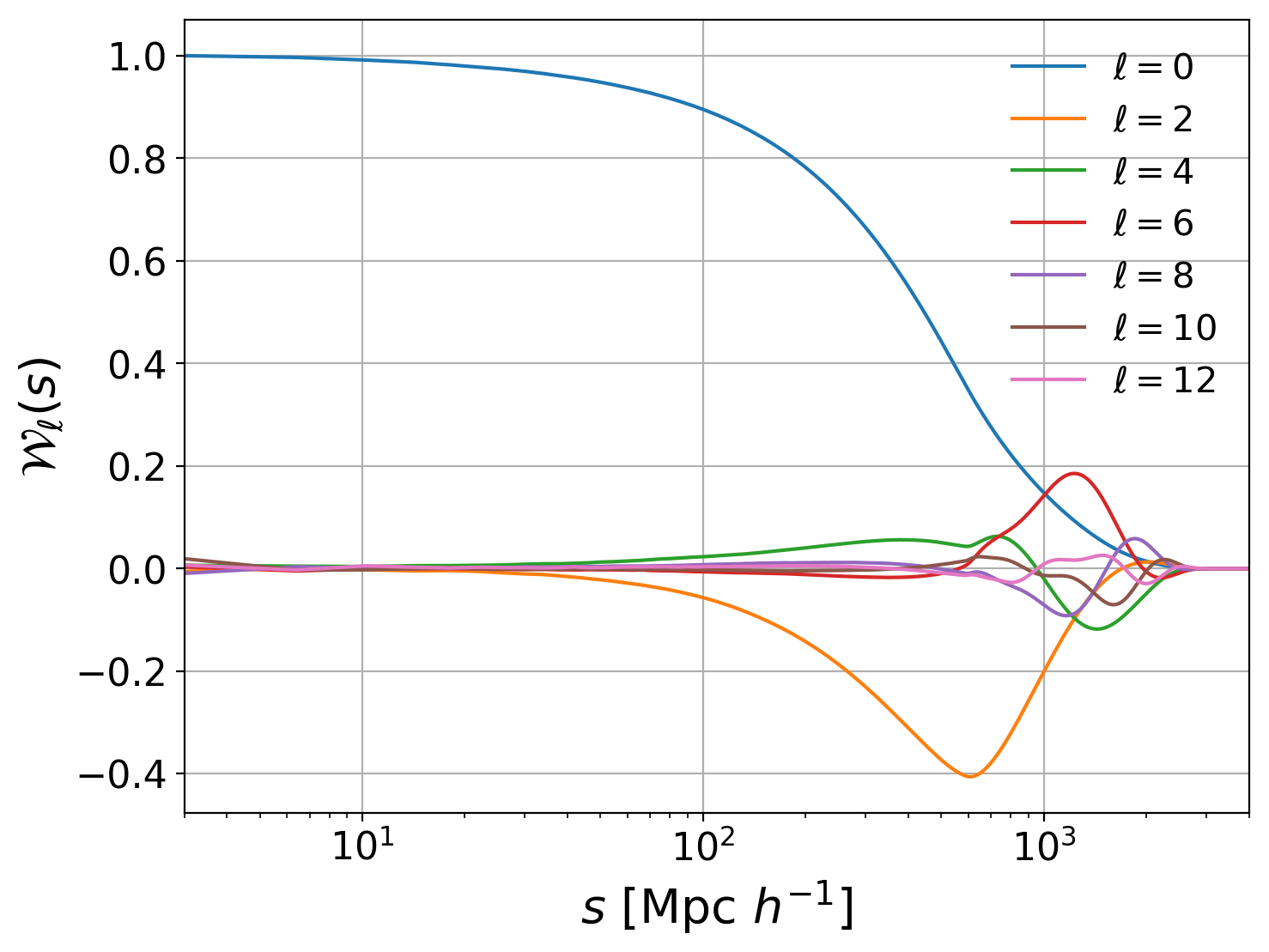}
\includegraphics[width=0.47\textwidth]{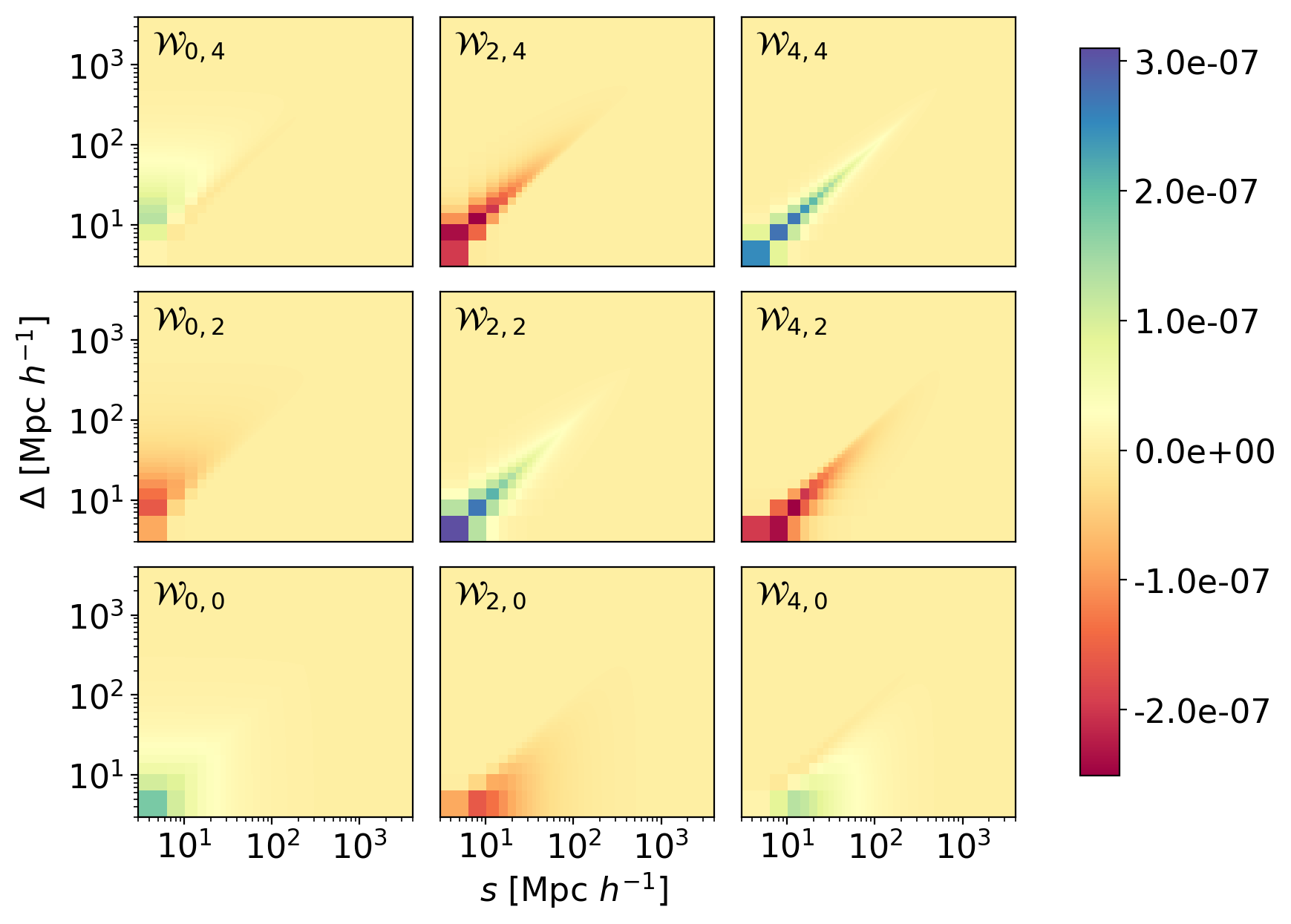}
\caption{Left: the window function multipoles $\window_{\ell}^{\delta,\delta}(s)$. Right: the window function multipoles $\window_{\ell p}^{\rad,\rad}(s,\Delta)$.}
\label{fig:window_functions}
\end{figure}

We show examples of $\window_{\ell}^{\delta,\delta}(s)$ and $\window_{\ell p}^{\rad,\rad}(s,\Delta)$ in \Fig{window_functions}. All window functions (for the global and radial integral constraints) could be accurately computed in $\sim 1500$ CPU hours.

\section{Radial integral constraint in RSD analyses}
\label{sec:analysis_radial}

In this section we discuss the impact of the radial integral constraint on Fourier-space RSD measurements, and show that \Eq{corr_cic}, along with practical details developed in \Sec{computing_window}, helps us recover an unbiased estimation of cosmological parameters. We consider three sets of mock catalogues: the \emph{baseline} mocks for measurements including only the global IC, and the \emph{binned} and \emph{shuffled} mocks for the radial one. These mocks are described in \Sec{mocks}, while details about power spectrum measurements and the RSD modelling are given in \Sec{analysis_methods}. Cosmological fits are performed in \Sec{fits_radial}. We discuss the potential impact of the radial integral constraint on other clustering analyses in \Sec{analysis_radial_discussion}.

\subsection{Mock catalogues}
\label{sec:mocks}

We work on a set of 84 high fidelity N-series mocks used for the LRG clustering mock challenge~\cite{Tinker2016} of the BOSS Data Release 12 (DR12)~\cite{Alam:1607.03155v1}.
These mocks were built from the 3 projections of 7 independent, periodic box realisations of side $2600 \Mpch$ at redshift $z=0.5$. The simulated cosmology is a flat $\Lambda$CDM model with $\Omega_{m} = 0.286$, $\Omega_{\Lambda} = 0.714$, $\sigma_{8} = 0.82$, $n_{s} = 0.96$ and $h = 0.7$. Simulations were run with the GADGET2 code~\cite{Springel2005:astro-ph/0505010v1}, and a HOD modelling was used to populate dark matter halos with galaxies, mimicking the observed clustering in data.

The precision matrix required for cosmological fits is built from 2048 Multidark Patchy mocks provided by the BOSS collaboration. These approximate mocks were calibrated on BigMultiDark simulations~\cite{Klypin2014:1411.4001v2}; halo abundance matching was applied to reproduce the 2- and 3-point clustering measurements and mocks at different redshifts were combined into light cones~\cite{Kitaura2015:1411.4001v2}. We use V6C catalogues, which were adjusted to reproduce the data clustering measurements. These mocks are also used to compute the standard deviation (blue shaded area) shown in figures~\ref{fig:mocks_binned_shuffled}, \ref{fig:mocks_angular_systematics} and~\ref{fig:mocks_density} and the error bars plotted in \Fig{residuals_radial}.

Both N-series and Multidark Patchy mocks were trimmed following the DR12 CMASS NGC (hereafter abbreviated by CMASS) selection function using the \texttt{make\_survey} software~\cite{White2013:1309.5532v2}. We impose integral constraints consistently on N-series and Multidark Patchy mocks in each series of cosmological fits. \emph{Baseline} mocks are obtained by using a random catalogue sampling the true survey selection function, normalised according to \Eq{alpha_global}, so that those mocks are impacted by the global integral constraint only. The \emph{binned} or \emph{shuffled} schemes are used to impose the radial integral constraint. In the \emph{shuffled} scheme, random redshifts are drawn from each mock data realisation. In the \emph{binned} scheme, random galaxies are weighted to reproduce the mock data radial distribution in comoving distance bins of size $\delta r = 2 \Mpch$.

\begin{figure}[t]
\centering
\includegraphics[width=0.45\textwidth]{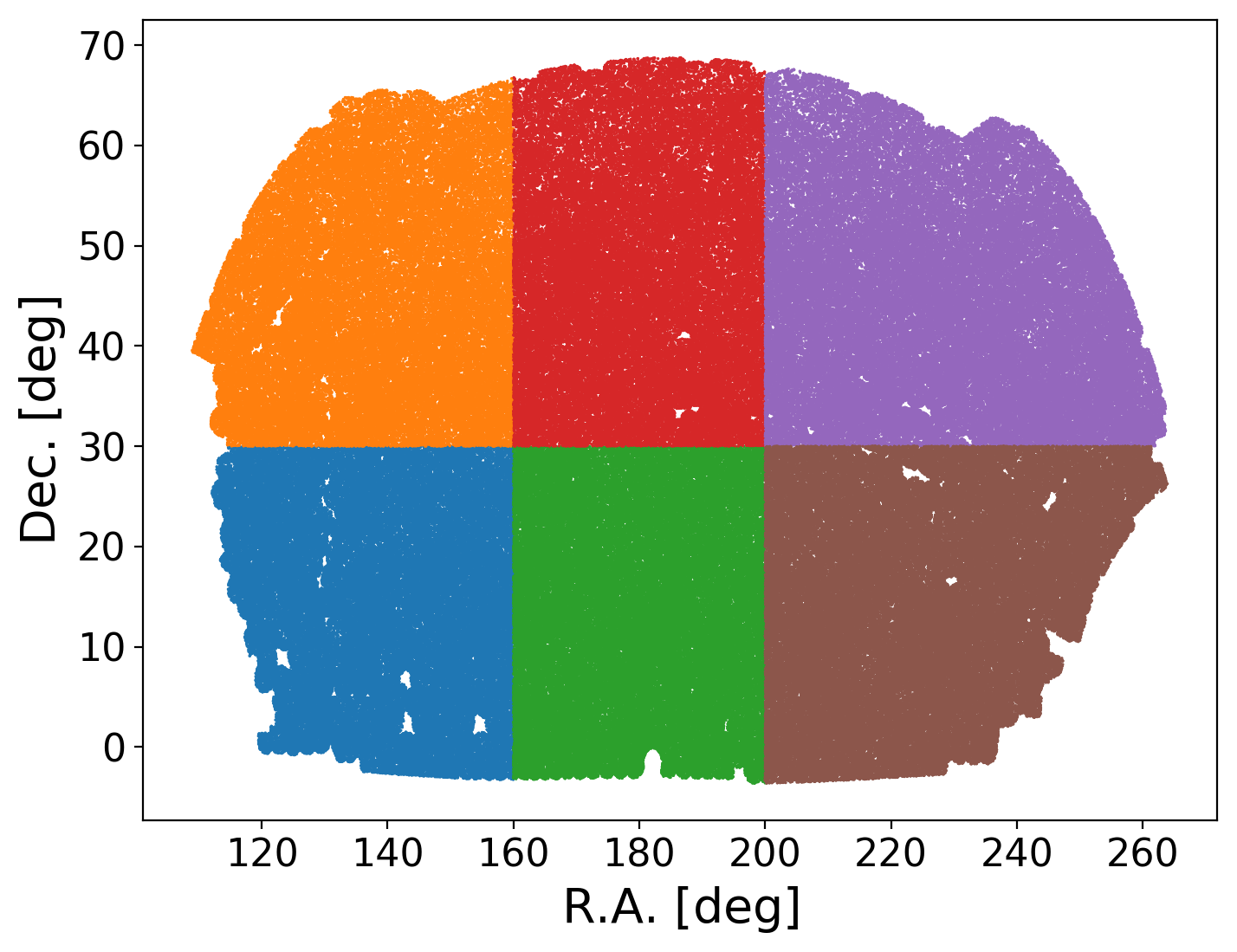}
\includegraphics[width=0.46\textwidth]{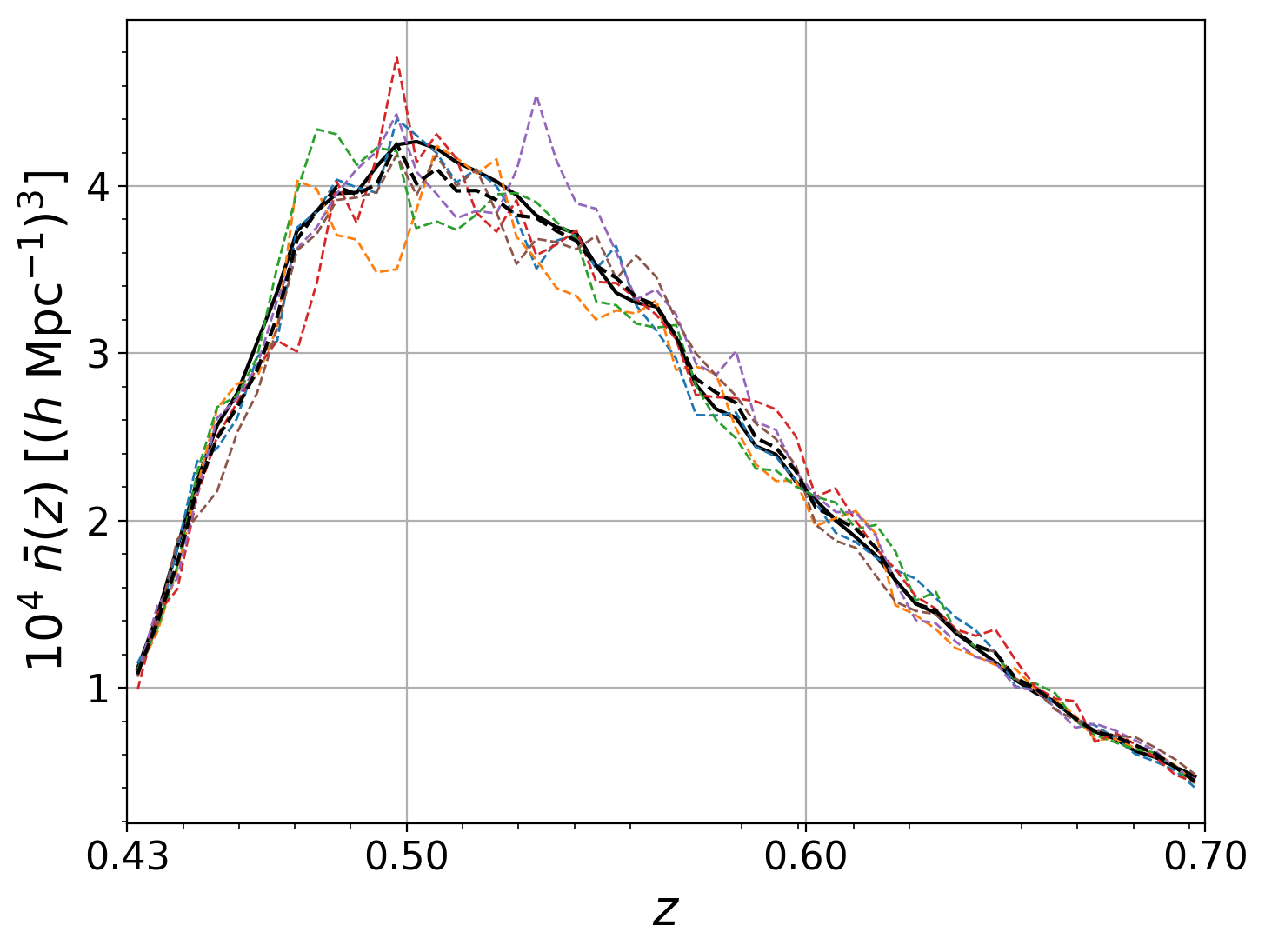}
\caption{Left: the CMASS footprint, divided in $6$ different chunks. Right: different estimates of the redshift density $n(z)$, using bins $\Delta z = 0.005$: the redshift density of the true survey selection function (continuous black curve), and, for one mock data realisation, $n(z)$ measured in the whole CMASS footprint (dashed black curve) and in the $6$ chunks individually. The scatter between the $n(z)$ estimates is due to noise and clustering.}
\label{fig:footprint}
\end{figure}

To enhance the radial integral constraint effect coming from the \emph{binned} or \emph{shuffled} schemes, we divide the CMASS footprint ($\simeq 7420 \deg^{2}$) in $6$ smaller chunks of size ranging from $\simeq  980\deg^{2}$ to $\simeq  1570\deg^{2}$ (see \Fig{footprint}), areas representative of the 4 chunks of the eBOSS ELG survey~\cite{Raichoor2017:1704.00338v1}. The \emph{binned} or \emph{shuffled} schemes are applied to the $6$ chunks separately, before these are recombined in a single catalogue used for the power spectrum measurement. This procedure, though not representative of the real BOSS DR12 analysis, allows us to test our modelling of the radial integral constraint in stringent conditions.

One can appreciate the difference between the true survey selection function (continuous black curve) and the ones measured from one mock realisation (dashed black curve for full CMASS and colored dashed curves for each chunk). Note that the radial selection function of future spectroscopic surveys like DESI~\cite{DESI2016:1611.00036v2} and Euclid~\cite{Laureijs2009:0912.0914v1} may vary on the sky; modelling this effect may indeed require to combine small patches with locally constant radial selection functions.

We apply FKP weights~\cite{Feldman1993:astro-ph/9304022v1} $w_{\mathrm{FKP}} = \frac{1}{1+n(z)P_{0}}$ to both mock data and randoms using $P_{0} = 20000 \Mpchc$. The redshift density $n(z)$, which also enters the normalisation of power spectrum measurements (see \Eq{power_spectrum_normalisation}), is calculated in bins of $\Delta z = 0.005$.

The redshift density is computed according to the type of mocks.

When the radial IC (\emph{binned} or \emph{shuffled} mocks) is imposed, the redshift density $n(z)$ is computed for each CMASS mock data realisation, as is done in an actual data analysis (see e.g.~\cite{Reid2015:1509.06529v2}). As we divide the CMASS footprint into $6$ chunks, one may want to measure $n(z)$ in the $6$ chunks separately (see \Fig{footprint}, right-panel). However, we show in \App{FKP_weights} that doing so significantly biases clustering measurements as FKP weights, using local $n(z)$, smooth out clustering. We therefore choose to measure $n(z)$ from the whole CMASS sample, making the bias due to FKP weights negligible. 

Similarly, when the global IC (\emph{baseline} mocks) is imposed, we compute $n(z)$ on the random catalogue accounting for the true survey selection function (see the continuous black curve on the left panel of \Fig{footprint}) but rescale it according to the weighted number of data for each mock realisation.

\subsection{Analysis methods}
\label{sec:analysis_methods}

We use the implementation of the Yamamoto estimator in the Python toolkit \texttt{nbodykit}~\cite{Hand2017:1712.05834v1} to compute mock data power spectra. We take a large box size of $4000 \Mpch$ to reduce sampling effects in low $k$-bins. The FKP field is interpolated on a $512^{3}$ mesh following the Triangular Shaped Cloud (TSC) scheme. The Nyquist frequency is thus $k \simeq 0.4 \hMpc$, more than twice larger than the maximum wavenumber used in our analysis. We employ the interlacing technique~\cite{Sefusatti2015:1512.07295v2} to mitigate aliasing effects. Power spectrum multipoles are measured in bins of $\Delta k = 0.01 \hMpc$, from $k = 0 \hMpc$.

Power spectrum multipoles are calculated on a discrete $k$-space grid, making the angular modes distribution irregular at large scales. We correct for this effect using the technique employed in~\cite{Beutler2016:1607.03150v1}. The correction is very small, given the large box size used to measure power spectra. Hankel transforms between power spectrum and correlation functions multipoles (e.g. \Eq{hankel}) are performed using the \texttt{FFTLog}~\cite{Hamilton2000:astro-ph/9905191v4} software. As in~\cite{Beutler2016:1607.03150v1} we only consider correlation function multipoles up to $\ell = 4$ in our calculations. We checked that adding $\xi_{6}(s)$ has a completely neligible impact on the model prediction.

Mock data are fitted by the RSD TNS~\cite{Taruya2010:1006.0699v1} model as implemented in~\cite{Beutler2016:1607.03150v1}. However, contrary to~\cite{Beutler2016:1607.03150v1}, we adopt a Lorentzian form for the Finger-of-God effect~\cite{Cole1995:astro-ph/9412062}: 
\begin{equation}
D_{\mathrm{FoG}}(k,\mu,\sigma_{v}) = \left(1+\frac{(k\mu\sigma_{v})^{2}}{2}\right)^{-2},
\end{equation}
with $\sigma_{v}$ the velocity dispersion.
A common practice of Fourier space clustering analyses is to remove the shot noise contribution from the power spectrum monopole measurement. Instead, we fit the power spectrum monopole with its shot noise and add up in the model the shot noise term $N$ measured from the data, including contributions from integral constraints as shown in \Sec{shot_noise}. As in~\cite{Beutler2016:1607.03150v1}, we let the constant galaxy stochastic term (see~\cite{McDonald2009:0902.0991v1}) free.

Power spectrum monopole, quadrupole and hexadecapole are fitted from $0.01 \hMpc$ to $0.15 \hMpc$. Fitted cosmological parameters are the logarithmic growth rate of structure $f$ and the Alcock-Paczynski~\cite{Alcock1979} parameters $\apar$ and $\aper$. As $f$ is almost completely degenerate with the power spectrum normalisation $\sigma_{8}$, we quote the combination $f\sigma_{8}$. We consider $4$ nuisance parameters: the linear and second order biases $b_{1}$ and $b_{2}$, the velocity dispersion $\sigma_{v}$ and $A_{g} = N_{g}/N$, with $N_{g}$ the constant galaxy stochastic term and $N$ the measured Poisson shot noise. Minimisations are performed using the algorithm \texttt{Minuit}~\cite{Minuit1975,iminuit}, taking large variation intervals for all parameters. We checked that the fitted parameters do not reach the input boundaries.

We use the fiducial BOSS DR12 cosmology in our analysis:
\begin{equation}
\begin{split}
h = 0.676, \,\, \Omega_{m} = 0.31, \,\, \Omega_{\Lambda} = 0.69, \,\, \Omega_{b}h^{2} = 0.022,\\
\sigma_{8} = 0.80, \,\, n_{s} = 0.97, \,\, \sum m_{\nu} = 0.06 \eV.  \qquad
\end{split}
\end{equation}
In all figures showing the power spectrum model alone (\ref{fig:integral_constraint_terms_global_radial}, \ref{fig:integral_constraint_effect_global_radial}), we use $f=0.75,\, b_{1}=2, \, b_{2}=1, \, \sigma_{v}=4 \Mpch$.

The radial integral constraint correction is consistently calculated using in $\epsilon_{\rad}(\vr,\vx)$ the same comoving distance bins of size $\delta r = 2 \Mpch$ as used in the \emph{binned} scheme. We account for the chunk-splitting of the CMASS footprint by adding the condition that $\vr$ and $\vx$ should belong to the same chunk for $\epsilon_{\rad}(\vr,\vx)$ to be non-zero.

Window function calculations are based on different estimations of the survey selection function, according to the type of mocks.

In the radial IC case (\emph{binned} or \emph{shuffled} mocks), to mimic an actual data analysis, the radial part of $W(\vr)$ should in principle be estimated on each realisation of the mocks, for which specific window function and integral constraint corrections should be derived. However, this would require a large computation time. Thus, in this work, window function estimations for the radial IC case are based on the selection function $n_{s}(\vr)$ calculated with a synthetic catalogue tuned to match the radial distribution (using the \emph{binned} scheme) in each chunk of one realisation of mock data. All illustrations of window functions (\Fig{window_functions}) and integral constraint corrections (figures~\ref{fig:integral_constraint_terms_global_radial}, \ref{fig:shot_noise_global_radial}, \ref{fig:integral_constraint_effect_global_radial}, \ref{fig:shot_noise_radialxangular}) are provided for this estimation of the survey selection function.

Similarly, in the global IC case (\emph{baseline} mocks), $W(\vr)$ entering window function calculations is taken to be the true selection function, which should be normalised using \Eq{alpha_global} for each realisation of the mocks, as done in a real data analysis. However, as for the radial IC case, we normalise the selection function on one realisation of the mocks only (the same realisation as in the radial IC case).

We checked that using in the model the estimate provided by \Eq{window_ns_ic} (with $\ic = \glo$ in the \emph{baseline} case, $\ic = \rad$ in the \emph{binned} or \emph{shuffled} cases) for $\window_{\ell}^{\delta,\delta}(s)$ has a negligible impact on the measurement of cosmological parameters in our analysis. Hence, we do not include this correction in our cosmological fits.

\subsection{Cosmological fits with the radial integral constraint}
\label{sec:fits_radial}

\begin{figure}[t]
\centering
\includegraphics[width=0.45\textwidth]{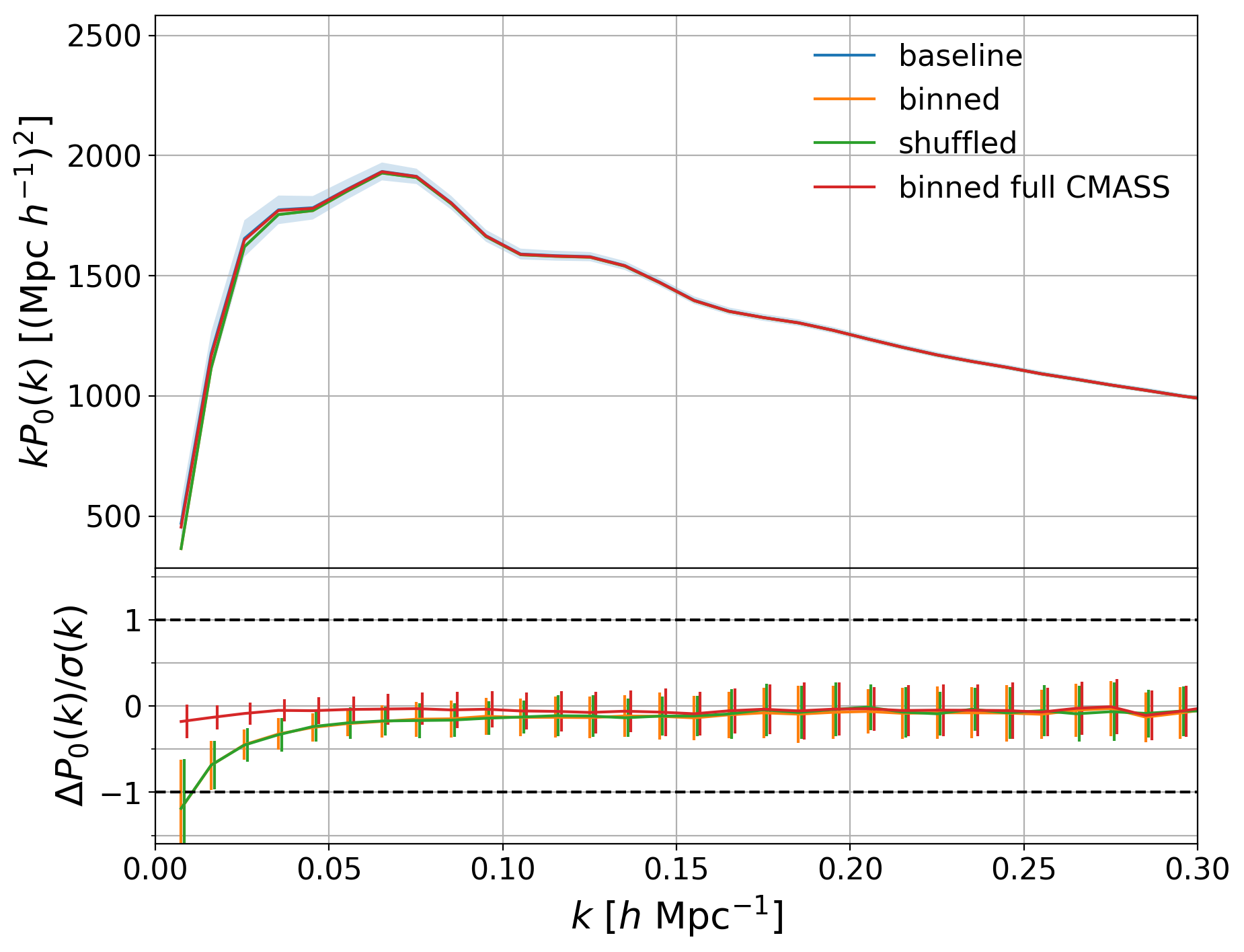}
\includegraphics[width=0.45\textwidth]{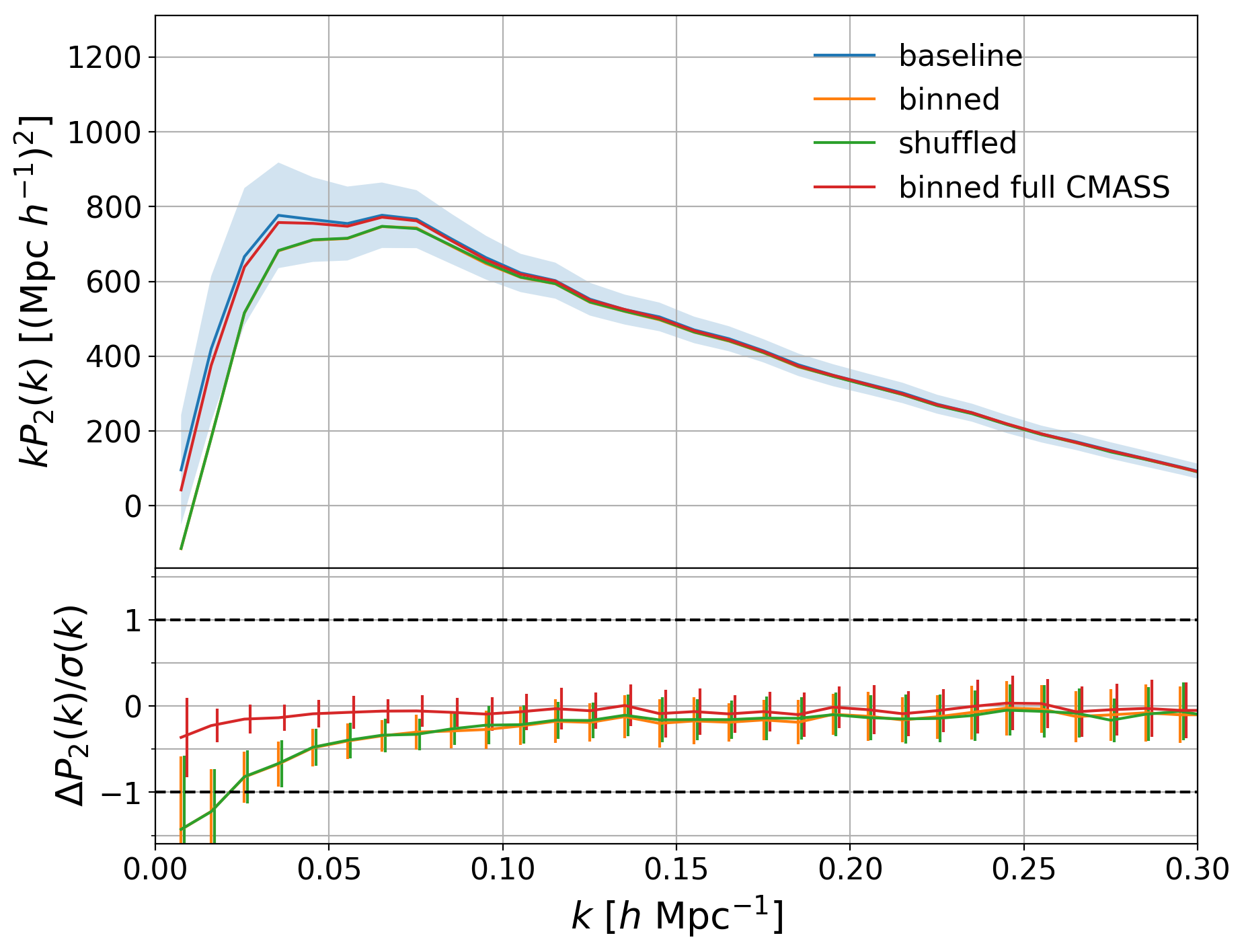}
\includegraphics[width=0.45\textwidth]{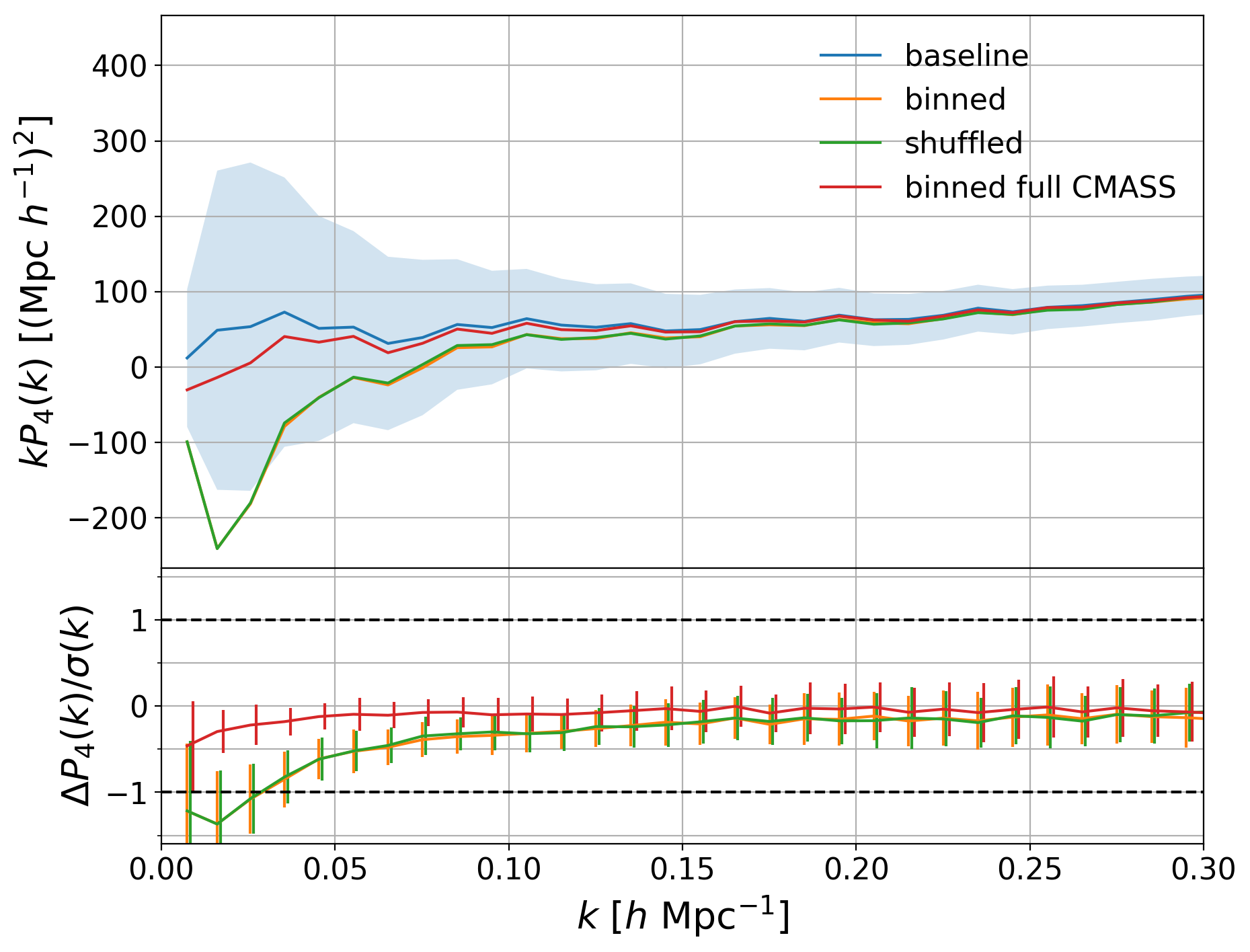}
\caption{Top panels: power spectrum multipoles (upper left: monopole, upper right: quadrupole, bottom: hexadecapole) measured from the 84 N-series mocks (see \Sec{mocks}), using three different ways to model the redshift distribution in the random catalogues. The true selection function is used in the \emph{baseline} (blue). The random redshift distribution of the \emph{binned} (orange) and \emph{shuffled} (green) schemes is inferred in $6$ separate chunks of each mock data realisation, as described in the text. For comparison, the red curve shows the effect of the \emph{binned} scheme applied to the full CMASS footprint. The blue shaded area represents the standard deviation of the mocks. Bottom panels: difference of the \emph{shuffled} and \emph{binned} schemes to the \emph{baseline}, with the standard deviation of the difference given by the error bars, normalised by the standard deviation of the mocks.
}
\label{fig:mocks_binned_shuffled}
\end{figure}

Figure~\ref{fig:mocks_binned_shuffled} illustrates the effect of tuning the radial selection function on data. Both the \emph{binned} scheme (using comoving distance bins of size $\delta r = 2 \Mpch$) and the \emph{shuffled} scheme result in a loss of power parallel to the line-of-sight, thus mostly affecting the power spectrum quadrupole and hexadecapole, compared to the \emph{baseline} relying on the true survey selection function. The effect of the radial integral constraint is significantly increased when it is imposed to the $6$ chunks separately (orange and red curves), which is the case considered in the following cosmological fits. We thus expect that neglecting the radial integral constraint would lead to a significant bias in the fitted cosmological parameters. Figure~\ref{fig:mocks_binned_shuffled} also shows that power spectrum measurements obtained with the \emph{binned} and \emph{shuffled} scheme are very similar. It is thus fair to account for e.g. the \emph{shuffled} scheme with the \emph{binned} scheme, for which we derived integral constraint corrections in \Sec{delta_radial}. We also checked that differences between these two schemes in the measurement of cosmological parameters are negligible (i.e. small compared to the uncertainty on the mean of the mocks).

\begin{figure}[t]
\centering
\includegraphics[width=0.45\textwidth]{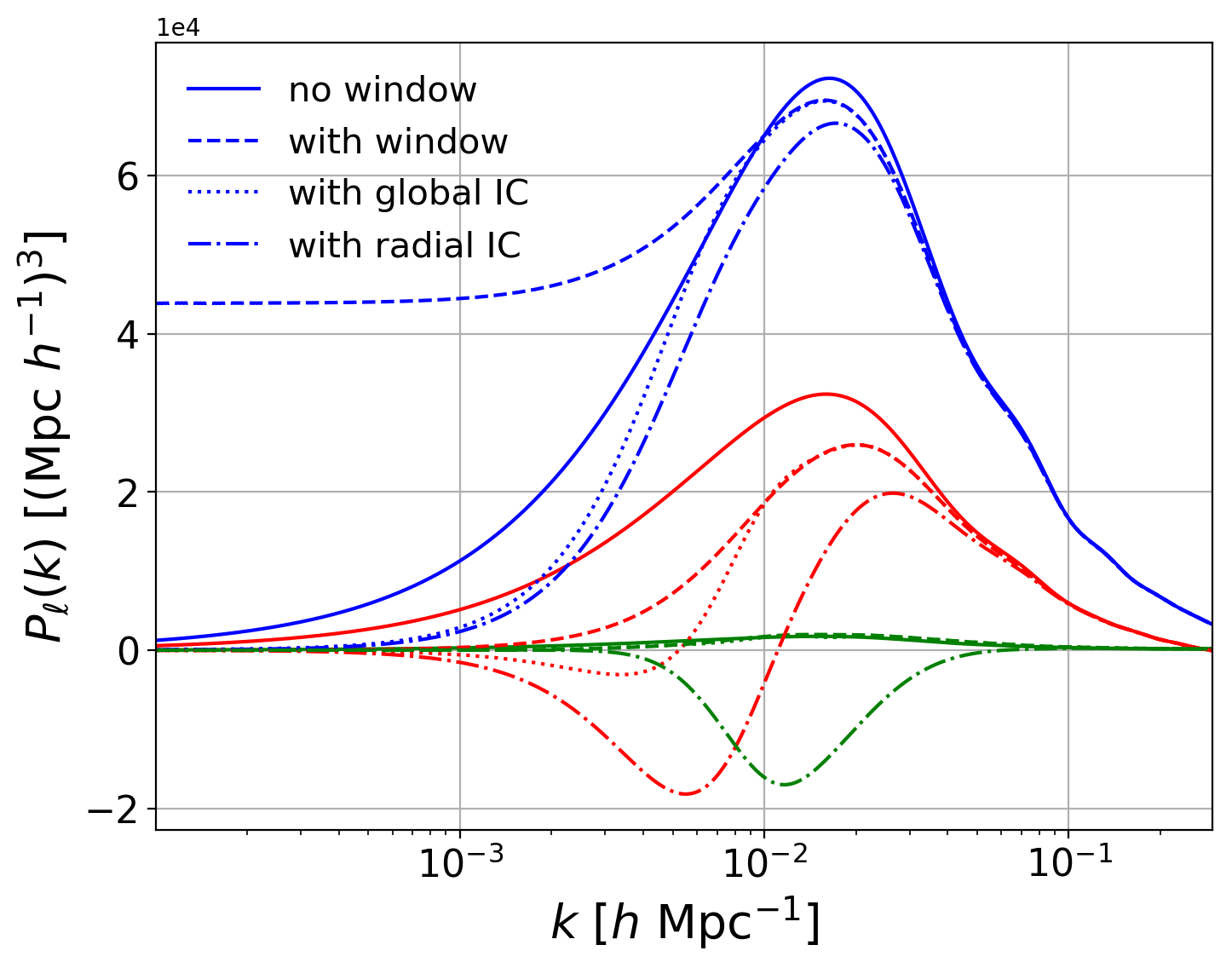}
\includegraphics[width=0.47\textwidth]{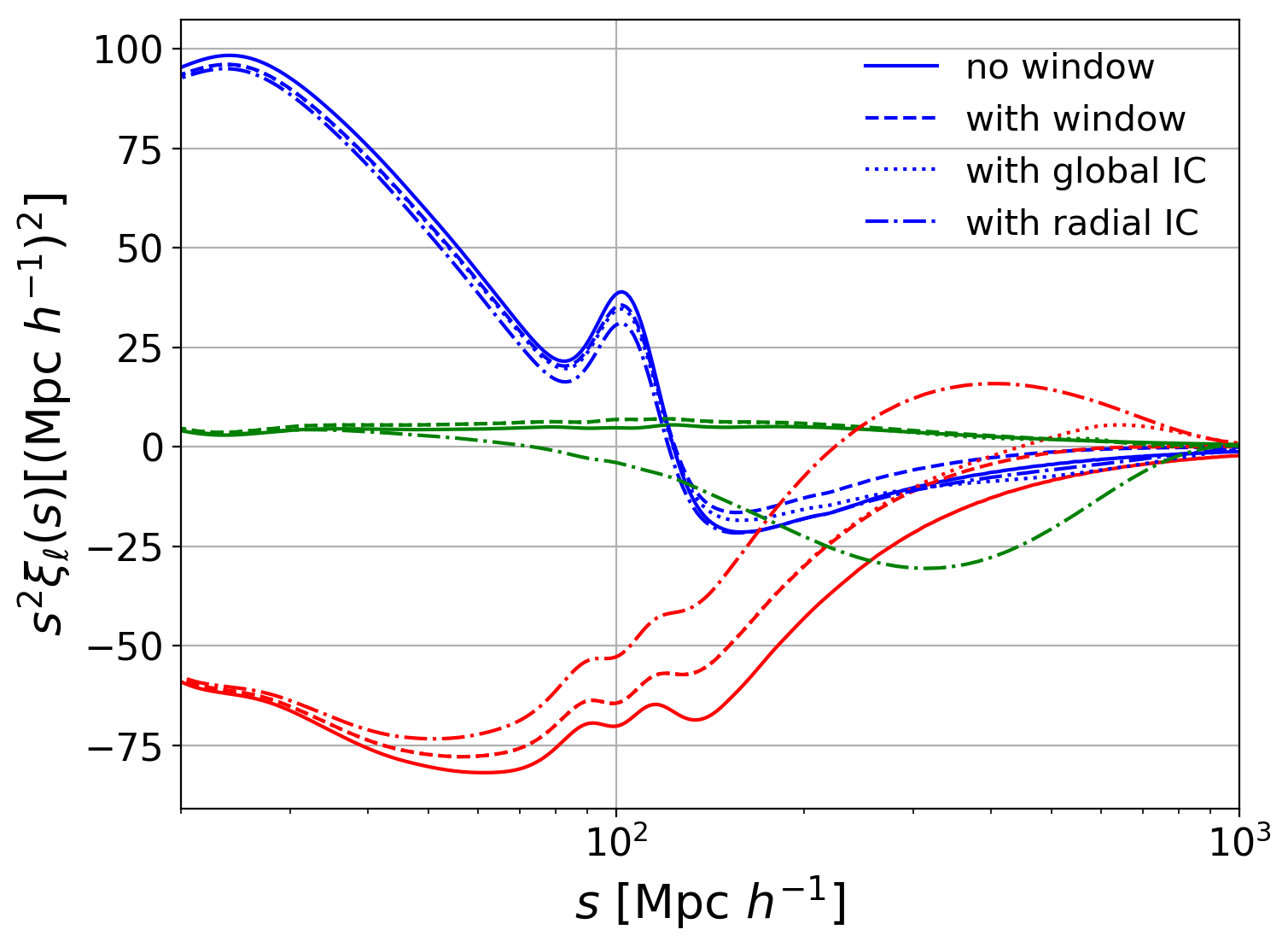}
\caption{Left: power spectrum multipoles (blue: monopole, red: quadrupole, green: hexadecapole) without (continuous line) and with the window function effect only (dashed lines), with the global integral constraint (dotted lines) and with the radial integral constraint (dash-dotted lines). Integral constraints ensure that the power spectrum monopoles reaches zero on large scales. Right: same, in configuration space. The window function effect is taken into account in the correlation function multipoles.
}
\label{fig:integral_constraint_effect_global_radial}
\end{figure}

Figure~\ref{fig:integral_constraint_effect_global_radial} displays the model for the global and radial integral constraint corrections corresponding to the CMASS footprint divided in 6 chunks. By construction, the power spectrum monopoles converge to zero at large scales when the global or radial IC is applied. We stress again that window functions are normalised according to \Sec{normalisation}, without using any low-$k$ nor low-$s$ limit. In the specific, illustrated case, the effect of the radial integral constraint is found to be large (comparable to the window function effect alone) in both Fourier and configuration space correlation functions. Note however that the latter prediction cannot be directly compared to measurements using e.g. the Landy-Szalay estimator~\cite{Landy1993} from which the window function effect is already removed.

\begin{figure}[t]
\centering
\includegraphics[width=0.45\textwidth]{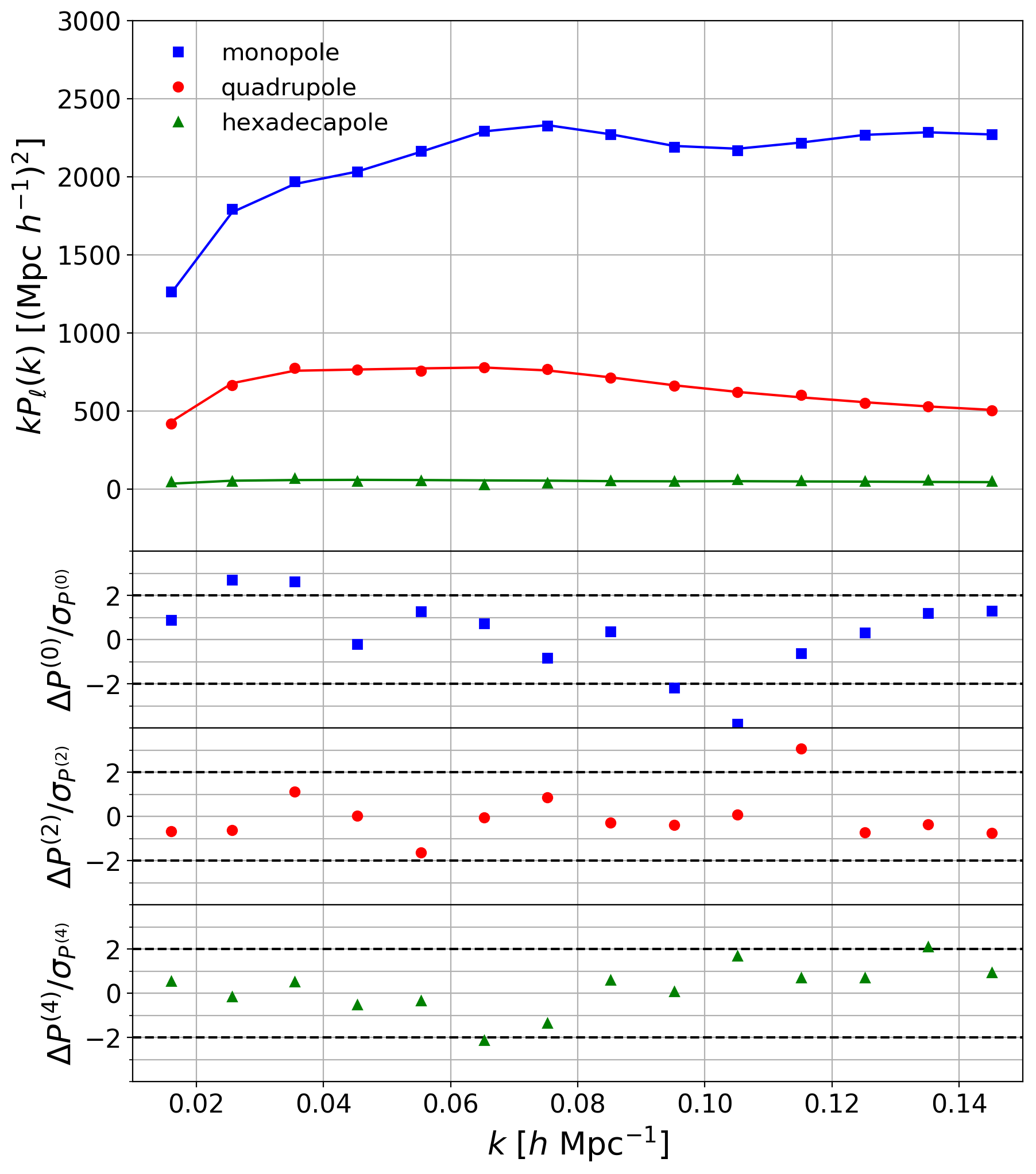}
\includegraphics[width=0.45\textwidth]{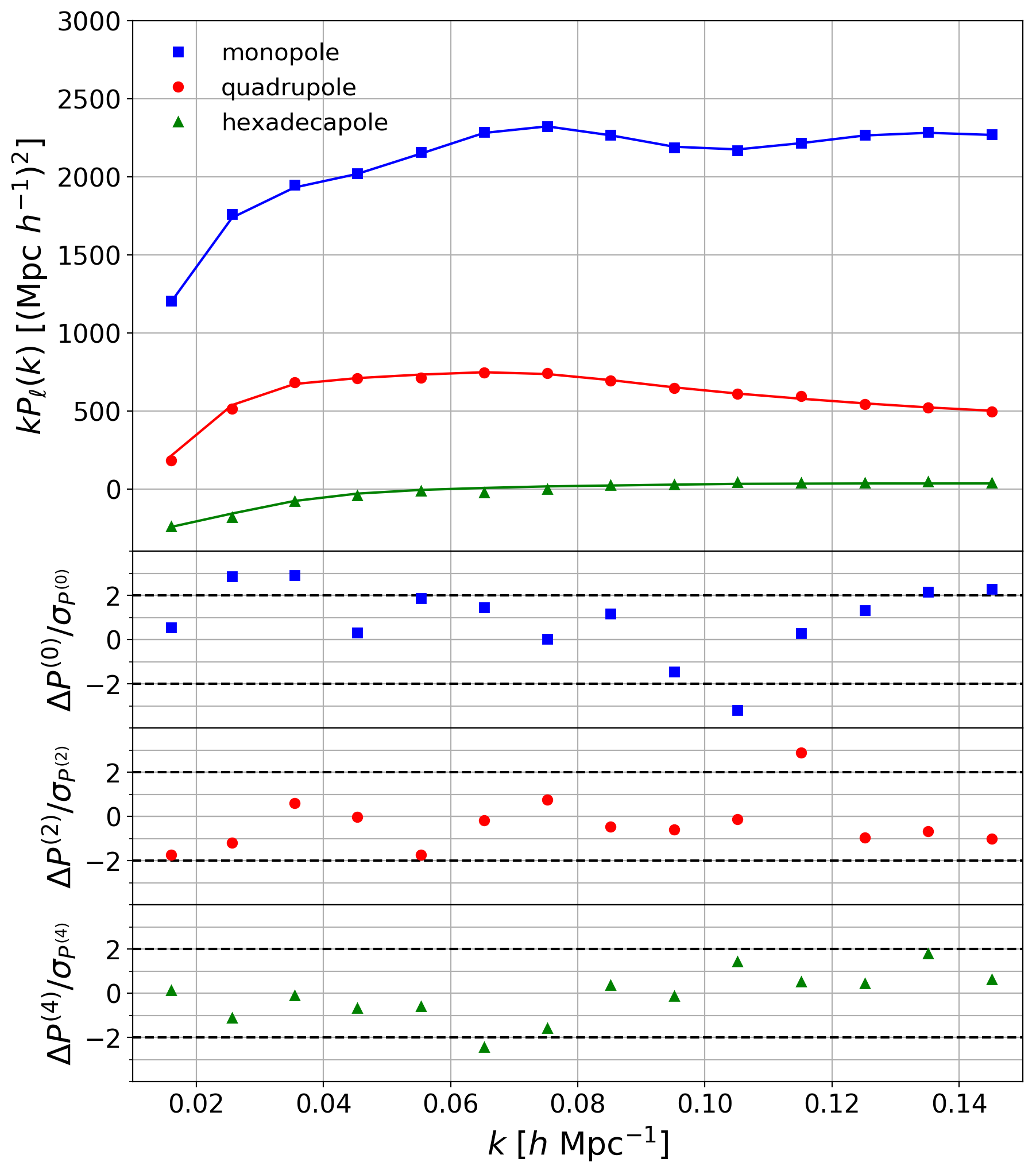}
\caption{Top panels: mean of the power spectrum multipoles (blue: monopole, red: quadrupole, green: hexadecapole) measured from the 84 N-series mocks (data points, including error bars) and the corresponding model predictions (continuous lines). Error bars give the uncertainty on the mean of the 84 mocks (standard deviation of the MultiDark Patchy mocks divided by $\sqrt{84}$) but are below the marker size. Bottom panels: residuals (difference of the measurements to the model divided by error bars). Left: the model is fitted to the mean of the \emph{baseline} mocks (see \Fig{mocks_binned_shuffled}), including the global integral constraint only. Right: the \emph{binned} scheme is applied to the mocks, and the radial integral constraint is included in the model (without refitting).}
\label{fig:residuals_radial}
\end{figure}

As a preview, \Fig{residuals_radial} illustrates the agreement between the mean of the 84 N-series mocks and the appropriate model in the \emph{baseline} (left panel) and the \emph{binned} scheme (right panel).
In the \emph{baseline} case, we fit the TNS model including the global integral constraint. The model performs well enough, given the high measurement precision (reduced $\chi^{2}=77.4/(42-7)=2.21$). Note that same cosmological and nuisance parameters are kept for the model in the right-hand plot, where the \emph{binned} scheme is applied to the mocks, and the radial integral constraint is included in the model. Here again, the agreement betwen the model and the mean of the mocks is correct (reduced $\chi^{2}=82.2/(42-7)=2.35$). We thus anticipate that our radial integral constraint correction will very well account for the \emph{binned} and \emph{shuffled} schemes in the cosmological fits.

\begin{figure}[t]
\centering
\includegraphics[width=0.5\textwidth]{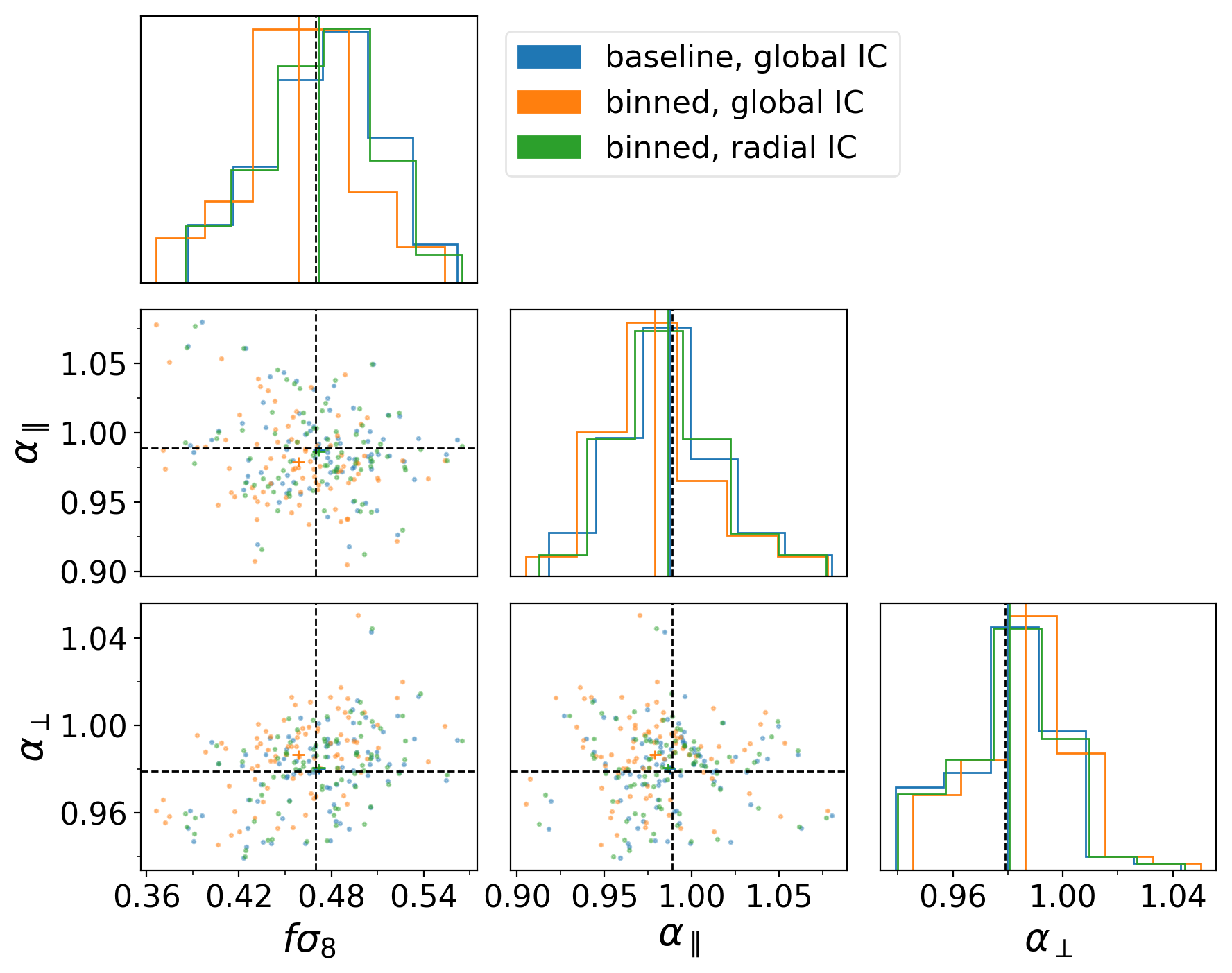}
\includegraphics[width=0.4\textwidth]{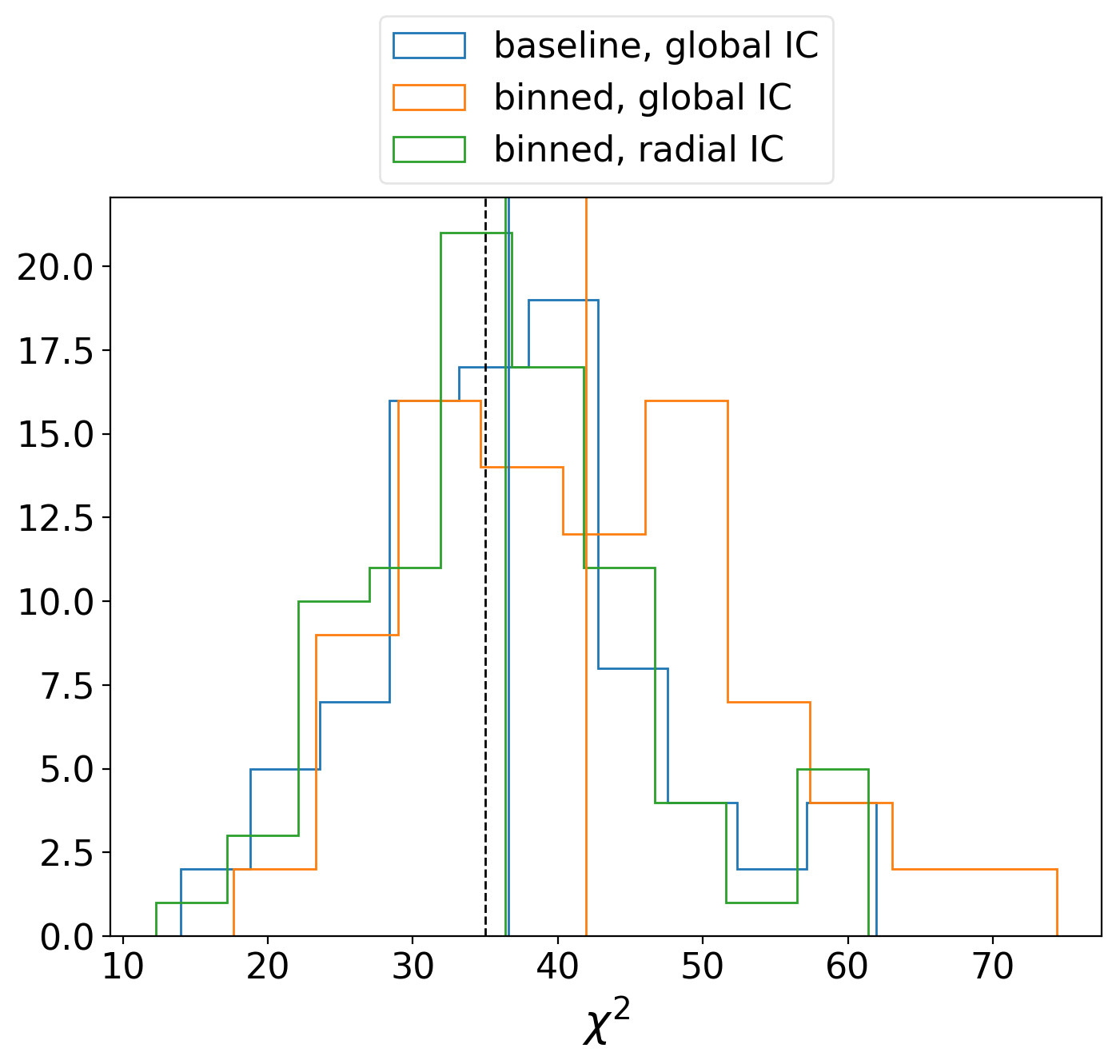}
\caption{Left: distributions of the cosmological parameters $f\sigma_{8}$, $\apar$, $\aper$ measured on the 84 N-series mocks. The \emph{baseline} (blue) uses the true selection function of the whole CMASS footprint. In orange, the \emph{binned} scheme is applied to the mocks. In green, the radial integral constraint is added to the model. Continuous lines give the mean of the 84 best fits; the size of the cross in the 2D plots is the standard deviation of the best fits divided by $\sqrt{84}$. Dashed lines show the expected values from the mock cosmology. Right: the corresponding $\chi^{2}$ distributions. The vertical dashed line shows the number of degrees of freedom ($42-7=35$).}
\label{fig:fits_radial}
\end{figure}

Figures~\ref{fig:fits_radial} and \Tab{fits_radial} present the cosmological fit results obtained in three different cases.

The \emph{baseline} cosmological fits (column~1 in \Tab{fits_radial}) are obtained with the power spectrum measurements on the \emph{baseline} mocks and with the global integral constraint applied in the model. The expected parameter values (column~4 in \Tab{fits_radial}), predicted from the mock fiducial cosmology, are recovered to the statistical uncertainty on the mean of the mocks.

Applying the \emph{binned} scheme to the mocks (see the orange curve in \Fig{mocks_binned_shuffled}) while modelling the global integral constraint only results in a bias (compared to the \emph{baseline} case) on all cosmological parameters of roughly $30 \%$ of the statistical error on one realisation (column~2 in \Tab{fits_radial}). The goodness-of-fit (probed by the $\chi^{2}$ distribution, \Fig{fits_radial}) is significantly degraded ($\Delta \chi^{2} \simeq 5$). 

The modelling of the radial integral constraint successfully removes the bias to better than the statistical uncertainty on the mean of the mocks (column~3 in \Tab{fits_radial}), and the goodness-of-fit is well recovered. No increase on cosmological parameter errors is detected. Similar results (not reported here) were obtained with the \emph{shuffled} scheme.

\begin{table}
\centering
\begin{tabular}{|c|c|c|c|c|}
\hline
& \begin{tabular}{c} Baseline \\ global IC \end{tabular} & \begin{tabular}{c} Binned \\ global IC \end{tabular} & \begin{tabular}{c} Binned \\ radial IC \end{tabular} & Expected\\
\hline
$\apar$ & $0.988 \pm 0.031$ & $0.979 \pm 0.032$ & $0.987 \pm 0.031$ & $0.989$ \\
$\aper$ & $0.980 \pm 0.018$ & $0.986 \pm 0.019$ & $0.980 \pm 0.018$ & $0.979$ \\
$f\sigma_{8}$ & $0.472 \pm 0.037$ & $0.459 \pm 0.037$ & $0.472 \pm 0.037$ & $0.470$ \\
\hline
$\chi^{2}$ & $36.6 \pm 9.7$ & $42 \pm 11$ & $36.4 \pm 9.6$ & $\mathrm{ndof} = 35$\\
\hline
\end{tabular}
\caption{Mean and standard deviation of the cosmological parameters fitted on the 84 N-series mocks, corresponding to \Fig{fits_radial}. Error bars should be divided by $\sqrt{84} \sim 10$ to obtain errors on the mean of the mocks.}
\label{tab:fits_radial}
\end{table}

\subsection{Discussion}
\label{sec:analysis_radial_discussion}

We showed that the tuning of the radial selection function on data may significantly bias RSD measurements. We recall that the radial IC effect has been purposedly enhanced to test the robustness of our predictions by applying the \emph{binned} scheme separately in $6$ small chunks cut into the CMASS footprint. After modelling of the induced radial integral constraint, the bias on $f\sigma_{8}$ and on Alcock-Paczynski parameters is well below the uncertainty on the mean of the mocks. More statistics would be required to test our modelling further, but we note that the obtained bias is already below the expected $\simeq 1 \%$ statistical uncertainty on $f\sigma_{8}$ provided by future galaxy surveys~\cite{DESI2016:1611.00036v2,Laureijs2009:0912.0914v1}. 

The red curve in \Fig{mocks_binned_shuffled} shows that the \emph{shuffled} (or \emph{binned}) scheme applied to the CMASS NGC footprint as a whole (i.e. without the division in $6$ chunks) induces a small radial integral constraint on clustering measurements, as already noted on SDSS DR7 and DR9 by~\cite{Samushia2012:1102.1014,Ross2012:1203.6499v3}. We checked that not accounting for the radial IC in the model results in a negligible shift of cosmological parameters (compared to the statistical uncertainty) $\Delta \apar = -0.003$, $\Delta \aper = 0.001$ and $\Delta f\sigma_{8} = -0.001$. Note that, contrary to our test case, the BOSS DR12 analysis used the NGC and SGC CMASS and LOWZ samples. These were divided in three overlapping redshift slices~\cite{Alam:1607.03155v1}, thus probably decreasing the radial integral constraint correction with respect to the pure window function effect. We thus infer that conclusions of the BOSS DR12 analysis are likely unaffected by the radial integral constraint effect.

Nonetheless, RSD measurements based on deep, pencil-like or small surveys such as the eBOSS ELG program~\cite{Raichoor2017:1704.00338v1} may suffer from the radial integral constraint, if not correctly accounted for. Also, analyses focusing on large scales, aiming at e.g. setting constraints on primordial non-Gaussianity~\cite{Ross2012:1208.1491v3,Castorina2019} may benefit from accounting for the radial IC.

\section{Angular integral constraint to mitigate angular systematics}
\label{sec:analysis_angular}

In this section we show that potential unknown angular systematics (\Sec{analysis_angular_problem}) can be mitigated using an angular integral constraint (\Sec{fits_angular}) which can be combined with the radial one (\Sec{fits_radialxangular}). Though we note in \Sec{multiplicative_systematics} that only the additive part of systematics is removed in this way, we emphasise that the technique described hereafter may be simply used as a consistency check in clustering analyses.

\subsection{Problem statement}
\label{sec:analysis_angular_problem}

As the radial selection function, the angular selection function of a spectroscopic survey can be difficult to evaluate because of residual photometric calibration errors or other potential photometric systematics. We consider these systematics to be completely unknown; that is, they are not and cannot be corrected by any photometric template in the analysis.

\begin{figure}[t]
\centering
\includegraphics[width=0.45\textwidth]{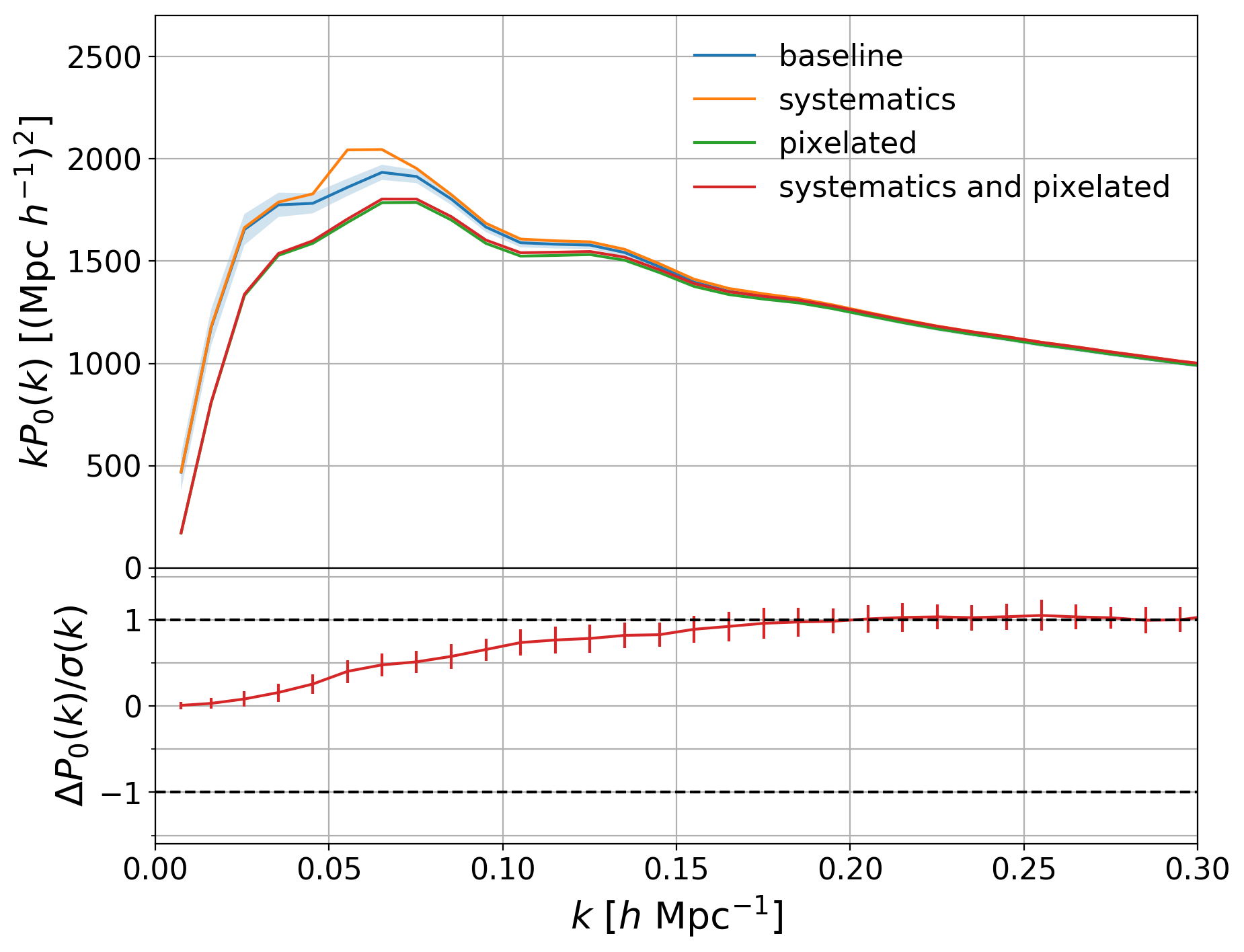}
\includegraphics[width=0.45\textwidth]{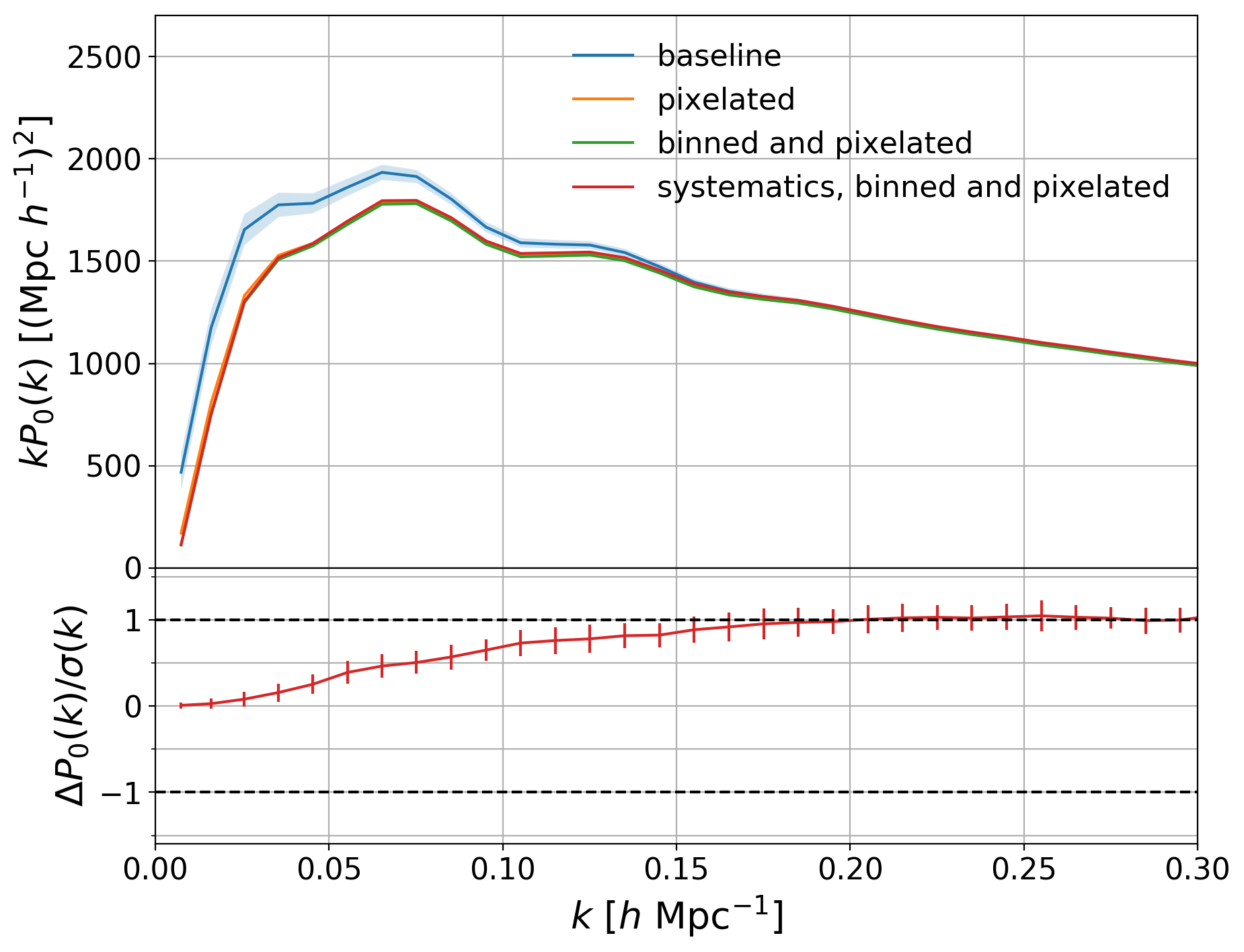}
\includegraphics[width=0.45\textwidth]{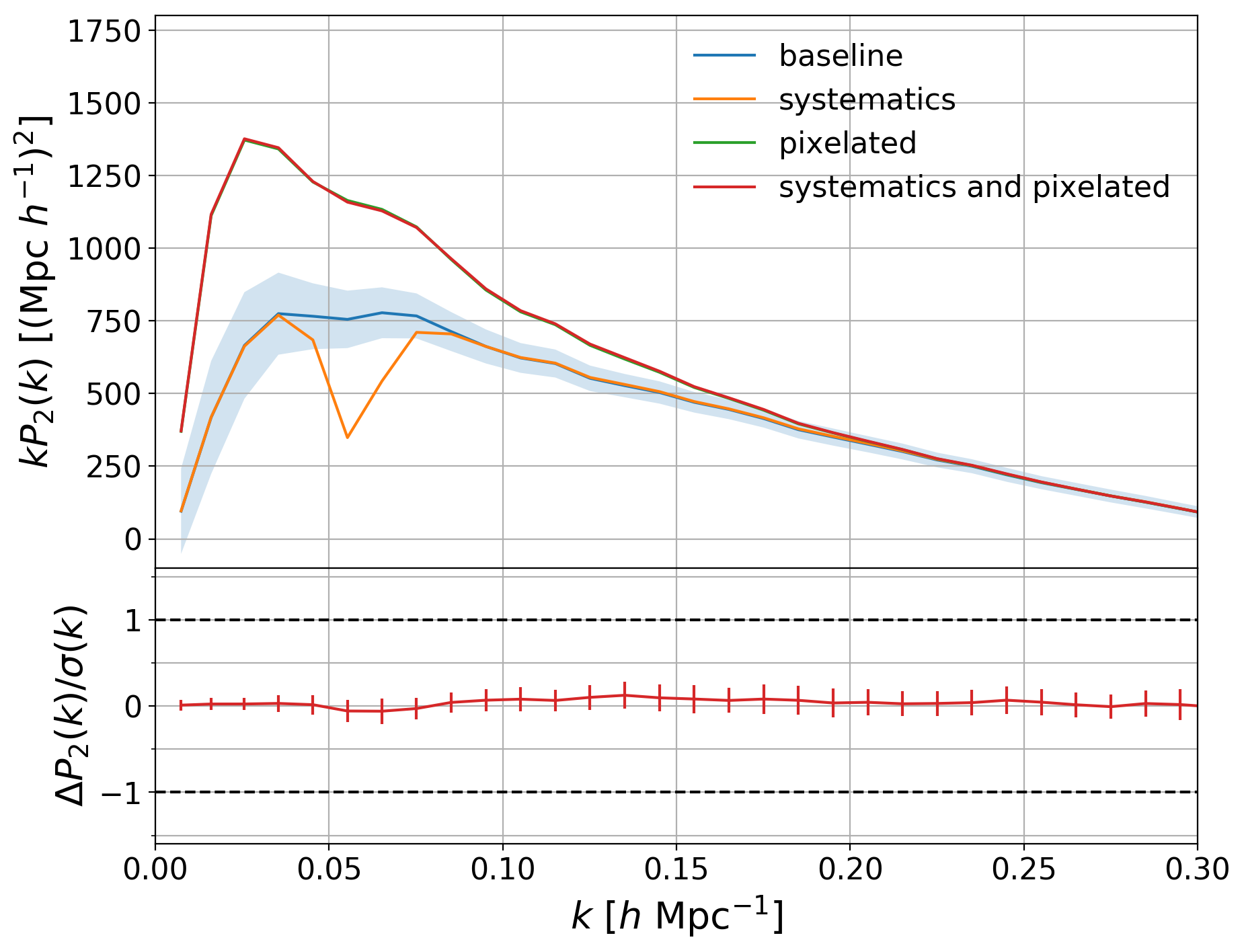}
\includegraphics[width=0.45\textwidth]{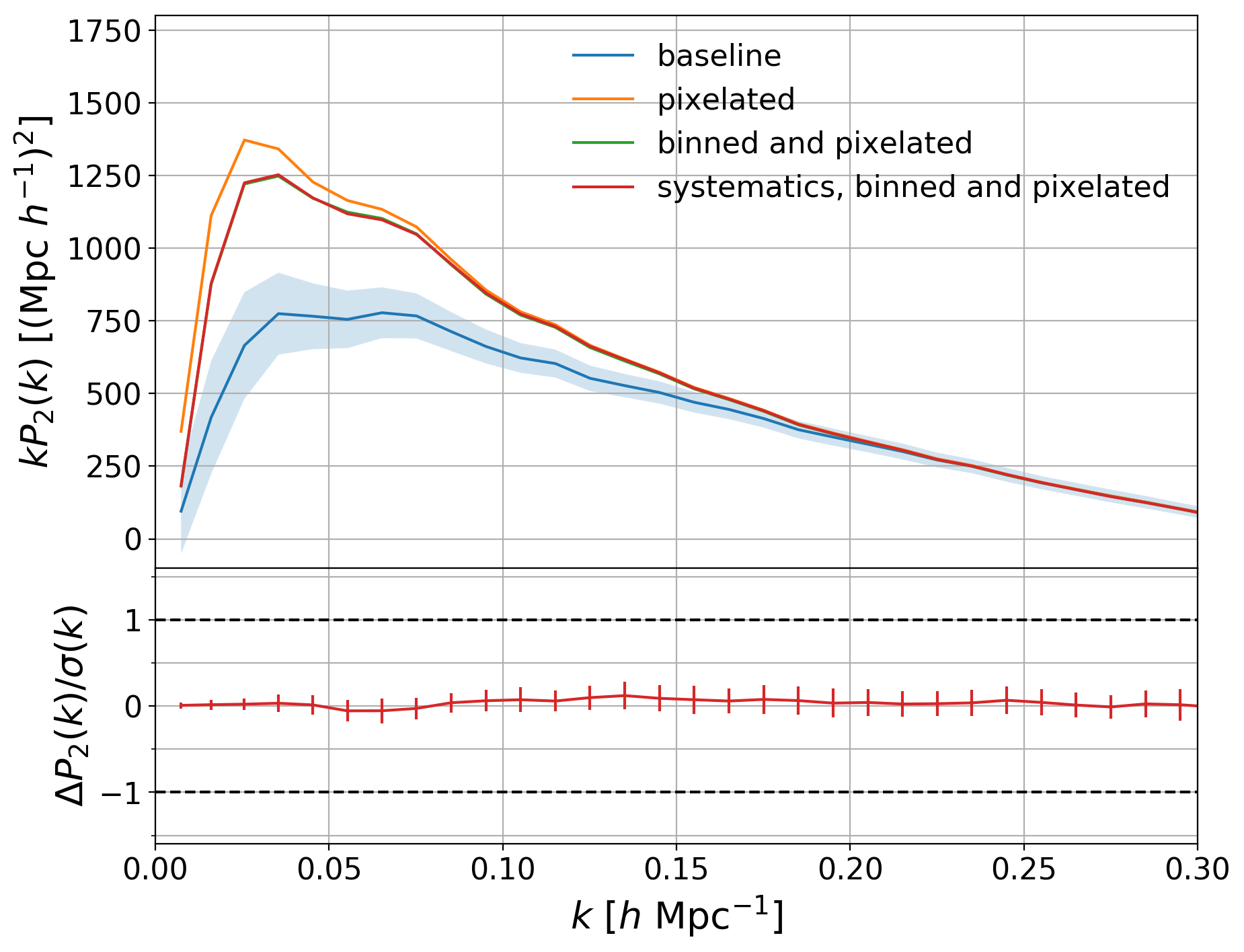}
\includegraphics[width=0.45\textwidth]{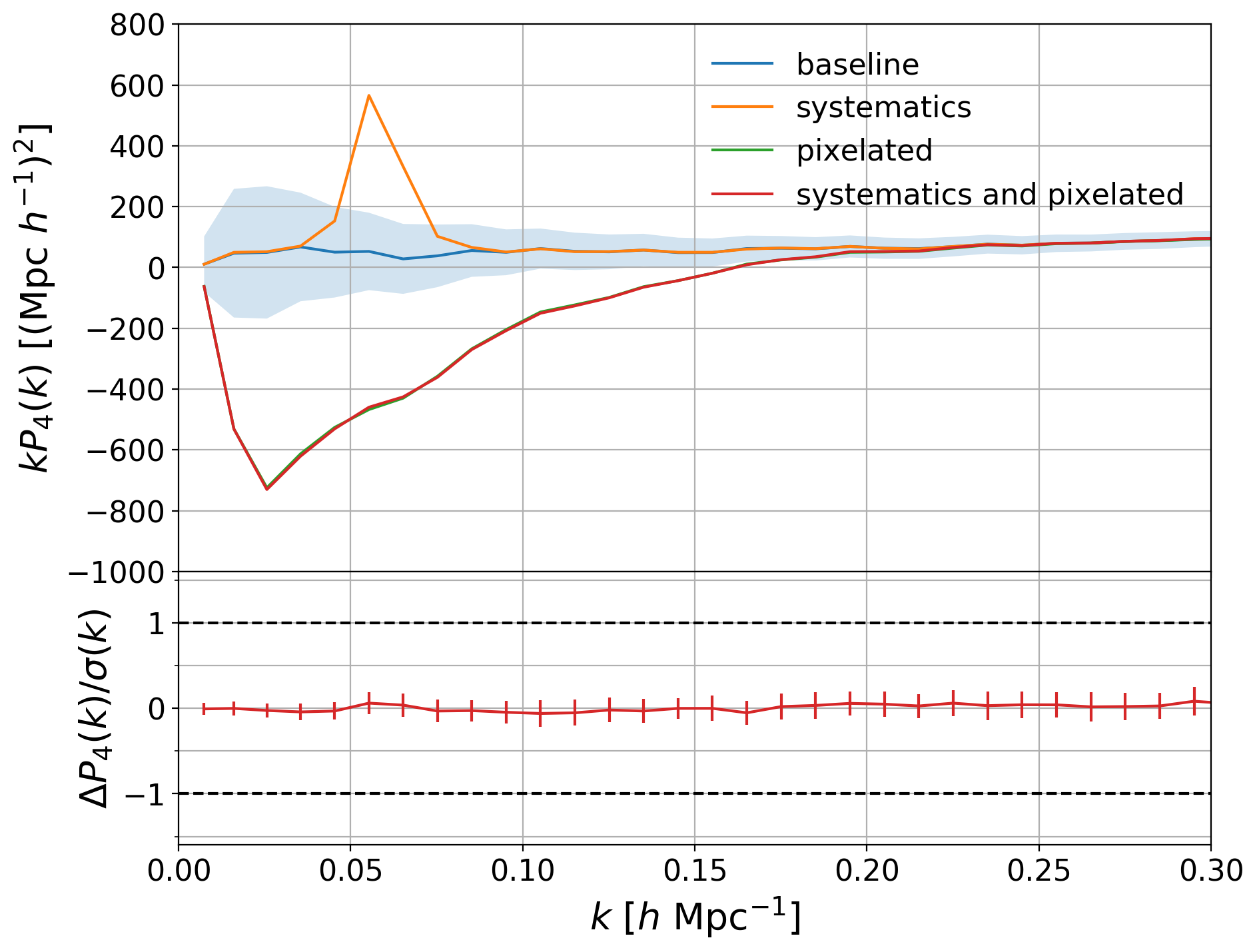}
\includegraphics[width=0.45\textwidth]{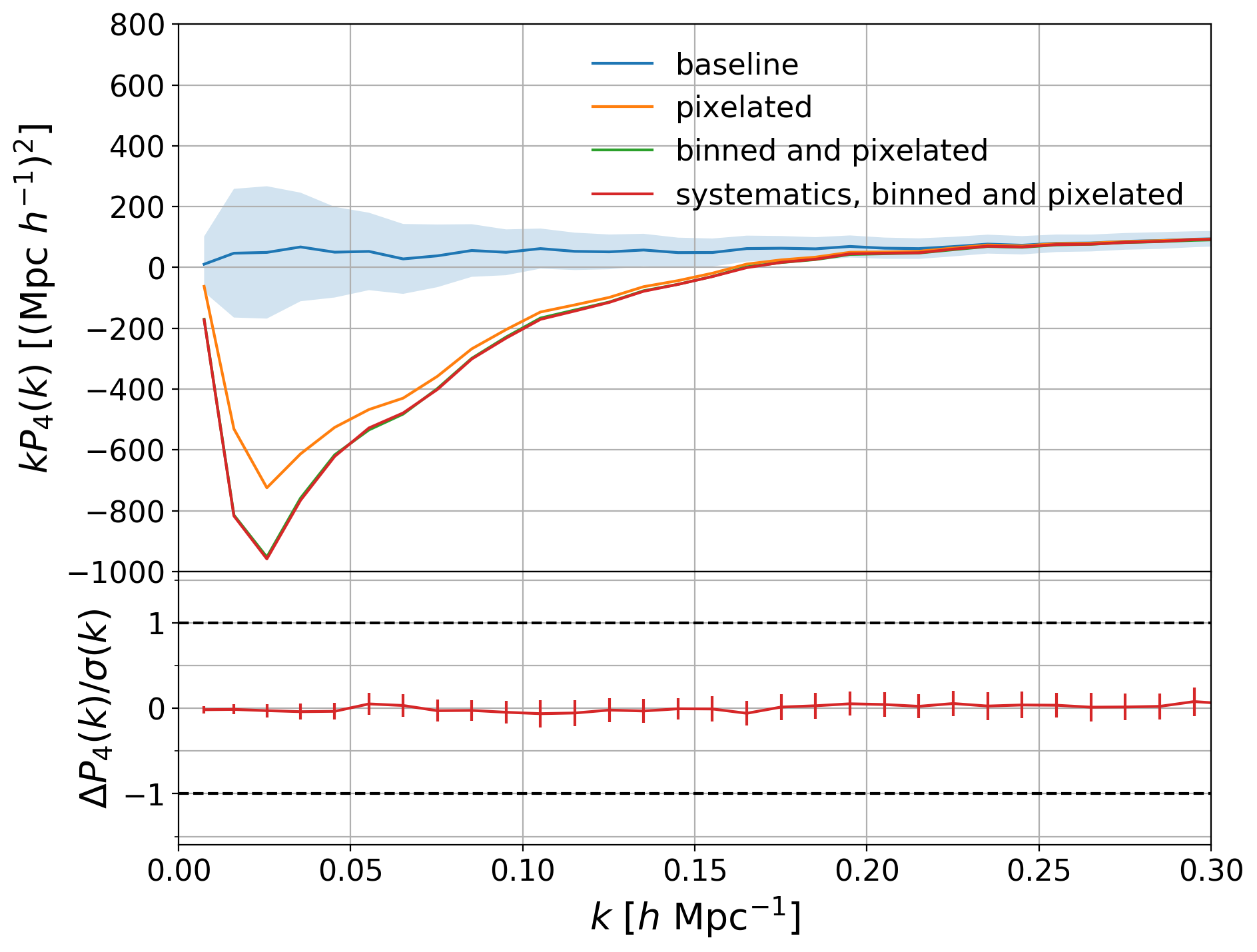}
\caption{
Left column, top panels: power spectrum multipoles (top: monopole, middle: quadrupole, bottom: hexadecapole) measured from the 84 N-series mocks: \emph{baseline} measurements (blue), with angular systematics (orange, \Eq{weight_systematics}), using the \emph{pixelated} scheme applied to uncontaminated (green) and contaminated (red) mocks (see text for details). Left column, bottom panels: difference of the \emph{pixelated} scheme with and without angular systematics, with the standard deviation of the difference given by the error bars, normalised by the standard deviation of the mocks. Right column, top panels: \emph{baseline} (blue), \emph{pixelated} scheme (orange, same as green in the left column), \emph{binned}  and \emph{pixelated} schemes applied to uncontaminated (green) and contaminated (red) mocks. Right column, bottom panels: normalised difference of the \emph{binned} and \emph{pixelated} scheme with and without angular systematics. In both columns, the blue shaded area represents the standard deviation of the mocks.
}
\label{fig:mocks_angular_systematics}
\end{figure}

As a case study, we inject photometric systematics into our mocks as a function of right ascension (R.A.) and declination (Dec.), using a weight:
\begin{equation}
w_{\mathrm{sys}} = 1 + 0.2 \sin{\left(\frac{2\pi}{10(\deg)} \mathrm{R.A.}(\deg) \right)} \sin{\left(\frac{2\pi}{5(\deg)} \mathrm{Dec} (\deg) \right)}.
\label{eq:weight_systematics}
\end{equation}
The amplitude of the systematics ($\pm 20\%$) is very large compared to the typical requirements of a target selection (e.g. $\pm 7.5 \%$ in \cite{Prakash2015:1508.04478v1}). $w_{\mathrm{sys}}$ has a drastic impact on power spectrum measurements as can be seen on the left-hand plots of \Fig{mocks_angular_systematics} (orange and blue curves). A standard RSD analysis, using scales $0.01 \hMpc < k < 0.15 \hMpc$ would be impossible. We will see that these systematics can be strongly reduced using a similar procedure as in \Sec{analysis_radial}.

\subsection{Angular integral constraint}
\label{sec:fits_angular}

We suggest to remove the contaminated modes by weighting randoms from the synthetic catalogue by $\sum_{i \in \mathrm{pixel}} w_{g,i} / \sum_{i \in \mathrm{pixel}} w_{s,i}$ in pixels (the \emph{pixelated} scheme), thus nulling the density fluctuations in each pixel. We use a \texttt{HEALPix}\footnote{\url{http://healpix.jpl.nasa.gov/}}~\cite{Gorski2005:astro-ph/0409513v1} map with $\mathrm{nside} = 64$ (pixel area of $\simeq 0.84 \deg^2$), for which the contaminated and uncontaminated mocks look similar (see red and green curves in \Fig{mocks_angular_systematics}, left). The \emph{pixelated} scheme completely mitigates angular systematics in the quadrupole and hexadecapole. However, a difference close to a scale factor and statistically significant at small scales remains in the monopole.

We model the \emph{pixelated} scheme by an angular integral constraint. A similar idea is developed in~\cite{Burden2017:1611.04635v2} to mitigate the impact of the DESI~\cite{DESI2016:1611.00036v2} fiber assignment.

Formulae are directly deduced from \Sec{ic_general} by replacing $\epsilon_{\ic}(\vx,\vy)$ by $\epsilon_{\ang}(\vx,\vy)$, being non-zero if $\vx$ and $\vy$ lie within the same pixel. Contrary to the radial IC case (\emph{binned} or \emph{shuffled} schemes), for the angular IC (\emph{pixelated} scheme) one has access to an estimate of the survey selection function accounting for known systematics (weighted synthetic catalogue), independent of the observed data. Thus, in the following, window function calculations are based on the true survey selection function, as in the \emph{baseline} case. The angular IC correction is calculated from the same window functions for both uncontaminated and contaminated \emph{pixelated} mocks, though in the latter case the normalisation of the survey selection function would be inferred from contaminated data in a practical analysis. However, we do not expect any systematic bias from this simplification, as introduced systematics are independent of the actual clustering.

\begin{figure}[t]
\centering
\includegraphics[width=0.5\textwidth]{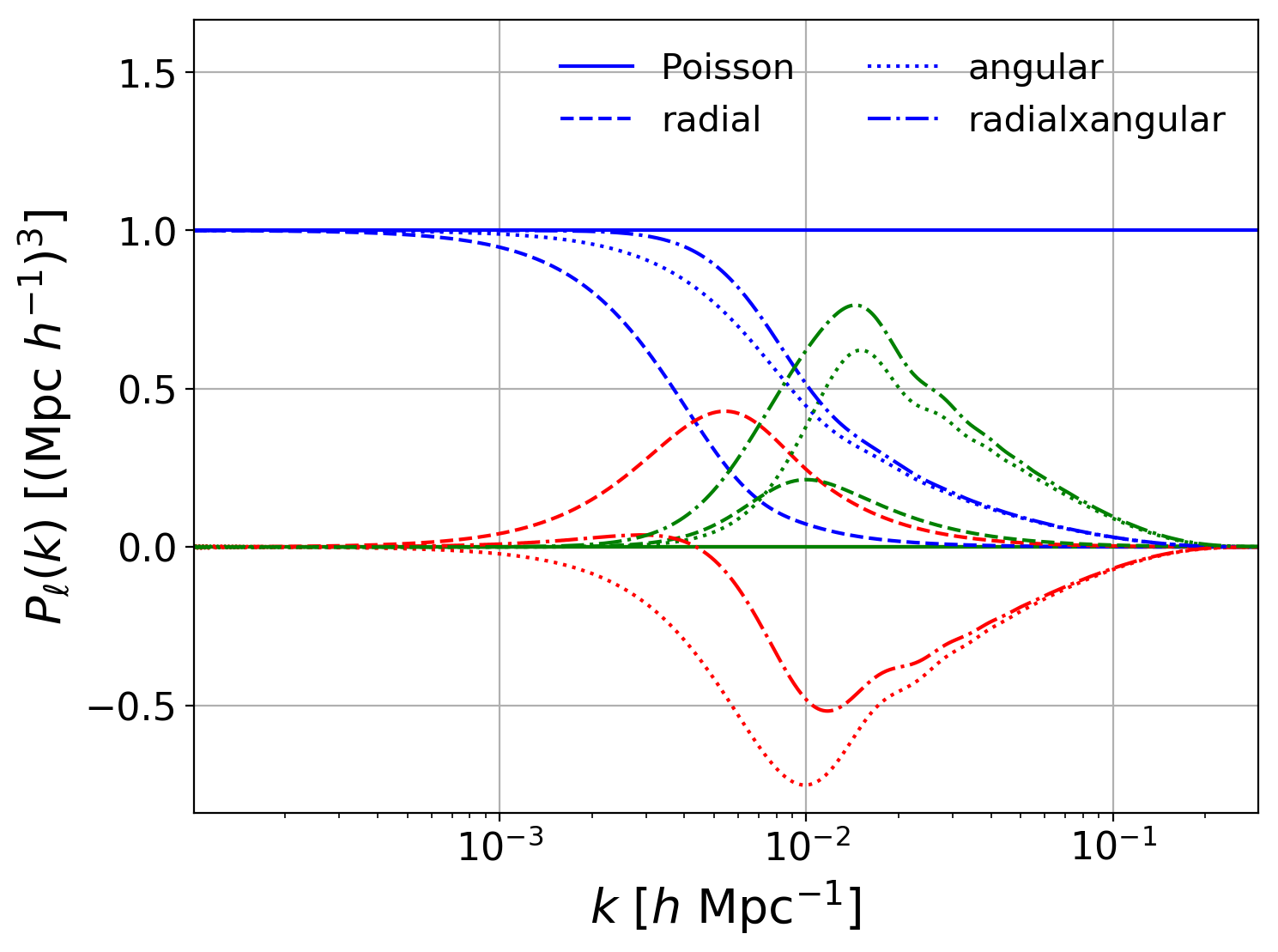}
\caption{Normalised shot noise contributions to the radial, angular, and combined (radial x angular) integral constraints (blue: monopole, red: quadrupole, green: hexadecapole).}
\label{fig:shot_noise_radialxangular}
\end{figure}

The shot noise contribution to the angular integral constraint is very large in the quadrupole and hexadecapole, as can be seen in \Fig{shot_noise_radialxangular}.

\begin{figure}[t]
\centering
\includegraphics[width=0.5\textwidth]{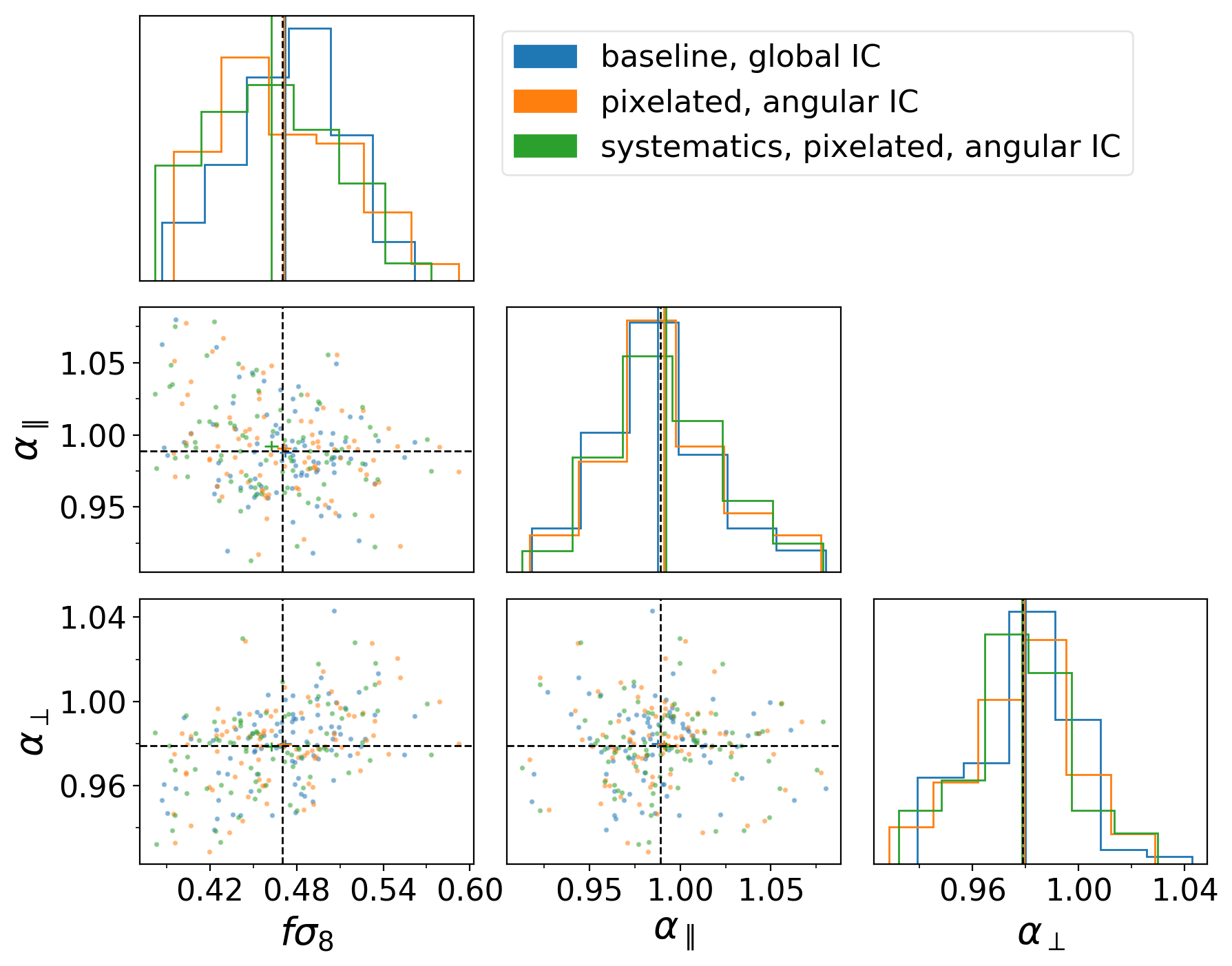}
\includegraphics[width=0.4\textwidth]{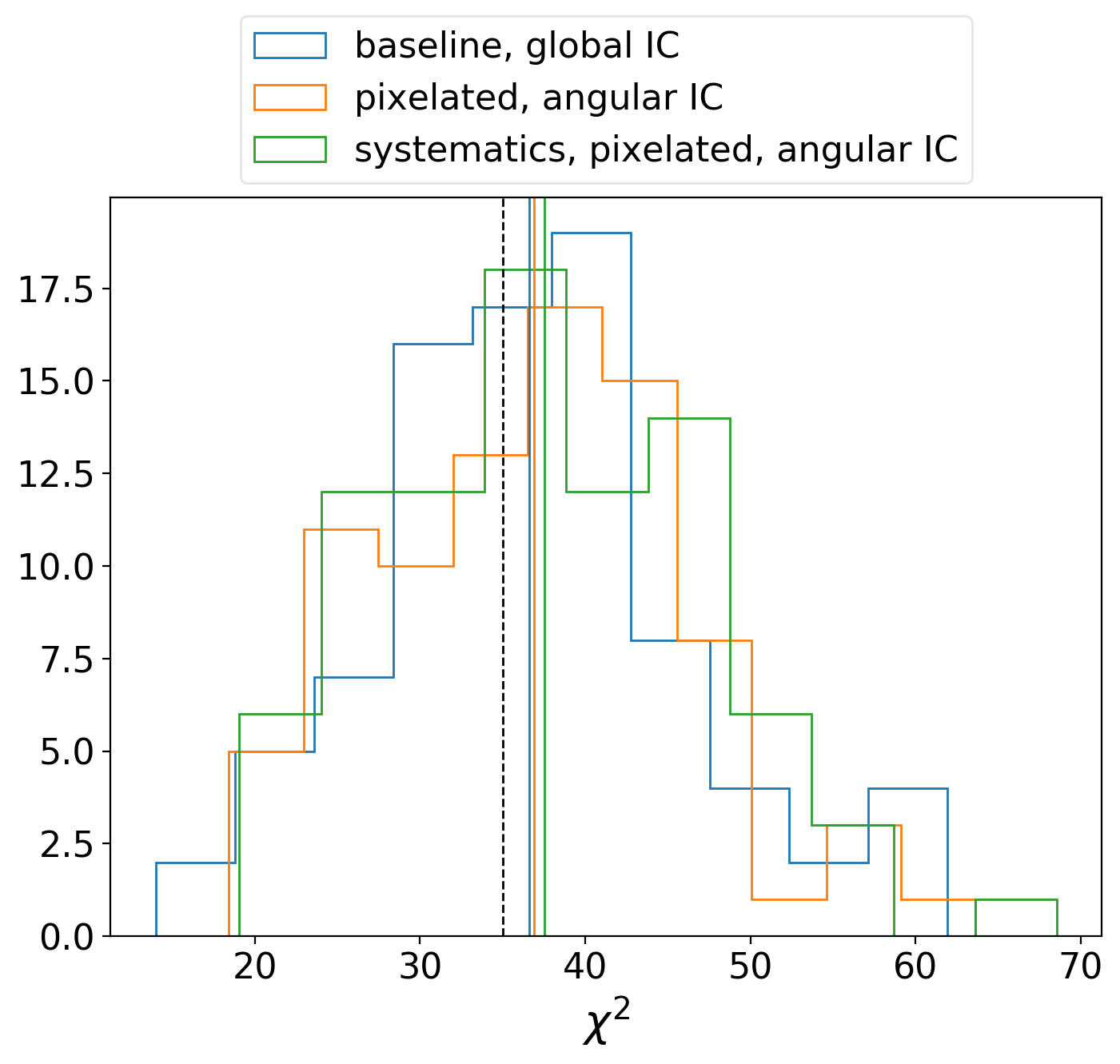}
\caption{Left: distributions of the cosmological parameters $f\sigma_{8}$, $\apar$, $\aper$ measured on the 84 N-series mocks. The \emph{baseline} (blue) uses the true selection function of the whole CMASS footprint. In orange, the \emph{pixelated} scheme is applied on the mocks and the angular integral constraint is used in the model. In green, angular systematics (\Eq{weight_systematics}) are added onto the mocks. Right: the corresponding $\chi^{2}$ distributions.}
\label{fig:fits_angular}
\end{figure}

In all this section, we use the same model and fitting $k$-range as in \Sec{analysis_radial}. Applying the \emph{pixelated} scheme to uncontaminated mocks, cosmological parameters are well recovered when modelling the angular integral constraint: as shown in \Fig{fits_angular} and \Tab{fits_angular}, differences with the \emph{baseline} case (same as in \Sec{analysis_radial}) are within the uncertainty on the mean of the mocks (column~2 in \Tab{fits_angular}). The error bar on $f\sigma_{8}$ increases by $22 \%$.

The magnitude of angular systematics makes it impossible to perform any relevant standard cosmological fit: we measure biases of $\Delta \apar = 0.043$, $\Delta \aper = -0.037$ and $\Delta f\sigma_{8} = -0.17$. On the contrary, the cosmological analysis remains possible when applying the \emph{pixelated} scheme to the contaminated mocks and the angular integral constraint in the model (column~3 in \Tab{fits_angular}). Alcock-Paczynski parameters are recovered within the statistical uncertainty on the mean of the mocks, and a bias of $\simeq 20 \%$ of the error on a single realisation can be seen on $f\sigma_{8}$. A lower bias on the growth rate would be expected with smaller, more realistic angular systematics.

\begin{table}
\centering
\begin{tabular}{|c|c|c|c|c|}
\hline
& \begin{tabular}{c} Baseline \\ global IC \end{tabular} & \begin{tabular}{c} Pixelated \\ angular IC \end{tabular} & \begin{tabular}{c} Systematics, pixelated \\ angular IC \end{tabular} & Expected\\
\hline
$\apar$ & $0.988 \pm 0.031$ & $0.991 \pm 0.032$ & $0.992 \pm 0.032$ & $0.989$ \\
$\aper$ & $0.980 \pm 0.018$ & $0.979 \pm 0.020$ & $0.979 \pm 0.020$ & $0.979$ \\
$f\sigma_{8}$ & $0.472 \pm 0.037$ & $0.471 \pm 0.045$ & $0.463 \pm 0.044$ & $0.470$ \\
\hline
$\chi^{2}$ & $36.6 \pm 9.7$ & $36.9 \pm 9.3$ & $37.5 \pm 9.7$ & $\mathrm{ndof} = 35$\\
\hline
\end{tabular}
\caption{Mean and standard deviation of the cosmological parameters fitted on the 84 N-series mocks, corresponding to \Fig{fits_angular}. Error bars should be divided by $\sqrt{84} \sim 10$ to obtain errors on the mean of the mocks.}
\label{tab:fits_angular}
\end{table}

\subsection{Combining radial and angular integral constraints}
\label{sec:fits_radialxangular}

One would probably like to combine the radial and angular integral constraints, to account for both unknown radial and angular selection functions. A prerequisite to applying the radial integral constraint is that the redshift distribution does not depend on the angular position on the sky in each chunk of the survey. Thus, both radial and angular integral constraints can be imposed at the same time since radial and angular selection functions are independent.

Right-hand plots of \Fig{mocks_angular_systematics} show the effects of the combined \emph{binned} and \emph{pixelated} schemes on N-series mocks. They add up in the hexadecapole, and partially cancel in the quadrupole. This is expected as $\Leg{2}$ is negative around $\mu = 0$ (where the angular integral constraint removes signal) and positive around $\mu = 1$ (where the radial integral constraint plays up). As with the \emph{pixelated} scheme alone (left-hand plots of \Fig{mocks_angular_systematics}), angular systematics are mitigated in the quadrupole and hexadecapole, but a multiplicative effect remains in the monopole.

Let us model the radial and angular integral constraints in a row:
\begin{equation}
\begin{split}
\delta^{\cic}(\vr) = W(\vr) & \left\lbrace \delta(\vr) - \int d^{3} x W_{\rad}(\vx)\epsilon_{\rad}(\vr,\vx)\delta(\vx) - \int d^{3} x W_{\ang}(\vx)\epsilon_{\ang}(\vr,\vx)\delta(\vx) \right. \\
& \left. + \int d^{3} x W_{\ang}(\vx)\epsilon_{\ang}(\vr,\vx)\int d^{3} y W_{\rad}(\vy)\epsilon_{\rad}(\vx,\vy)\delta(\vy) \right\rbrace.
\end{split}
\label{eq:delta_radialxangular_cic}
\end{equation}
Since the radial and angular parts of $W(\vr)$ are independent, the last term is just the integral of the density contrast over the whole (chunk) footprint, i.e. the global integral constraint. Thus, the two integral constraints commute. Then, building up the correlation function multipoles of the observed density fluctuations~\eqref{eq:delta_radialxangular_cic}, we find 16 terms:
\begin{subequations}
\begin{align}
\label{eq:combined_ic_radial}
\xi_{\ell}^{\cic}(s) & = \xi_{\ell}^{\mathrm{c}}(s) - IC_{\ell}^{\delta,\rad}(s) - IC_{\ell}^{\rad,\delta}(s) + IC_{\ell}^{\rad,\rad}(s) \\
\label{eq:combined_ic_angular}
& - IC_{\ell}^{\delta,\ang}(s) - IC_{\ell}^{\ang,\delta}(s) + IC_{\ell}^{\ang,\ang}(s) \\
\label{eq:combined_ic_global}
& + IC_{\ell}^{\delta,\glo}(s) + IC_{\ell}^{\glo,\delta}(s) + IC_{\ell}^{\glo,\glo}(s) \\
\label{eq:combined_ic_radialxangular}
& - IC_{\ell}^{\glo,\rad}(s) - IC_{\ell}^{\rad,\glo}(s) - IC_{\ell}^{\glo,\ang}(s) - IC_{\ell}^{\ang,\glo}(s) \nonumber \\
& + IC_{\ell}^{\rad,\ang}(s) + IC_{\ell}^{\ang,\rad}(s).
\end{align}
\end{subequations}
Terms~\eqref{eq:combined_ic_radial},~\eqref{eq:combined_ic_angular} and~\eqref{eq:combined_ic_global} correspond to the radial, angular and global integral constraints, while terms~\eqref{eq:combined_ic_radialxangular} are the cross-integral constraints, given by a formula similar to equations~\eqref{eq:ic_delta_ic} and~\eqref{eq:ic_ic_ic}, with:
\begin{equation}
\begin{split}
\window_{\ell p}^{i,j}(s,\Delta) &= \frac{(2\ell+1)(2p+1)}{(4\pi)^2} \int d\Os \int d\OD \int d^{3} x W(\vx)W(\vx+\vs) \\
& \int d^{3} y W_{i}(\vy) W_{j}(\vy+\vD) \Leg{\ell}(\loss\cdot\hat{\vs}) \Leg{p}(\losD\cdot\hat{\vD}) \epsilon_{i}(\vx,\vy) \epsilon_{j}(\vx+\vs,\vy+\vD).
\end{split}
\end{equation}

As in \Sec{analysis_methods}, window function calculations are based on the survey selection function radially-tuned on one realisation of mock data using the \emph{binned} scheme. As in \Sec{fits_angular}, the same window functions are used to calculate the angular IC correction for both uncontaminated and contaminated \emph{pixelated} mocks.

\begin{figure}[t]
\centering
\includegraphics[width=0.5\textwidth]{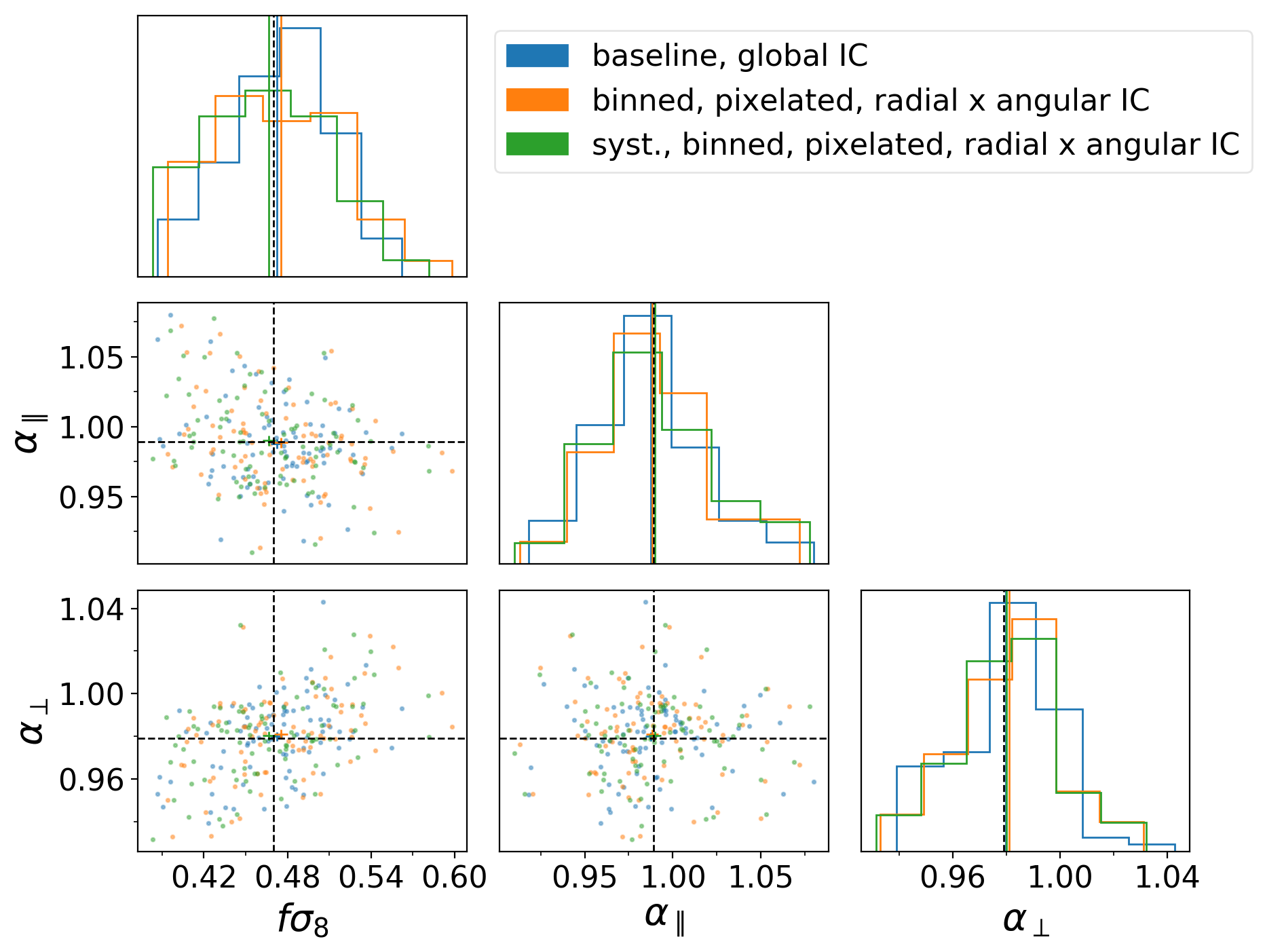}
\includegraphics[width=0.4\textwidth]{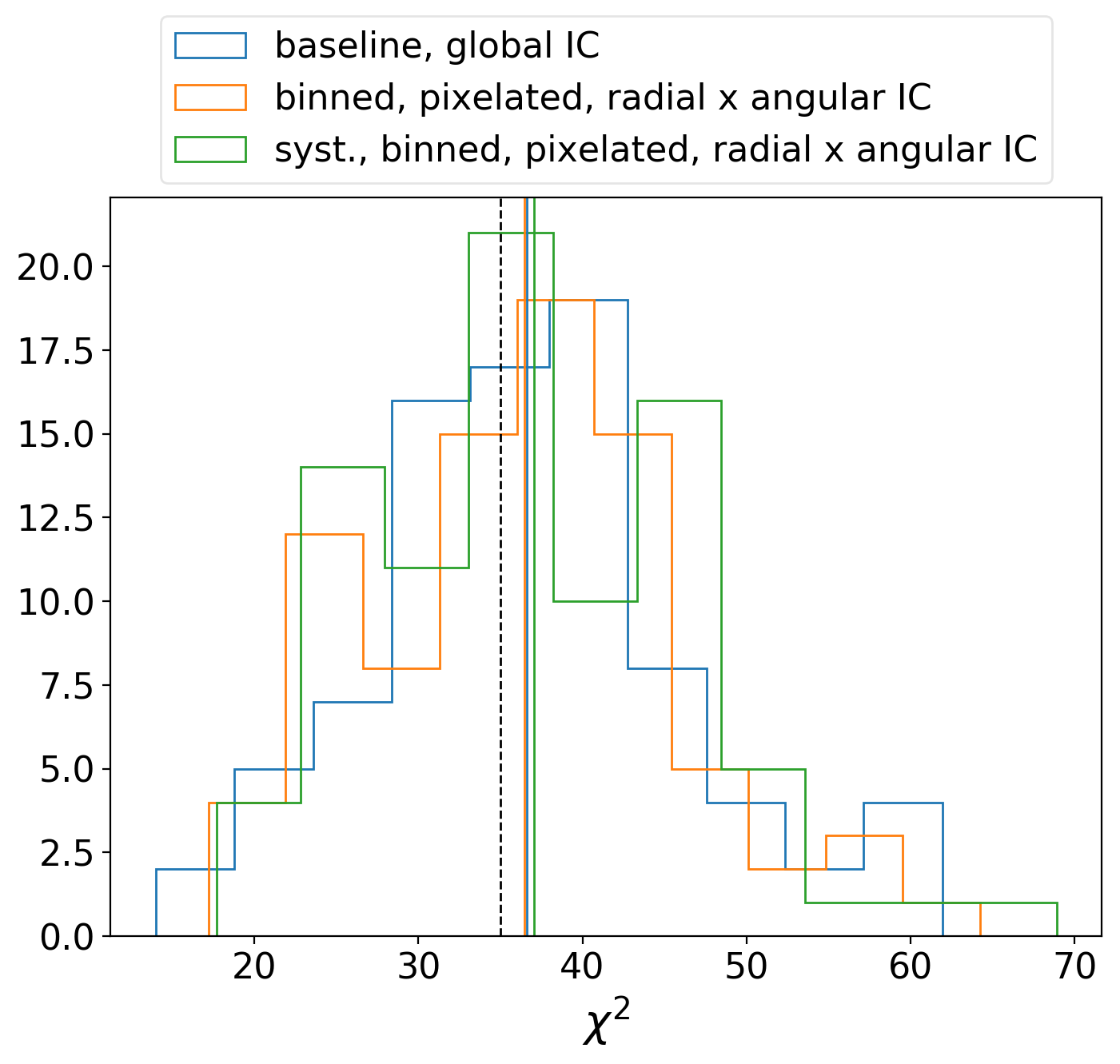}
\caption{Same as \Fig{fits_angular}, with the radial and angular integral constraints combined.}
\label{fig:fits_radialxangular}
\end{figure}

As shown in \Fig{fits_radialxangular} and \Tab{fits_radialxangular}, the combined (radial x angular) integral constraint accounts well for the \emph{binned} and \emph{pixelated} schemes in the mocks: cosmological parameters are recovered within the uncertainty on the mean of the mocks (column~2 in \Tab{fits_radialxangular}). Cosmological fits of contaminated mocks are not further degraded by adding the radial integral constraint on top of the angular one. As in \Sec{fits_angular}, a bias of $12 \%$ to $20 \%$ of the error on a single realisation can be seen on $f\sigma_{8}$ (comparing column~3 to columns~2 and~1).

\begin{table}
\centering
\begin{tabular}{|c|c|c|c|c|}
\hline
& \begin{tabular}{c} Baseline \\ global IC \end{tabular} & \begin{tabular}{c} Binned, pixelated \\ radial x angular IC \end{tabular} & \begin{tabular}{c} Systematics, binned, pixelated \\ radial x angular IC \end{tabular} & Expected\\
\hline
$\apar$ & $0.988 \pm 0.031$ & $0.988 \pm 0.032$ & $0.990 \pm 0.032$ & $0.989$ \\
$\aper$ & $0.980 \pm 0.018$ & $0.981 \pm 0.020$ & $0.980 \pm 0.020$ & $0.979$ \\
$f\sigma_{8}$ & $0.472 \pm 0.037$ & $0.475 \pm 0.044$ & $0.467 \pm 0.043$ & $0.470$ \\
\hline
$\chi^{2}$ & $36.6 \pm 9.7$ & $36.5 \pm 9.3$ & $37.0 \pm 9.6$ & $\mathrm{ndof} = 35$\\
\hline
\end{tabular}
\caption{Mean and standard deviation of the cosmological parameters fitted on the 84 N-series mocks, corresponding to \Fig{fits_radialxangular}. Error bars should be divided by $\sqrt{84} \sim 10$ to obtain errors on the mean of the mocks.}
\label{tab:fits_radialxangular}
\end{table}

\subsection{Caveat: multiplicative systematics}
\label{sec:multiplicative_systematics}

We emphasise that the angular (pixel) integral constraint can only mitigate the additive part of angular systematics. Let us call $c(\vr)$ the contamination signal, assumed constant over a pixel. First, let us suppose $c$ to be purely additive. Applying the \emph{pixelated} scheme, we would measure the power spectrum of the density fluctuations:
\begin{equation}
\delta^{\cic}(\vr) = W(\vr) \left\lbrace 1+\delta(\vr)+c(\vr) \right\rbrace - W(\vr) \frac{\int d^{3} x W(\vx) \left\lbrace 1+\delta(\vx)+c(\vx) \right\rbrace \epsilon_{\ang}(\vr,\vx)}{\int d^{3} x W(\vx) \epsilon_{\ang}(\vr,\vx)}.
\end{equation}
Then, as $c(\vr)$ is constant over a pixel, $c(\vx) \epsilon_{\ang}(\vr,\vx) = c(\vr) \epsilon_{\ang}(\vr,\vx)$ and:
\begin{equation}
\delta^{\cic}(\vr) = W(\vr) \left\lbrace \delta(\vr) - \int d^{3} x W_{\ang}(\vx)\delta(\vx) \epsilon_{\ang}(\vr,\vx) \right\rbrace.
\end{equation}
As expected, $c(\vr)$ disappears from the analysis.

Now, let us consider $c(\vr)$ to be multiplicative (as implemented in our contamination model~\eqref{eq:weight_systematics}):
\begin{subequations}
\begin{align}
\delta^{\cic}(\vr) &= W(\vr) c(\vr) \left\lbrace 1+\delta(\vr) \right\rbrace - W(\vr) \frac{\int d^{3} x W(\vx) c(\vx) \left\lbrace 1+\delta(\vx) \right\rbrace \epsilon_{\ang}(\vr,\vx)}{\int d^{3} x W(\vx) \epsilon_{\ang}(\vr,\vx)} \\
&= W(\vr) c(\vr) \left\lbrace \delta(\vr) - \int d^{3} x W_{\ang}(\vx)\delta(\vx) \epsilon_{\ang}(\vr,\vx) \right\rbrace,
\end{align}
\end{subequations}
i.e. $c(\vr)$ multiplies the selection function $W(\vr)$. However, by definition, $c(\vr)$ is unknow and therefore cannot be taken into account in $W(\vr)$. The resulting multiplicative systematics can explain the observed bias on $f\sigma_{8}$ (compensating for a higher $b_{1}\sigma_{8}$). A possible way to alleviate this effect may consist in estimating the angular survey selection function (including $c(\hat{\vr})$) with the \emph{pixelated} scheme. This would induce a bias which can be estimated using the method presented in \Sec{approaching_true_selection}. We report the reader to e.g.~\cite{Shafer2014:1410.0035v2} for a fully coherent treatment of unknown multiplicative systematics.

We showed that one could mitigate potential, unknown, systematics at the price of a small increase of statistical errors by applying an angular integral constraint in the data. Such a technique may be helpful to reduce the impact of poorly understood angular target density variations with photometric conditions, or can be considered as a regular consistency test to check for remaining angular systematics.

\section{Conclusions}
\label{sec:conclusions}

This paper revisited the notion of integral constraints which we showed to be useful in clustering analyses to account for biases related to calibrating the survey selection function partly on data.

We presented a general formalism to account for integral constraints that completes the existing models for the global integral constraint~\cite{Peacock1991:10.1093/mnras/253.2.307,Beutler2014:1312.4611v2,Wilson2015:1511.07799v2}. We discussed the window function normalisation and the shot noise contribution to the integral constraints and included wide-angle effects following~\cite{Beutler2018:1810.05051v2}. Our formalism was used to model the radial integral constraint effect induced by inferring the radial selection function from actual data.

We indeed noticed that the common practice of drawing random redshifts based on data may significantly bias large-scale clustering measurements. This effect can be particularly large compared to the window function effect if the survey is composed of several patches whose radial selection functions must be treated separately. In particular, its impact is likely to be large in analyses focusing on large scales, e.g. dedicated to primordial non-Gaussianity~\cite{Ross2012:1208.1491v3,Castorina2019}.

Applying our modelling of the radial integral constraint to a RSD analysis, the bias induced on cosmological parameters was shown to be successfully reduced. Though more statistics would be required to push our validation further, the potentially remaining bias is already below the expected statistical uncertainty of future galaxy surveys.

As a further application, we showed that we could similarly apply an angular integral constraint to help mitigating unknown angular systematics. This angular integral constraint can be combined to the radial one, as required if the radial selection function is also estimated from the data. Though ony additive systematics can be fully accounted for, our scheme performs well enough with multiplicative systematics on a BOSS CMASS-like survey.

Further developments may concern the use of a fully analytic selection function, as it can often be split into radial and angular components. This would enable a faster estimation of the window functions required in our analysis.

We noted that a potential bias can emerge when the selection function used in the window function calculations for integral constraint corrections is estimated from the data itself. We suggested a way to estimate this bias, which remained very subdominant in our analysis. A workaround for future analyses may consist in predicting the radial selection function from first principles (e.g. from the luminosity function), without relying on the observed data, while marginalising on possible unknowns in the cosmological fits.

\acknowledgments

We are very grateful to the anonymous referee for their useful comments. We thank Etienne Burtin, Will J. Percival and Ashley J. Ross for encouraging us in our study of the radial integral constraint. We are grateful to Florian Beutler and Emanuele Castorina for discussions about wide-angle corrections.
Power spectrum multipoles were measured using \texttt{nbodykit}~\cite{Hand2017:1712.05834v1}. Minimisations were performed with the \texttt{MINUIT} algorithm~\cite{Minuit1975} via the iminuit~\cite{iminuit} Python interface. Some of our results required the use of the \texttt{HEALPix} package~\cite{Gorski2005:astro-ph/0409513v1}. 
We acknowledge support from the P2IO LabEx (ANR-10-LABX-0038) in the framework "Investissements d'Avenir" (ANR-11-IDEX-0003-01) managed by the Agence Nationale de la Recherche (ANR, France).


\bibliographystyle{JHEP}
\bibliography{references}{}



\newpage

\appendix

\section{FKP weights}
\label{app:FKP_weights}

We apply FKP weights~\cite{Feldman1993:astro-ph/9304022v1} to both mock data and randoms. They are calculated according to:
\begin{equation}
w_{\mathrm{FKP}} = \frac{1}{1+n(z)P_{0}},
\end{equation}
using $P_{0} = 20000 \Mpchc$, which should be representative of the value of the power spectrum at the scales of interest for the clustering analysis. These weights require an estimation of the true redshift density $n(z)$ in absence of clustering. If $n(z)$ is computed from the data itself, using narrow bins in $z$, clustering overdensities may be smoothed out along the line-of-sight.

\begin{figure}[t]
\centering
\includegraphics[width=0.45\textwidth]{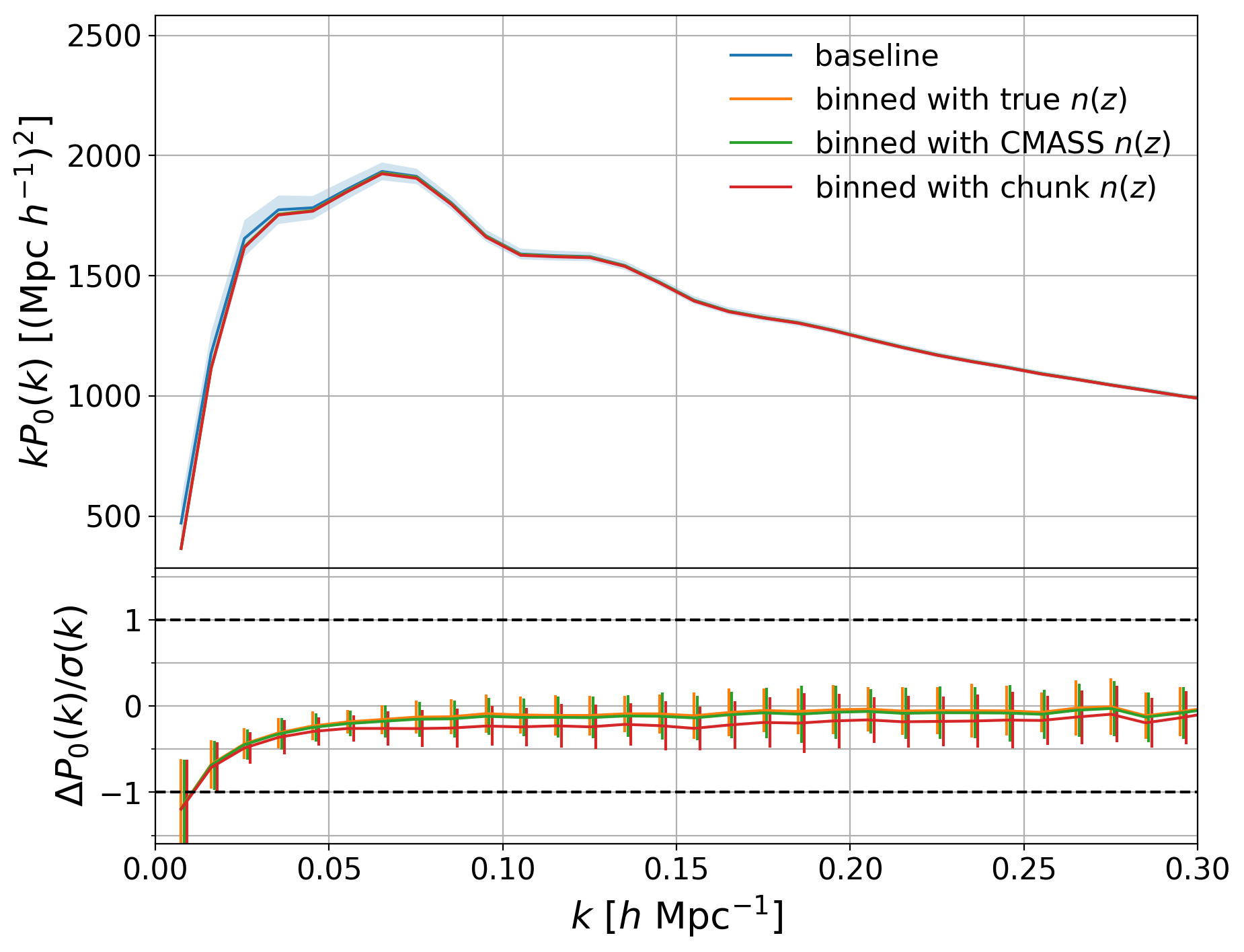}
\includegraphics[width=0.45\textwidth]{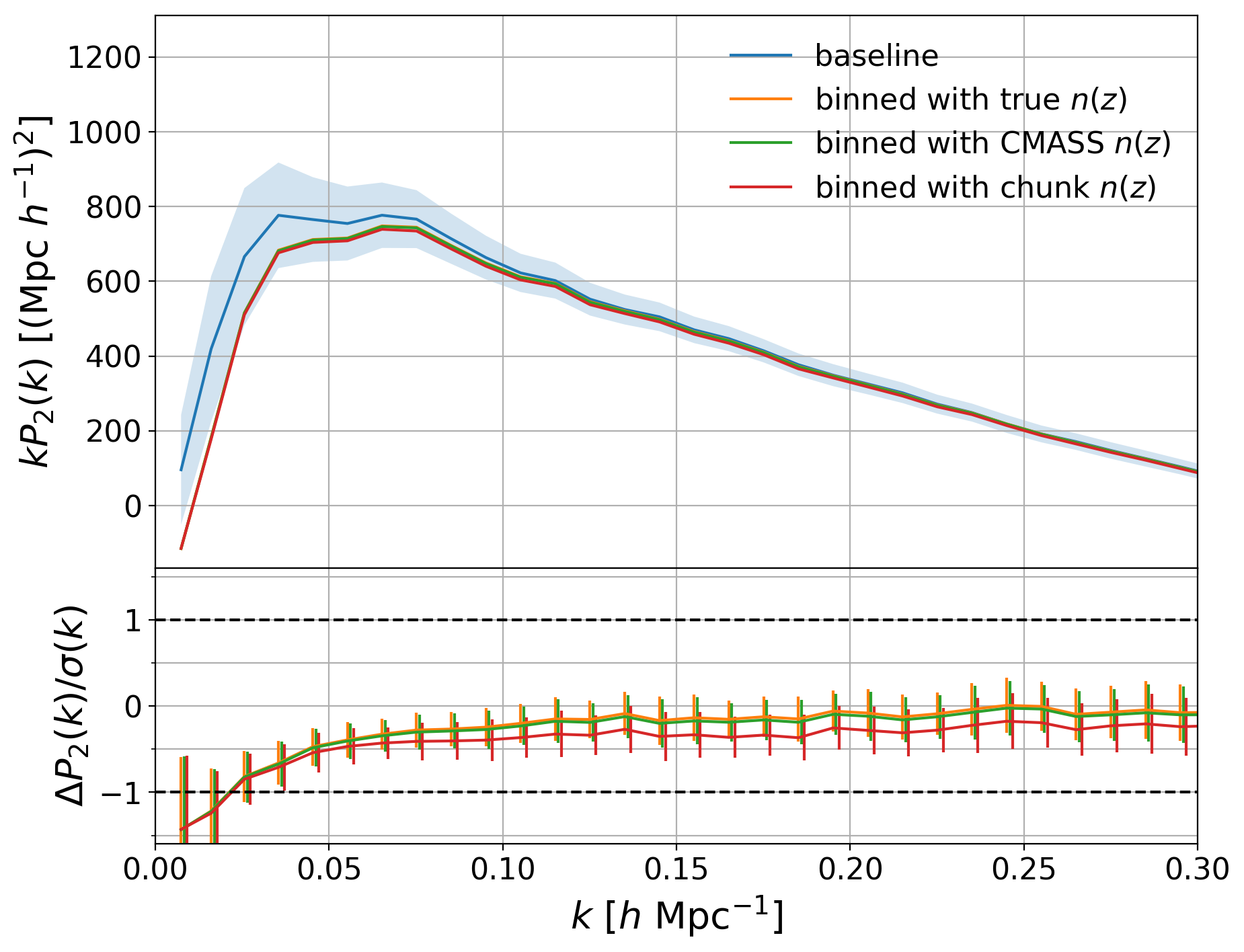}
\includegraphics[width=0.45\textwidth]{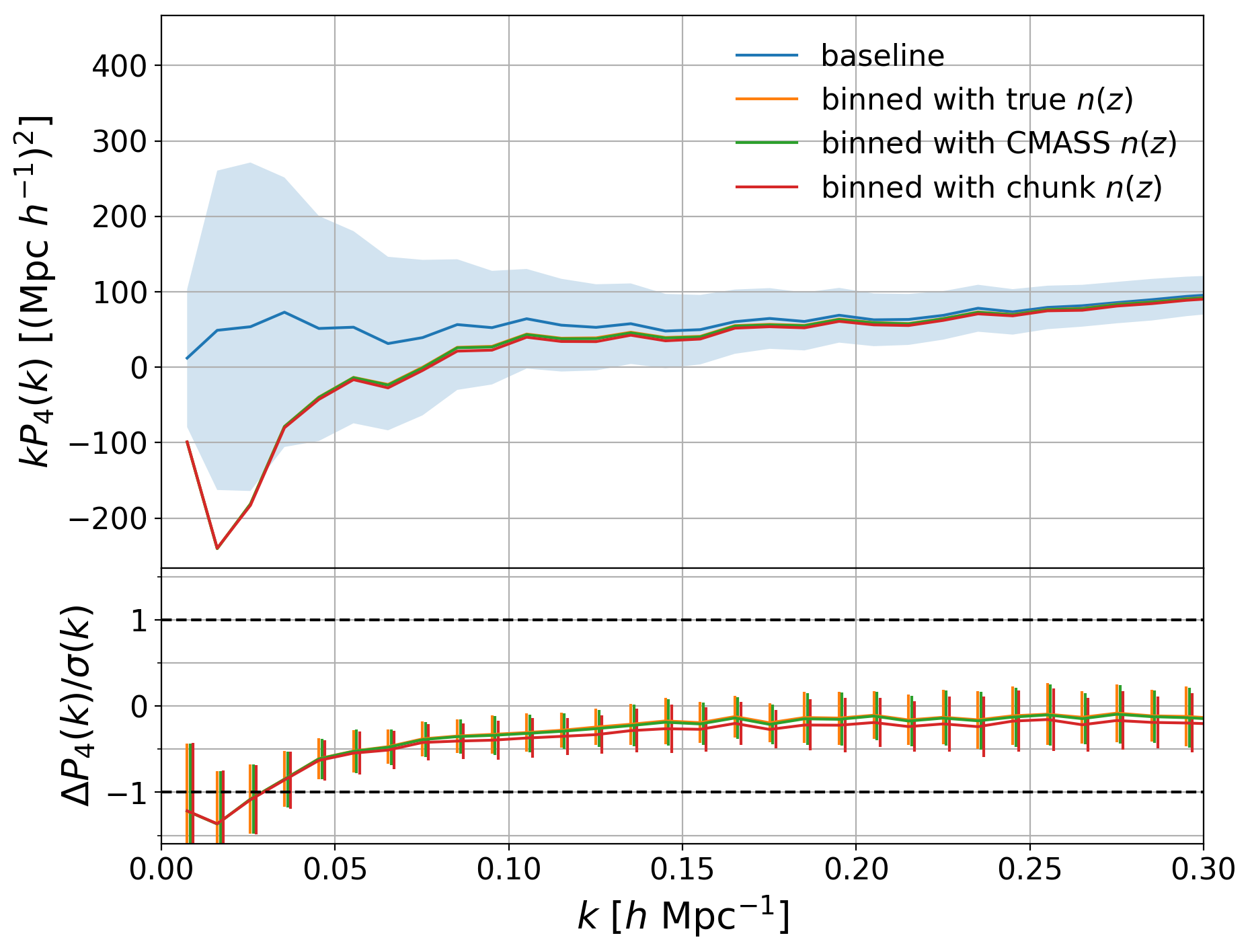}
\caption{Top panels: power spectrum multipoles (upper left: monopole, upper right: quadrupole, bottom: hexadecapole) obtained using the \emph{binned} scheme and different $n(z)$ estimations. In orange, we use the true density $n(z)$ from the radial selection function (rescaled for each realisation of the mocks, as explained at the end of \Sec{mocks}). In green, $n(z)$ is estimated from each full CMASS mock realisation. In red, $n(z)$ is measured from each mock realisation, in the $6$ chunks separately. We recall the \emph{baseline} in blue, obtained with the true selection function. The blue shaded area represents the standard deviation of the mocks. Bottom panels: difference of the \emph{binned} scheme with the different (true, full CMASS, chunk) $n(z)$ estimations to the \emph{baseline}, with the standard deviation of the difference given by the error bars, normalised by the standard deviation of the mocks.}
\label{fig:mocks_density}
\end{figure}

As mentioned in \Sec{mocks}, the redshift density $n(z)$ is computed in bins $\Delta z = 0.005$. Figure~\ref{fig:mocks_density} shows power spectrum measurements obtained with the \emph{binned} scheme, with different $n(z)$ estimations. Measuring $n(z)$ from the full CMASS sample induces a negligible bias with respect to taking the true $n(z)$. However, measuring $n(z)$ separately in each of the $6$ chunks dividing the CMASS footprint, as would be natural to do, leads to an additional bias on all scales of all multipoles. Though the density $n(z)$ also enters the power spectrum normalisation (see \Eq{power_spectrum_normalisation}), the main effect comes from the smoothing of the clustering along the line-of-sight, as can be seen from the loss of power in the quadrupole and hexadecapole. This effect is multiplicative and would be difficult to take into account.

A simple way to prevent FKP weights from biasing clustering measurements would be to fit a simple spline to the redshift distribution in wide redshift bins to reduce the correlation between $w_{\mathrm{FKP}}$ and the density field. As stated in \Sec{mocks}, we choose for simplicity to estimate $n(z)$ from each data realisation, using the full CMASS sample (instead of the $6$ different chunks), making the bias from $w_{\mathrm{FKP}}$ almost invisible. 

\section{Modelling wide-angle effects}
\label{app:wide-angle}

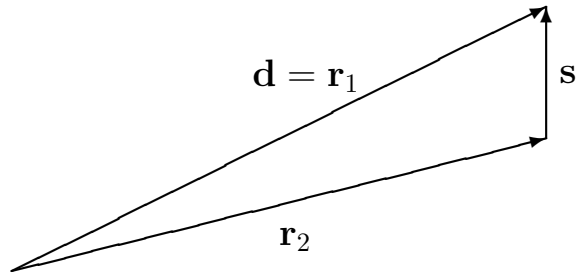
\begin{figure}
\begin{center}
\begin{picture}(210,100)
\Large
\thicklines
\put(0,0){\vector(4,1){200}}
\put(0,0){\vector(2,1){200}}
\put(200,50){\vector(0,1){50}}
\put(90,70){$\vd=\vrn1$}
\put(100,10){$\vrn2$}
\put(205,71){$\vs$}
\end{picture}
\end{center}
\caption{The assumed geometry and angles. The two galaxies lie at $\vrn1$ and $\vrn2$, with the separation vector $\vs=\vrn1-\vrn2$. The end-point line-of-sight is used: $\vd = \vrn1$.}
\label{fig:line_of_sight}
\end{figure}

Wide-angle effects arise when a single line-of-sight must be chosen for a galaxy pair~\cite{Castorina2017:1709.09730v2}. They can be accounted for by expanding both correlation and survey window functions in powers of $x=s/d$, with $s$ the separation of a pair of galaxies and $d$ its comoving distance to the observer. To allow a considerable reduction of computation time, the Yamamoto power spectrum estimator~\cite{Yamamoto2005:astro-ph/0505115v2} uses one galaxy of a pair as line-of-sight: the so-called end-point line-of-sight (see \Fig{line_of_sight}). As shown in~\cite{Beutler2018:1810.05051v2}, this choice leads to detectable wide-angle effects, mainly in odd multipoles.
The window-convolved power spectrum multipoles with respect to the end-point ($\mathrm{ep}$) line-of-sight can be written~\cite{Beutler2018:1810.05051v2}:

\begin{equation}
P_{\ell}^{\mathrm{c}}(k) = 4 \pi (-i)^{\ell} \int s^{2} ds j_{\ell}(ks) \sum_{p,q,n} A_{\ell p}^{q} \frac{2\ell+1}{2q+1} s^{n} \xi_{p}^{\ep,(n)}(s) \window_{q}^{\delta,\delta,\ep,(n)}(s)
\label{eq:power_c_ep}
\end{equation}
where:\footnote{Note that we now use the same triangle definition as in~\cite{Castorina2017:1709.09730v2} (see \Fig{line_of_sight}) $\vs = \vrn1-\vrn2$, which explains the minus signs in the selection functions.}
\begin{equation}
\window_{\ell}^{\delta,\delta,\ep,(n)}(s) = \frac{2\ell+1}{4\pi} \int d\Os \int d^{3} x x^{-n} W(\vx)W(\vx-\vs) \Leg{\ell}(\hat{\vx}\cdot\hat{\vs}).
\end{equation}
$\xi_{p}^{\ep,(n)}$ can be predicted within linear pertubation theory~\cite{Castorina2017:1709.09730v2,Beutler2018:1810.05051v2}.

Similar expansions can be derived for $\tilde{IC}_{\ell}^{i,j}(k), \, (i,j) \in \{(\delta,\ic),(\ic,\delta),(\ic,\ic)\}$. 
The first cross-term $(\delta,\ic)$ reads:
\begin{equation}
\begin{split}
\tilde{IC}_{\ell}^{\delta,\ic}(k) & = \frac{2\ell+1}{4\pi} \int d\Ok \int d^{3}\rn1 \int d^{3}\rn2 \int d^{3}\rn3 e^{-i\vk\cdot(\vrn1-\vrn2)} \aver{\delta(\vrn1)\delta(\vrn3)} \\ 
& W(\vrn1) W(\vrn2) W_{\ic}(\vrn3) \Leg{\ell}(\hat{\vk}\cdot\losrn1) \epsilon_{\ic}(\vrn2,\vrn3)
\end{split}
\end{equation}
with:
\begin{equation}
\aver{\delta(\vrn1)\delta(\vrn3)} = \sum_{n,p} \left(\frac{\Delta}{\rn1}\right)^{n} \xi_{p}^{\ep,(n)}(\Delta) \Leg{p}(\losrn1\cdot\hat{\vD}), \qquad \vD = \vrn1 - \vrn3.
\end{equation}
Using the Rayleigh plane wave expansion:
\begin{equation}
e^{-i\vk\cdot\vs} = \sum_{q=0}^{\infty} (-i)^{q} (2q+1) j_{q}(ks) \Leg{q}(\hat{\vk}\cdot\hat{\vs})
\end{equation}
and
\begin{equation}
\int \frac{d\Ok}{4\pi} \Leg{\ell}(\hat{\vk}\cdot\losrn1) \Leg{q}(\hat{\vk}\cdot\hat{\vs}) = \frac{\delta_{\ell q}}{2\ell+1} \Leg{\ell}(\losrn1\cdot\hat{\vs}),
\end{equation}
one gets:
\begin{equation}
\begin{split}
\tilde{IC}_{\ell}^{\delta,\ic}(k) &= (2\ell+1) (-i)^{\ell} \int d^{3}s j_{\ell}(ks) \sum_{p,n} \int d^{3}\Delta \Delta^{n} \xi_{p}^{\ep,(n)}(\Delta) \\
& \int d^{3}\rn1 \rn1^{-n} W(\vrn1) W(\vrn1-\vs) W_{\ic}(\vrn1-\vD) \Leg{\ell}(\losrn1\cdot\hat{\vs}) \Leg{p}(\losrn1\cdot\hat{\vD})\epsilon_{\ic}(\vrn1-\vs,\vrn1-\vD)
\end{split}
\end{equation}
with $\vs = \vrn1 - \vrn2$. $\tilde{IC}_{\ell}^{\delta,\ic}(k)$ is simply equal to the Hankel transform of a term similar to \Eq{ic_delta_ic}:
\begin{equation}
\tilde{IC}_{\ell}^{i,j}(k) = 4 \pi (-i)^{\ell} \int s^{2} ds j_{\ell}(ks) \int \Delta^{2} d\Delta \sum_{p,n} \frac{4 \pi}{2p+1} \Delta^{n} \xi_{p}^{\ep,(n)}(\Delta) \window_{\ell p}^{i,j,(n)}(s,\Delta)
\label{eq:ic_ep_ij}
\end{equation}
with $(i,j)=(\delta,\ic)$, if we define:
\begin{equation}
\begin{split}
\window_{\ell p}^{\delta,\ic,(n)}(s,\Delta) & = \frac{(2\ell+1)(2p+1)}{(4\pi)^2} \int d\Os \int d\OD \\
& \int d^{3} x x^{-n} W(\vx)W(\vx-\vs)W_{\ic}(\vx-\vD) \Leg{\ell}(\hat{\vx}\cdot\hat{\vs}) \Leg{p}(\hat{\vx}\cdot\hat{\vD}) \epsilon_{\ic}(\vx-\vs,\vx-\vD).
\end{split}
\label{eq:window_ep_delta_ic}
\end{equation}
The second cross-term $(\ic,\delta)$ is not trivially equal to the first one for even multipoles. Indeed:
\begin{equation}
\begin{split}
\tilde{IC}_{\ell}^{\ic,\delta}(k) & = \frac{2\ell+1}{4\pi} \int d\Ok \int d^{3}\rn1 \int d^{3}\rn2 \int d^{3}\rn3 e^{-i\vk\cdot(\vrn1-\vrn2)} \aver{\delta(\vrn2)\delta(\vrn3)} \\ 
& W(\vrn1) W(\vrn2) W_{\ic}(\vrn3) \Leg{\ell}(\hat{\vk}\cdot\losrn1) \epsilon_{\ic}(\vrn1,\vrn3)
\end{split}
\end{equation}
gives:
\begin{equation}
\begin{split}
\tilde{IC}_{\ell}^{\ic,\delta}(k) & = (2\ell+1) (-i)^{\ell} \int d^{3}s j_{\ell}(ks) \sum_{p,n} \int d^{3}\Delta  \Delta^{n} \xi_{p}^{\ep,(n)}(\Delta) \\
& \int d^{3} \rn2 \rn2^{-n} W(\vrn2+\vs) W(\vrn2) W_{\ic}(\vrn2-\vD) \Leg{\ell}(\losrn1\cdot\hat{\vs}) \Leg{p}(\losrn2\cdot\hat{\vD})\epsilon_{\ic}(\vrn2+\vs,\vrn2-\vD)
\end{split}
\end{equation}
with $\vs = \vrn1 - \vrn2$, $\vD = \vrn2 - \vrn3$. Taking the opposite $\vs \rightarrow -\vs$ and defining:
\begin{equation}
\begin{split}
\window_{\ell p}^{\ic,\delta,(n)}(s,\Delta) & = \frac{(2\ell+1)(2p+1)}{(4\pi)^2} \int d\Os \int d\OD \\
& \int d^{3} x x^{-n} W(\vx)W(\vx-\vs)W_{\ic}(\vx-\vD) \Leg{\ell}(\widehat{\vx-\vs}\cdot(-\hat{\vs})) \Leg{p}(\hat{\vx}\cdot\hat{\vD}) \epsilon_{\ic}(\vx-\vs,\vx-\vD)
\end{split}
\label{eq:window_ep_ic_delta}
\end{equation}
we obtain the integral constraint correction \Eq{ic_ep_ij}, with $(i,j)=(\ic,\delta)$. Let us move to the last term $(\ic,\ic)$:
\begin{equation}
\begin{split}
\tilde{IC}_{\ell}^{\ic,\ic}(k) & = \frac{2\ell+1}{4\pi} \int d\Ok \int d^{3}\rn1 \int d^{3}\rn2 \int d^{3}\rn3 \int d^{3}\rn4 e^{-i\vk\cdot(\vrn1-\vrn2)} \aver{\delta(\vrn3)\delta(\vrn4)} \\
& W(\vrn1) W(\vrn2) W_{\ic}(\vrn3) W_{\ic}(\vrn4) \Leg{\ell}(\hat{\vk}\cdot\losrn1) \epsilon_{\ic}(\vrn1,\vrn3) \epsilon_{\ic}(\vrn2,\vrn4)
\end{split}
\end{equation}
gives:
\begin{equation}
\begin{split}
\tilde{IC}_{\ell}^{\ic,\ic}(k) & = (2\ell+1) (-i)^{\ell} \int d^{3}s j_{\ell}(ks) \sum_{p,n} \int d^{3}\Delta \Delta^{n} \xi_{p}^{\ep,(n)}(\Delta) (2\ell+1) \int d^{3}\rn1 W(\vrn1) W(\vrn1-\vs) \\
& \int d^{3}\rn3 \rn3^{-n} W_{\ic}(\vrn3) W_{\ic}(\vrn3-\vD)
\Leg{\ell}(\losrn1\cdot\hat{\vs}) \Leg{p}(\losrn3\cdot\hat{\vD})\epsilon_{\ic}(\vrn1,\vrn3) \epsilon_{\ic}(\vrn1-\vs,\vrn3-\vD)
\end{split}
\end{equation}
with $\vs = \vrn1 - \vrn2$, $\vD = \vrn3 - \vrn4$. This is \Eq{ic_ep_ij} with $(i,j)=(\ic,\ic)$ if we define:
\begin{equation}
\begin{split}
\window_{\ell p}^{\ic,\ic,(n)}(s,\Delta) & = \frac{(2\ell+1)(2p+1)}{(4\pi)^2} \int d\Os \int d\OD \int d^{3} y W(\vy)W(\vy-\vs) \\
& \int d^{3} x x^{-n} W_{\ic}(\vx) W_{\ic}(\vx-\vD) \Leg{\ell}(\hat{\vy}\cdot\hat{\vs}) \Leg{p}(\hat{\vx}\cdot\hat{\vD}) \epsilon_{\ic}(\vy,\vx) \epsilon_{\ic}(\vy-\vs,\vx-\vD).
\end{split}
\label{eq:window_ep_ic_ic}
\end{equation}
In particular, results for the global integral constraint are obtained by replacing $\ic \rightarrow \glo$ and $\epsilon_{\ic} \rightarrow 1$. Then, \Eq{ic_glo_glo} is recovered at any order $n$ if one takes $\window_{\ell}^{\delta,\delta,(0)}(s)$ for $\window_{\ell}^{\delta,\delta}(s)$, as stressed out by~\cite{Beutler2018:1810.05051v2}.

Contrary to the line-of-sight definition in $\window_{\ell}^{\delta,\delta,\ep,(n)}(s)$ (and the first line-of-sight in $\window_{\ell p}^{i,j,(n)}(s,\Delta)$), which should be the same as in the power spectrum (or correlation function) estimator, the line-of-sight connecting $\xi_{p}^{\ep,(n)}(\Delta)$ to $\window_{\ell p}^{i,j,(n)}(s,\Delta)$ is a purely practical choice. Our calculations use the end-point line-of-sight, but the derivation with any line-of-sight $\vd$ is straightforward by replacing $x^{-n}$ in eq.~\eqref{eq:window_ep_delta_ic}, \eqref{eq:window_ep_ic_delta} and~\eqref{eq:window_ep_ic_ic} by $d^{-n}$, changing the arguments of Legendre polynomials accordingly and taking the corresponding $\xi_{p}^{\vd,(n)}(\Delta)$ in \Eq{ic_ep_ij}. In particular, taking the mid-point line-of-sight makes first-order ($n=1$) wide-angle corrections vanish.

\begin{figure}[t]
\centering
\includegraphics[width=0.45\textwidth]{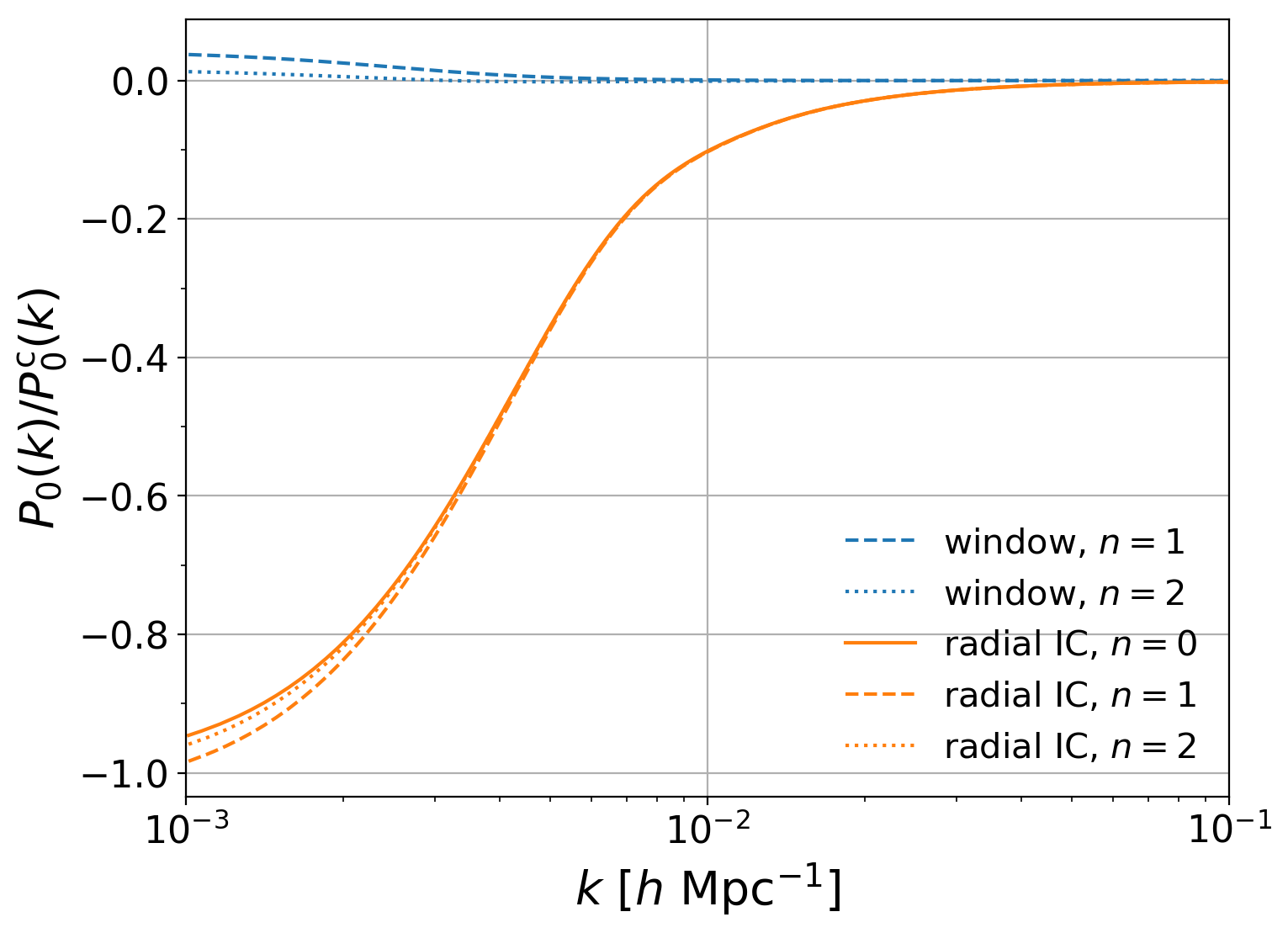}
\includegraphics[width=0.45\textwidth]{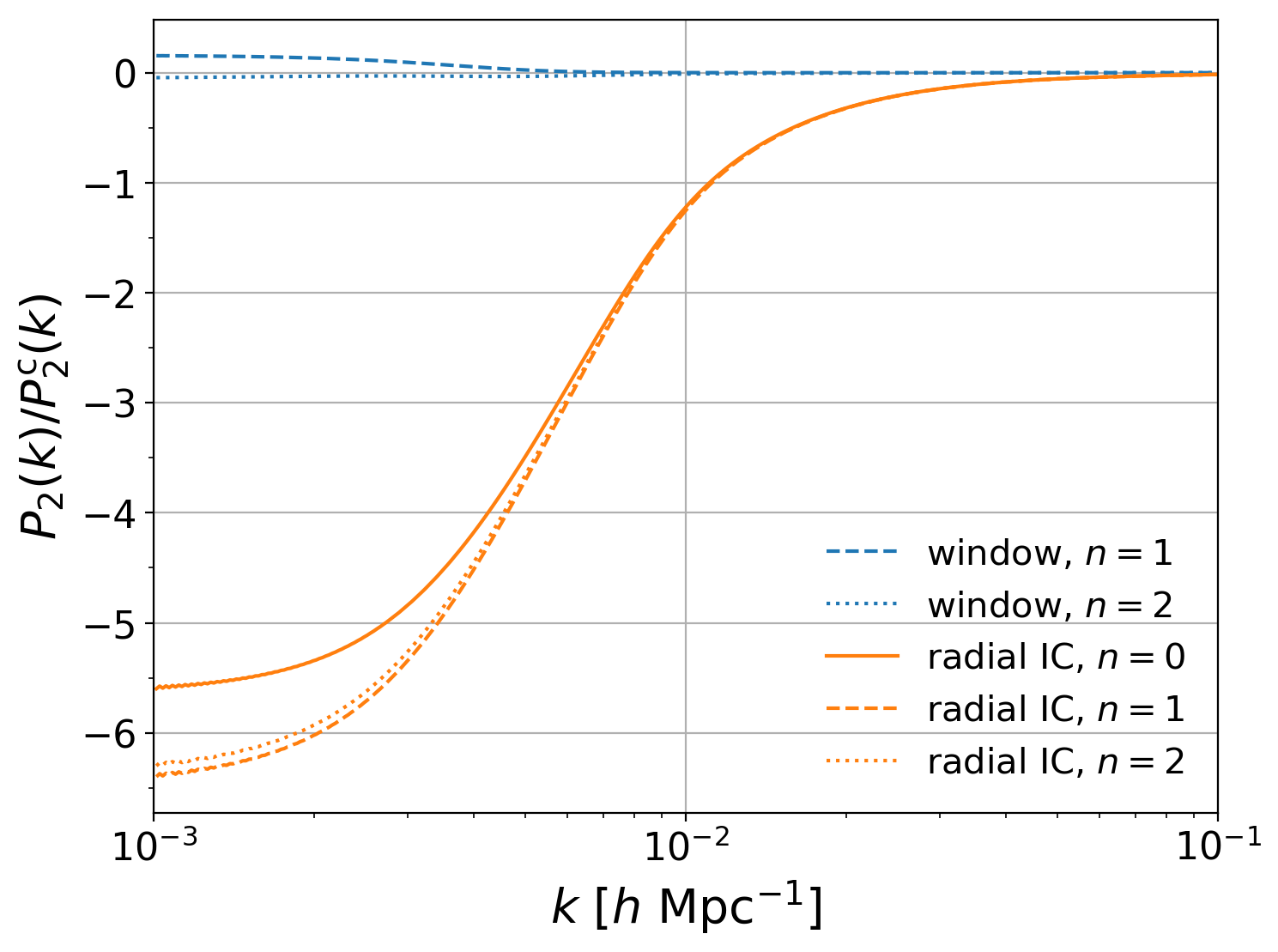}
\includegraphics[width=0.45\textwidth]{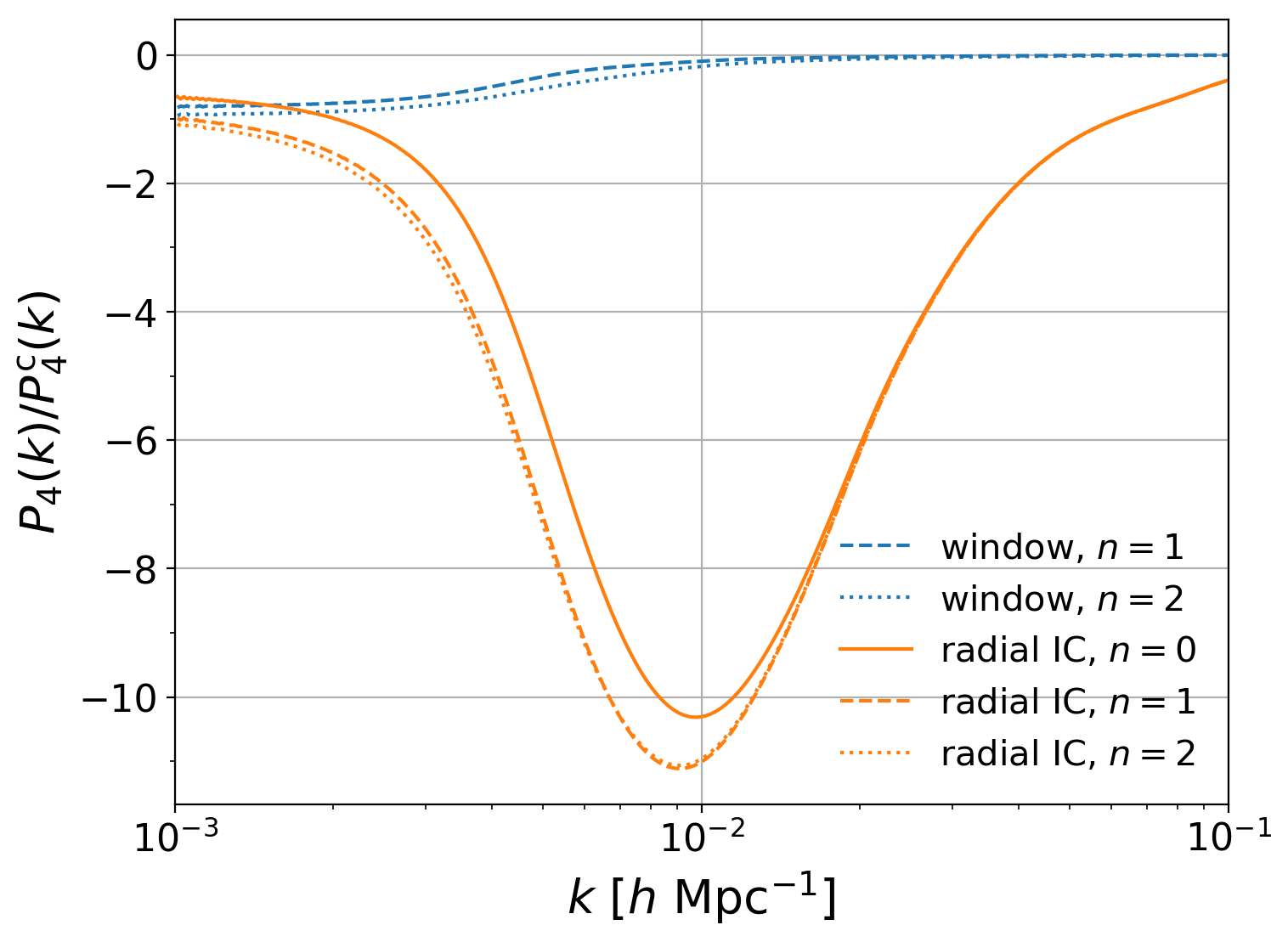}
\caption{Ratio of wide-angle corrections to the convolved power spectrum multipoles $P_{\ell}^{\mathrm{c}}(k)$ (upper left: monopole, upper right: quadrupole, bottom: hexadecapole) at zeroth order. In blue are shown wide-angle contributions to the convolved power spectrum monopole up to order $n=1,2$ (dashed and dotted lines respectively). Corrections for the radial integral constraint up to order $n=0,1,2$ (continuous, dashed and dotted lines respectively) are plotted in orange.}
\label{fig:wide-angle_corrections}
\end{figure}

For clarity, we have used the end-point line-of-sight as second line-of-sight in $\window_{\ell p}^{i,j,(n)}(s,\Delta)$ throughout the paper. With this choice, by definition, the integral constraint corrections completely cancel with the convolved power spectrum monopole $P_{0}^{\mathrm{c}}(k)$ on large scales. This would only be asymptotically true (as the wide-angle correction order $n \rightarrow \infty$) if a different line-of-sight definition was used in $\window_{\ell}^{\delta,\delta,(n)}(s)$ and $\window_{\ell p}^{i,j,(n)}(s,\Delta)$. We show wide-angle contributions to the convolved power spectrum multipoles and the radial integral constraint in \Fig{wide-angle_corrections}. They are significant for $k \lesssim 10^{-2} \hMpc$. However, they remain small compared to the radial integral constraint correction. Then, for simplicity, we do not include these wide-angle corrections ($n \ge 1$) in our analysis.
However, wide-angle contributions may dominate over radial integral constraint corrections for a large survey with a constant radial selection function.

\end{document}